\documentclass[a4paper,11pt]{article}
\usepackage{jheppub} 
\usepackage[T1]{fontenc} 
\usepackage{lmodern}
\usepackage{mathrsfs}
\usepackage{multirow}
\usepackage{array}

\title{\boldmath Two body final states production in electron-positron annihilation and their contributions to $(g-2)_{\mu}$}

\author[a,b]{Shi-Jia Wang,}
\emailAdd{wsj107625@hnu.edu.cn}
\author[a,b,1]{Zhen Fang }
\emailAdd{zhenfang@hnu.edu.cn}
\author[a,b,1]{and Ling-Yun Dai \note{Corresponding author.}}
\emailAdd{dailingyun@hnu.edu.cn}

\affiliation[a]{School of Physics and Electronics, Hunan University, Changsha 410082, China}
\affiliation[b]{Hunan Provincial Key Laboratory of High-Energy Scale Physics and Applications, Hunan University, Changsha 410082, China}

\abstract{In this paper, we study the processes of $e^{+}e^{-}$ annihilation into two body final states, either two pseudoscalar mesons or one meson with a photon. The hadronic vacuum polarization form factors are calculated within the framework of resonance chiral theory in the energy region of $E \lesssim 2$ GeV, with final state interactions taken into account. 
A joint analysis on the processes of $e^{+}e^{-} \rightarrow \pi^{+}\pi^{-}$, $K^{+}K^{-}$, $K_{L}^{0}K_{S}^{0}$, $\pi^{0}\gamma$, and $\eta\gamma$ has been performed, and the latest experimental data are included. Based on the vacuum polarization form factors of these processes, we estimate their contributions to the lowest order of anomalous magnetic moment of the muon, $(g-2)_\mu$. Combined with other contributions from hadronic vacuum polarization and other interactions from the standard model, the discrepancy between theoretical prediction and experimental measurement is $\Delta a_{\mu}=(24.1\pm5.4)\times 10^{-10}$, i.e., 4.5$\sigma$. }

\begin{document}
\allowdisplaybreaks[4]
\maketitle
\flushbottom

\section{Introduction}
\label{sec:intro}

Quantum Chromodynamics (QCD) is the fundamental theory of the strong interaction of quarks and gluons, which plays an essential role in studying hadron dynamics. However, because of rapid increase of the running coupling constants at long distance and color confinement, applying QCD to tackle the low-energy interactions of hadrons is challenging. Effective field theories (EFT) of QCD are proposed to study the dynamics of hadrons in the low-energy region. Chiral perturbation theory (ChPT) \cite{Weinberg:1978kz,Gasser:1983yg} is such an EFT that works in the low-energy region ($E \ll M_{\rho}$) without resonance states appearing. The pseudoscalar mesons are filled in an octet representation according to the chiral symmetry breaking of QCD. It supplies a powerful theoretical framework \cite{Weinberg:1978kz,Gasser:1983yg} to deal with low-energy hadron decays and scatterings, e.g., $\pi\pi$ scatterings and weak pion decays. 
Nevertheless, ChPT is a nonrenormalizable theory, as its power counting is based on momenta and masses of pseudoscalars. As a result, the number of unknown couplings will increase if one wants to refine the analysis by considering higher-order corrections. This makes it difficult to predict reliably in the high-energy region. Therefore, in the middle energy region where resonances appear ($M_{\rho} \leq E \leq 2$ GeV), neither QCD nor ChPT works. Resonance chiral theory (RChT) \cite{Ecker:1988te, Ecker:1989yg,Cirigliano:2006hb,Kampf:2006yf,Portoles:2010yt,Kampf:2011ty} is proposed to study physics in this middle-energy region, with the resonances included as new dynamical degrees of freedom of the EFT and filled in the $SU(3)$ octets.
The power counting is accomplished by large $N_{c}$ (the number of colors) expansions.  Also, one should keep in mind that, upon resonance integration, ChPT is recovered. The chiral countings on the light pseudoscalars would be restored, and the chiral low energy constants (LECs) shift their values between these two theories. Besides, the unknown couplings in RChT can be fixed by matching the Green functions of RChT with that of QCD in the high-energy region and by fitting the experimental data.

In 2021, the Fermilab National Accelerator Laboratory (FNAL) presented its first results of the anomalous magnetic moment for the positive muon ($a_{\mu} \equiv (g-2)_\mu/2$). Combined with the previous measurements of both muon and antimuon \cite{Charpak:1962zz,Bailey:1968rxd,BAILEY19791,Muong-2:2006rrc}, the new experimental average is $a_{\mu}({\rm Exp})=116592061(41)\times 10^{-11}$ (0.35 ppm) \cite{Muong-2:2021ojo}. This increases the tension between theoretical prediction from the standard model (SM) and experimental measurement, with a discrepancy of 4.2~$\sigma$, implying the emergence of new physics beyond standrad model (SM). 
The SM prediction can be separated into four parts: contributions from Quantum Electrodynamics (QED), electroweak interactions (EW), hadronic vacuum polarization (HVP), and hadronic light-by-light scatterings (HLBL). The contributions from QED and EW dominate but have only minor uncertainties \cite{Aoyama:2020ynm}, and the primary source of uncertainty is from the contributions of hadron interactions, HVP and HLBL. HVP is related to the $e^{+}e^{-}$ annihilation with the production of pseudoscalar mesons, giving the most significant hadronic contribution. The high-energy part ($E \geq 2$ GeV) can be estimated from perturbative QCD, while the low-energy part is difficult to be fixed.  In the past decades, data-driven method \cite{Colangelo:2018mtw,Davier:2019can,Keshavarzi:2019abf}  gave an overall estimation of the HVP  by focusing on the data directly\footnote{Notice that for the HVP contributions, the prediction from lattice QCD \cite{Ce:2022kxy,Borsanyi:2020mff,Alexandrou:2022amy} and that from data-driven method have a significant discrepancy. The reason is still unknown. Besides, the HVP contributions estimated from $\tau$ decays would be much closer to that of lattice QCD \cite{Miranda:2020wdg}.  }, but it lacks a systematic theoretical tool to deal with interactions of hadrons. For recent progress trying to refine HVP contributions, we refer to Refs.~\cite{Hoid:2020xjs,Benayoun:2021ody,Yi:2021ccc,Hoferichter:2021wyj,Colangelo:2022jxc}. Furthermore, as has been pointed out by Ref.~\cite{Qin:2020udp}, the most important contributions come from the $\rho,\omega,\phi$ region, and there are not enough high statistics datasets in these energy regions. Some of the datasets even contradict each other.  Thus, it is necessary to construct a theoretical tool that can extend the EFT to the middle energy region ($M_{\rho} \leq E \leq 2$ GeV) and fits the datasets well. This is realized by generalizing the RChT to the higher-energy region with heavier resonances included \cite{Dai:2013joa, Qin:2020udp}. Here we will follow the strategy and include more processes to refine our analysis of almost all the contributions with two body final states.
It is worth pointing out that a similar approach via the hidden local symmetry type of Lagrangians is proposed and applied in phenomenology studies up to $E=1.05$~GeV. See e.g., Refs.~\cite{Benayoun:2012wc,Benayoun:2015gxa}. Note that the pion vector form factor is studied within RChT in Refs.~\cite{Rosell:2004mn,Masjuan:2008fv}, too.

As is known, the processes of $e^{+}e^{-}$ annihilation into two-body final states contribute most to the HVP. Therefore, we will focus on the processes of $e^{+}e^{-} \rightarrow \pi^{+}\pi^{-}$, $K^{+}K^{-}$, $K_{L}^{0}K_{S}^{0}$, $\pi^{0}\gamma$, and $\eta\gamma$. The first two processes have been studied in our previous work \cite{Qin:2020udp}. Here we include $e^{+}e^{-} \rightarrow K_{L}^{0}K_{S}^{0}$ to give a complete analysis on the $\pi\pi-K\bar{K}$ coupled channels. The $\rho-\omega$ mixing mechanism has been modified a bit to be more consistent with that of Ref.~\cite{Gasser:1982ap}. 
Besides, there are new experimental measurements for the process of $e^{+}e^{-} \rightarrow K^{0}_{L}K^{0}_{S}$, which should be included. 
At earlier times, the measurements from CMD-2 \cite{Akhmetshin:2002vj}, SND \cite{Achasov:2006bv}, OLYA \cite{Ivanov:1982cr}, and DM1 \cite{Mane:1980ep} lack enough information around the $\phi$ resonance region. The measurements of CMD-2 in 1996 \cite{CMD-2:2003gqi} and SND in 2001 \cite{Achasov:2000am} were partly improved, but they were still not precise enough. 
In the past decade, the measurements by SND in 2016 \cite{CMD-3:2016nhy} and BABAR \cite{BaBar:2014uwz} in 2014 supply high statistics cross sections in the $\phi$ resonance region and make a precise analysis possible. There is also a recent measurement in the high-energy region of [2.00-3.08]~GeV by BESIII \cite{BESIII:2021yam}. 
For the last two processes of $e^{+}e^{-} \rightarrow \pi^{0}\gamma$ and $\eta\gamma$, they have not been studied in our approach before, and we take them into the analysis for completeness. On the experimental side, in 2000, SND started a measurement firstly in the energy region of [0.99-1.03] GeV for the process of $e^{+}e^{-} \rightarrow \pi^{0}\gamma$ \cite{Achasov:2000zd}. It is updated in the energy region of [0.60-0.97]~GeV in 2003 \cite{Achasov:2003ed},  in the energy region of [0.6-1.35] GeV in 2016 \cite{SND:2016drm}, and in the energy region of [1.075-2.0] GeV in 2018 \cite{Achasov:2018ujw}. CMD-2 performed a measurement for this process in 2004 \cite{CMD-2:2004ahv}, too. For the process of $e^{+}e^{-} \rightarrow \eta \gamma$, there are only four datasets: That of CMD-2 in 2001 \cite{CMD-2:2001dnv}, CMD-2 again in 2004 \cite{CMD-2:2004ahv}, SND in 2006 \cite{Achasov:2006dv}, with a big difference from the data of CMD-2 in 2001 and 2004, and SND  in 2014 \cite{Achasov:2013eli} in the energy region of [1.07-2.0] GeV.
Since all of these processes are combined together and analyzed based on the same framework of RChT, it would be expected that the coupling constants are well determined. Also, the cross sections around the $\rho-\omega$ region are very sensitive to the masses and widths of these two resonances. A combined analysis would help to fix these resonance parameters, and the prediction on HVP with the framework of RChT would be reliable.

The paper is organized as follows:
First, we give an overview of the theoretical framework  based on RChT to calculate the amplitudes of the electron-positron annihilation processes with two-body final states, as shown in Sec.~\ref{Sec:II}. With these amplitudes, we give two analyses, one is from the threshold up to 1.1~GeV, and the other is up to 2.3~GeV, with the latest experimental datasets fitted and the unknown couplings fixed. See Sec.~\ref{Sec:III}. With the obtained hadronic vacuum polarization form factors, we predict the leading order HVP contribution to muon g-2 in Sec.~\ref{Sec:IV}. Finally, a summary is given in Sec.~\ref{Sec:V}.


\section{Theoretical framework}
\label{Sec:II}
\subsection{Construction of the chiral effective Lagrangians}
As discussed above, ChPT \cite{Weinberg:1978kz,Gasser:1983yg} is a well-established EFT to describe the interactions of light pseudoscalars in the low-energy region. 
The pseudoscalar mesons have small masses, and they are regarded as Goldstone bosons generated by the spontaneous symmetry breaking of the chiral group $G=SU(3)_{L} \times SU(3)_{R}$ that is down to the subgroup $SU(3)_{V}$. By nonlinear realization, the dynamical variables of the pseudoscalar octet ($\pi,K,\eta$) can be represented by
\begin{eqnarray}
u(\phi)=\exp \{ \frac{i}{\sqrt{2}F} \Phi \} , \nonumber
\end{eqnarray}
where $F$ is the pion decay constant, given as $F \approx 92.2$MeV \cite{Zyla:2020zbs}.  
$\Phi$ is the $SU(3)$ matrix of light pseudoscalar octet. The mixing mechanism of $\eta-\eta'$ states is defined through the single angle mixing scheme \footnote{How well the double angles mixing scheme \cite{Leutwyler:1997yr, Kaiser:1998ds,Guo:2015xva,Gao:2022xqz} would improve the model is unknown,  but the single angle mixing scheme (SAMS) describes the current data well and has only one parameter. Also, the SAMS can give good solutions in some other analyses, e.g., $e^+e^-\to\eta\pi\pi$ \cite{Arteaga:2022xxy}, and $\eta'\to\pi\pi\gamma$ \cite{Dai:2017tew}.}, in terms of the mixing angle $\theta_{P}$, 
\begin{eqnarray}
\left (   
\begin{array}{c}
 \eta  \\
 \eta'
\end{array}
\right ) &=
\left (
\begin{array}{cc}
\cos\theta_{P} & -\sin\theta_{P} \\
\sin\theta_{P} & \cos\theta_{P}
\end{array}
\right )
\left (
\begin{array}{c}
 \eta_{8} \\
 \eta_{0}
\end{array}
\right )\,.
\end{eqnarray}
The other degrees of freedom in the EFT are about the resonances. They are filled in the octet or singlet, taking into account their transformation under the subgroup $SU(3)_V$.  
The octet and singlet resonances are degenerate in the large $N_{c}$ limit so that they can be filled in the nonet,
\begin{eqnarray}
\label{eq:R}
R=\sum^{8}_{i=1}\frac{\lambda_{i}}{\sqrt{2}}R_{i}+\frac{R_{0}}{\sqrt{3}} \,.
\end{eqnarray}
Here we focus on vector meson resonances, of which the $\omega-\phi$ are mixed with an angle $\theta_{V}$, 
\begin{eqnarray}
\label{eq:V;thetaV}
\left (
\begin{array}{c}
V^{8} \\
V^{0}
\end{array}
\right ) &= 
\left (
\begin{array}{cc}
\cos\theta_{V} & \sin\theta_{V} \\
-\sin\theta_{V} & \cos\theta_{V}
\end{array}
\right )
\left (
\begin{array}{c}
\phi \\
 \omega
\end{array}
\right ) .
\end{eqnarray}
Following Ref.~\cite{Qin:2020udp}, the isospin symmetry breaking caused by $\rho-\omega$ mixing is considered, too. We adopt the momentum-dependent mixing mechanism as
\begin{eqnarray}
\label{eq:rho,omega}
\left (
\begin{array}{c}
 \left \vert \bar{\rho}^{0}  \right \rangle  \\
 \left \vert \bar{\omega}    \right \rangle
\end{array}
\right ) &=& 
\left (
\begin{array}{cc}
\cos\delta  &  \sin\delta \frac{M_{V}\Gamma_{\rho}}{M_{V}^{2}-s+iM_{V}\Gamma_{\rho}} \\
\sin\delta \frac{M_{V}\Gamma_{\rho}}{M_{V}^{2}-s-iM_{V}\Gamma_{\rho}}      & \cos \delta
\end{array}
\right )
\left (
\begin{array}{c}
 \left \vert \rho^{0}  \right \rangle  \\
 \left \vert \omega    \right \rangle
\end{array}
\right ) \nonumber \\ [2mm]
&\equiv &
\left (
\begin{array}{cc}
\cos\delta  &  -\sin\delta^{\omega}(s) \\
\sin\delta^{\rho}(s)  & \cos \delta
\end{array}
\right )
\left (
\begin{array}{c}
 \left \vert \rho^{0}   \right \rangle  \\
 \left \vert \omega     \right \rangle
\end{array}
\right ) \,. 
\end{eqnarray}
Here one can set $M_V=M_\rho$ as usually done in RChT. This mixing mechanism will be applied in all the processes of our analysis.   
The $(M_V^2-s)$ parts in the off-diagonal terms now have an opposite sign with that of Ref.~\cite{Qin:2020udp},
to make sure that Eq.~(\ref{eq:rho,omega}) will return back to that of Ref.~\cite{Gasser:1982ap} in the non-relativistic limit. As in Ref.~\cite{Qin:2020udp}, the small quantities $\Gamma_\omega/\Gamma_\rho$ and $(M_\rho-M_\omega)/M_V$ are ignored.

With the fields defined above, the Lagrangians to be used can be written as 
\begin{equation} \label{eq:rchtl1}
  {\cal L}_{\mathrm{RChT}} \, = \, {\cal L}^{\rm GB} \, + \, {\cal L}^{\mathrm{V}}_{\mathrm{kin}} \, + \, {\cal L}_{\rm int}^{\rm V-GB} \, ,
  \end{equation}
where $GB$ represents the lightest pseudoscalar mesons, $\pi, K, \eta, \eta'$. The relevant Lagrangians of these pseudoscalars are taken from ChPT at the lowest order 
\begin{equation}
\mathcal{L}_{(2)}^{\mathrm{GB}}=\frac{F^{2}}{4}\left\langle u_{\mu}u^{\mu}+\chi_{+}\right\rangle  \label{eq:ChPT}
\end{equation}
with $\langle \dots \rangle$ the trace in the ${\rm SU}(3)$ flavour space. 
The subscript \lq 2'  in the bracket is of $\mathcal{O}(p^2)$ in the chiral counting. 
The chiral operators such as $u_\mu$ and $\chi_+$ can be found in Ref.~\cite{Scherer:2002tk}.
Notice that when integrating out one resonance in the ${\cal L}^{\rm V-GB}$, one would obtain chiral terms of light pseudoscalars with chiral counting no lower than $\mathcal{O}(p^2)$ \cite{Dai:2019lmj}. Hence, ${\cal L}^{\rm V-GB}$ would be no less than $\mathcal{O}(p^4)$ in pure chiral counting, and there is no double-counting problem here.  
For the processes of $e^+e^-$ annihilating into $\pi^0\gamma$ and $\eta\gamma$, the leading contribution to ${\cal L}^{\rm GB}$ is from Wess-Zumino-Witten (WZW) anomaly, which is of
odd-intrinsic-parity at the chiral counting ${\cal O}(p^4)$ \cite{Wess:1971yu,Witten:1983tw}. The explicit expression relevant to this work is
\begin{equation}
  L^{\rm GB}_{(4)}=-\frac{\sqrt{2}N_{C}}{8\pi^{2}F}\epsilon_{\mu\nu\rho\sigma}\left\langle \Phi\partial^{\mu} v^{\nu}\partial^{\rho} v^{\sigma}\right\rangle  \label{eq:anomaly}
\end{equation}
with $v^{\nu}$ the external vector current.
The ${\cal L}^{\mathrm{V}}_{\mathrm{kin}}$ is the kinetic term of the vector resonance field 
\begin{equation} \label{eq:vkin}
\mathcal{L}_{\mathrm{kin}}^{\mathrm{V}}=-\frac{1}{2}\left\langle \nabla^{\lambda}V_{\lambda\mu}\nabla_{\nu}V^{\nu\mu}\right\rangle +\frac{1}{4}M_{V}^{2}\left\langle V_{\mu\nu}V^{\mu\nu}\right\rangle \,.
\end{equation}
The ${\cal L}_{\rm int}^{\rm V-GB}$ is the interaction Lagrangian involved with vector resonances and light pseudoscalars, ${\cal L}_{\rm int}^{\rm V-GB}={\cal L}_{(2)}^{\rm V}+{\cal L}_{(4)}^{\rm V}+{\cal L}_{(2)}^{\rm VV}$. One has 
\begin{eqnarray} \label{eq:lv4}
  {\cal L}^{\mbox{\tiny V}}_{(2)} &= &\frac{F_V}{2 \sqrt{2}} \, \langle V_{\mu \nu} f_+^{\mu \nu} \rangle + i \frac{G_V}{\sqrt{2}} \, \langle V_{\mu \nu} u^{\mu} u^{\nu} \rangle \, ,\nonumber\\
  {\cal L}^{\mbox{\tiny V}}_{(4)}& = &\sum_{j=1}^7 \frac{c_j}{M_V} \, {\cal O}_{\mbox{\tiny VJP}}^j \, ,\quad
  {\cal L}^{\mbox{\tiny VV}}_{(2)} = \sum_{j=1}^4 d_j {\cal O}_{\mbox{\tiny VVP}}^j \,. \nonumber
 \end{eqnarray}
For details of the Lagrangians, we refer to Ref.~\cite{Dai:2013joa}. By integrating out vector resonances, the Lagrangian ${\cal L}^{\mbox{\tiny V}}_{(2)}$ will contribute at least $\mathcal{O}(p^4)$  and the Lagrangians  $ {\cal L}^{\mbox{\tiny V}}_{(4)}$ and ${\cal L}^{\mbox{\tiny VV}}_{(2)}$ will contribute at least $\mathcal{O}(p^6)$.  
  
The interaction Lagrangians discussed above are only for the lightest multiplet of the vector resonances, which contains the dynamics below roughly 1 GeV. To extend our form factors up to $E_{CM}\sim 2.3$~GeV, one needs to include two sets of heavier vector resonance multiplets ($V'_{\mu\nu}$ and $V''_{\mu\nu}$). Here we follow Refs.~\cite{Dai:2013joa, Qin:2020udp}, applying the extension to the Breit-Wigner propagator
\begin{eqnarray}
\label{eq:BW;extend}
\frac{1}{M_{V}^{2}-x} \rightarrow \frac{1}{M_{V}^{2}-x}+\frac{\beta^{'}_{X}}{M_{V'}^{2}-x}+\frac{\beta^{''}_{X}}{M_{V''}^{2}-x}\,,
\end{eqnarray}
where the subscript \lq $X$' represents the label of a different process. They are: 
$\beta^{',''}_{\pi\pi}$, $\beta^{',''}_{KK}$, $\beta^{',''}_{K^{0}_{L}K^{0}_{S}}$, $\beta^{',''}_{\pi^{0}\gamma}$, $\beta^{',''}_{\eta\gamma}$.
Notice that Eq.~(\ref{eq:BW;extend}) implies the assumption that the Lagrangians $\mathcal{L}_{\rm int}^{{\rm V}^{'}{\rm -GB}}$ and $\mathcal{L}_{\rm int}^{{\rm V}^{''}{\rm -GB}}$ have similar forms as $\mathcal{L}_{\rm int}^{\rm V-GB}$. Consequently, the amplitudes/formfactors will have similar forms as that calculated by $\mathcal{L}_{\rm int}^{\rm V-GB}$, with only different couplings (e.g., $F_V'$, $G_V'$, $c_j'$, and $d_j'$) and propagators of the resonances. The contributions of heavier states can be absorbed into the $\beta^{'}_{X}$, $\beta^{''}_{X}$ and propagators of $V'$, $V''$ without destructing the structure of the amplitudes/formfactors, which has been successfully applied in the phenomenology analyses \cite{Dai:2013joa, Qin:2020udp}.

\subsection{Formulas of cross-sections and hadronic vacuum polarization form factors }
\begin{figure}[htbp]
\centering 
\includegraphics[width=0.5\textwidth,height=0.15\textheight]{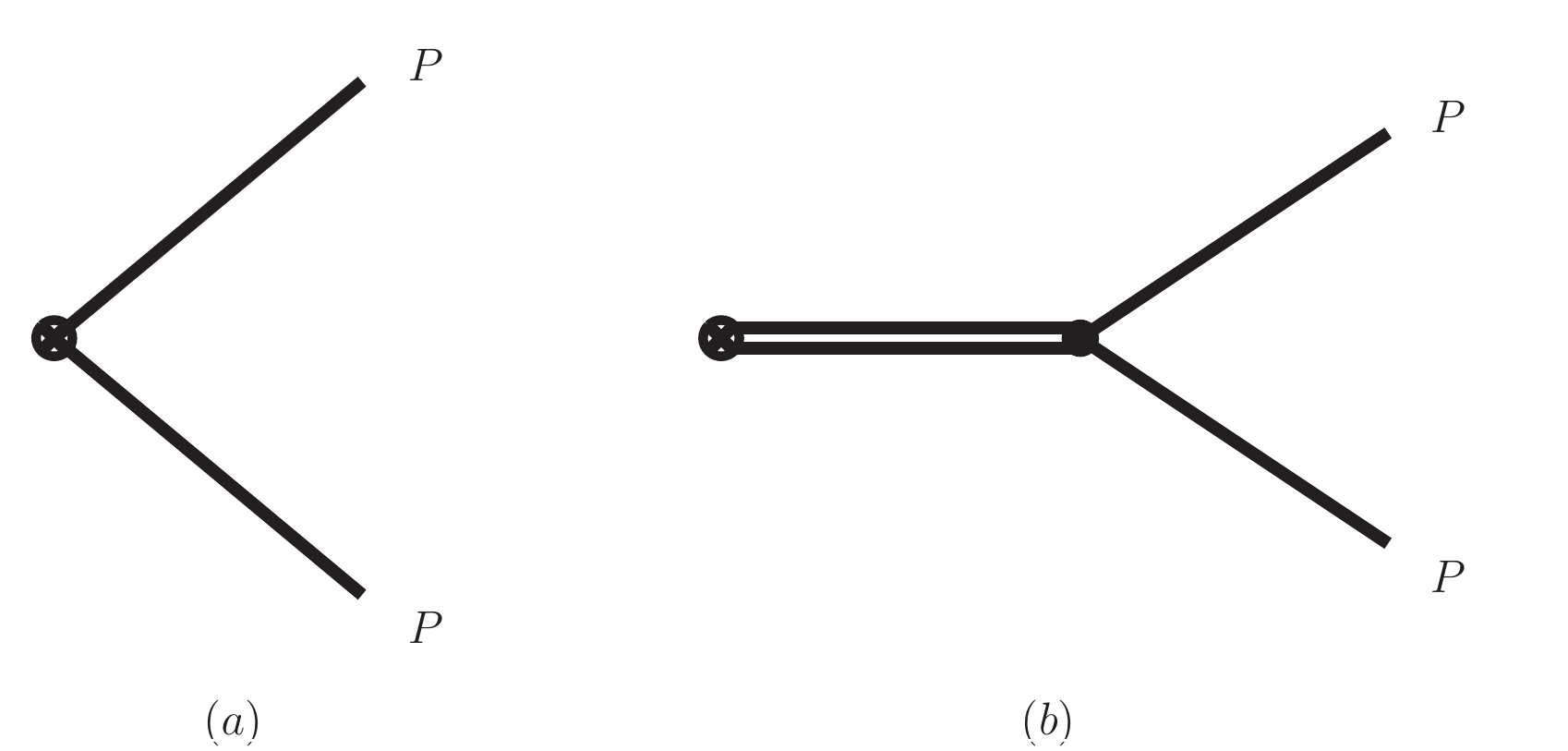}
\caption{\label{Fig:FDPP}Feynman diagrams contributing to the hadronization of the vector current in the processes of $ e^{+}e^{-} \rightarrow P P$, with $PP=\pi^+\pi^-$, $K^+K^-$, and $K^0_L K^0_S$, respectively. The intermediate particle is the vector resonances, $\rho(770)$, $\omega(782)$, $\phi(1020)$ and the relevant heavier states. }
  \end{figure}

The amplitudes of $e^{+}e^{-} \rightarrow \pi^{+}\pi^{-}, K^{+}K^{-}, K_{L}^{0}K_{S}^{0}$ are driven by the electromagnetic current
\begin{eqnarray}
\label{eq:Fv}
&& \left\langle P_{1}(p_{1})P_{2}(p_{2}) \vert (\mathcal{V}^{3}_{\mu}+\mathcal{V}^{8}_{\mu}/\sqrt{3})e^{i\mathcal{L}_{QCD}} \vert 0 \right\rangle = (p_{1}-p_{2})_{\mu} F^{PP}_{V}(Q^{2})
\end{eqnarray}
with $\mathcal{V}^{i}_{\mu} = \bar{q}\gamma_{\mu}(\lambda^{i}/2)q$, $Q = p_{1}+p_{2}$ is relevant to the energy in the center of mass frame, $E_{CM} \equiv \sqrt{Q^{2}}$ and $PP= \pi^{+}\pi^-, K^{+}K^-, K_{L}^{0}K_{S}^{0}$,  respectively. The cross-sections of $e^{+}e^{-} \rightarrow \pi^{+}\pi^{-}$, $K^{+}K^{-}$ and $K_{L}^{0}K_{S}^{0}$ are given by
\begin{eqnarray}
\label{eq:cs}
\sigma_{e^+e^-\to PP} = \alpha_{e}^{2} \frac{\pi}{3Q^{2}}(1-4\frac{m_{P}^{2}}{Q^{2}})^{3/2} \vert F^{PP}_{V }\vert ^{2}.
\end{eqnarray}
The Feynman diagrams to calculate the form factors of $F^{PP}_V$ are shown in Fig.~\ref{Fig:FDPP}.
Note that for  $ e^{+}e^{-} \rightarrow K_{L}^{0}K_{S}^{0}$, only the diagram of Fig.~\ref{Fig:FDPP} (b) contributes. 
The form factors in the ideal mixing case can be written as 
\begin{eqnarray}
F^{\pi^+\pi^-}_{V}&=&1+\frac{F_V G_V Q^2}{F^2(M_\rho^2-Q^2)} \,, \nonumber   \\
F^{K^+K^-}_V&=&\frac{F_V G_V M_{\rho}^{2}}{2F^2(M_{\rho}^{2}-Q^{2})}+\frac{F_V G_V M_{\omega}^{2}}{6F^2(M_{\omega}^{2}-Q^{2})}+\frac{F_V G_V M_{\phi}^{2}}{3F^2(M_{\phi}^{2}-Q^{2})} \,, \nonumber \\
F^{K^0_L K^0_S}_{V}&=&-\frac{F_{V}G_{V}M_\rho^{2}}{2F^{2}(M_{\rho}^{2}-Q^{2})}+\frac{F_{V}G_{V}M_\omega^{2}}{6F^{2}(M_{\omega}^{2}-Q^{2})}+\frac{F_{V}G_{V}M_\phi^{2}}{3F^{2}(M_{\phi}^{2}-Q^{2})}  \,, \label{Eq:FPP;ideal}
\end{eqnarray} 
The detailed results of the vector form factors $F_{V}^{\pi^+\pi^-}$, $F_{V}^{K^+K^-}$, and $F_{V}^{K_{L}^{0}K_{S}^{0}}$, including the mixing angles, heavier states, Final state interactions, and QCD high energy constraints are given in Appendix \ref{app:Fv}.

\begin{figure}[htbp]
\centering 
\includegraphics[width=1\textwidth]{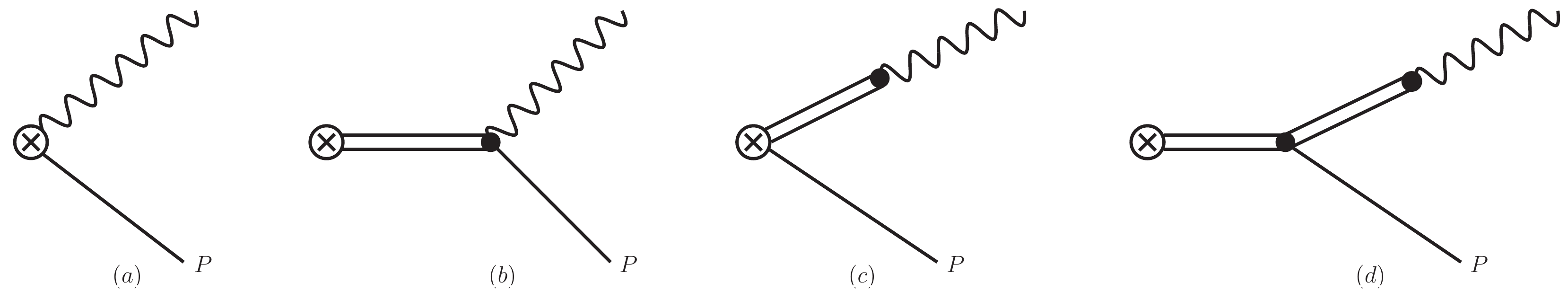}
\caption{\label{Fig:FD}Feynman diagrams contributing to the hadronization of the vector current in the scattering processes of $ e^{+}e^{-} \rightarrow P \gamma$, with $P=\pi^{0}, \eta $. The \lq V' are the vector resonances, $\rho(770)$, $\omega(782)$, $\phi(1020)$ and the relevant heavier states. Fig.~\ref{Fig:FD}
(a) is from the WZW anomaly, given in Eq.~(\ref{eq:anomaly}).  Fig.~\ref{Fig:FD} (b,c,d) are those in which the vector resonances are involved. }
  \end{figure}

The amplitude for the process  of $e^{+}e^{-} \rightarrow P\gamma$ is driven by the hadronization of the electromagnetic current, in terms of transition form factors \cite{Hoferichter:2014vra}:
\begin{eqnarray}
\label{eq:EM;transit}
\left \langle P(Q-k)\gamma(k) \vert (\mathcal{V}^{3}_{\mu}+\mathcal{V}^{8}_{\mu}/\sqrt{3})e^{i\mathcal{L}_{QCD}} \vert 0 \right \rangle= F_{P\gamma^{*}\gamma}(Q^{2})\epsilon_{\mu\nu\rho\sigma}\epsilon_\gamma^{\nu}(k)Q^{\rho}k^{\sigma} \,, 
\end{eqnarray} 
where one has $P=\pi^{0}, \eta$. With these form factors, one obtains the cross-sections
\begin{equation}
  \sigma_{e^{+}e^{-}\rightarrow P \gamma}(Q^{2})=\frac{2\pi^{2}\alpha^{3}}{3}\frac{(Q^{2}-M_{P}^{2})^{3}}{Q^{6}}\vert  F_{P\gamma^{*}\gamma}(Q^{2}) \vert^{2} \,.
\end{equation}
The Feynman diagrams to calculate the form factors $F_{P\gamma^*\gamma}$ are shown in Fig.~\ref{Fig:FD}.
At low energies, $E_{CM} \ll M_{\rho}$, the hadronization of the vector current is dominantly driven by the anomaly. See Fig.~\ref{Fig:FD} (a). At higher energies, $E_{CM} \gtrsim  M_\rho$, vector resonances appear and would dominate the dynamics of the hadronization. See Figs.~\ref{Fig:FD} (b)-(d). The form factors $F_{P\gamma^*\gamma }$ in the \lq ideal' case are listed below, where $\eta'$, $\rho-\omega$ mixing, and heavier vector resonances $V^{'('')}$ are ignored and $\theta_V=\arcsin{1/\sqrt{3}}$.  One has
\begin{equation}
F_{P\gamma^*\gamma }(Q^2) = F^{P}_{a} + F^{P}_{b} + F^{P}_{c} + F^{P}_{d} \,, \label{eq:Fv;P}
\end{equation}
with the form factors for the process of $e^+e^-\to \pi^0\gamma$ given by
\begin{eqnarray}
F^{\pi}_{a}&=&\frac{N_{C}}{12\pi^{2}F} \,, \nonumber   \\
F^{\pi}_{b}&=&\frac{2\sqrt{2}F_{V}}{3FM_{V}}\bigg( \frac{1}{M_{\rho}^{2}-Q^{2}} +\frac{1}{M_\omega^2-Q^{2}}\bigg)\big[ Q^{2}(-c_{1}+c_{2}+c_{5}-2c_{6})+m_{\pi}^{2}(c_{1}+c_{2}+8c_{3}-c_{5}) \big]  \,, \nonumber   \\
F^{\pi}_{c}&=&\frac{2\sqrt{2}F_{V}}{3FM_{V}}\bigg( \frac{1}{M_{\rho}^{2}} +\frac{1}{M_{\omega}^{2}}\bigg) \big[ Q^{2}(c_{1}-c_{2}+c_{5})+m_{\pi}^{2}(c_{1}+c_{2}+8c_{3}-c_{5}) \big]  \,, \nonumber   \\
F^{\pi}_{d}&=&-\frac{4F_{V}^{2}}{3F}\bigg(  \frac{1}{M_{\omega}^{2}(M_{\rho}^{2}-Q^{2})} +\frac{1}{M_{\rho}^{2}(M_{\omega}^{2}-Q^{2})}\bigg)  \big[ d_{3}Q^{2}+m_{\pi}^{2}(d_{1}-d_{3}+8d_{2}) \big]  \,, \label{eq:Fv;P;pi}
\end{eqnarray}
and the ones for the process of $e^+e^-\to \eta\gamma$ given by
\begin{eqnarray}
F^{\eta}_{a}&=&\frac{N_{C}}{12\sqrt{3}\pi^{2}F} \,,\nonumber   \\
F^{\eta}_{b}&=&\frac{2\sqrt{2}F_{V}}{\sqrt{3}FM_{V}}\bigg(  \frac{1}{M_{\rho}^{2}-Q^{2}}+\frac{1}{9(M_{\omega}^{2}-Q^{2})} \bigg) \big[ Q^{2}(-c_{1}+c_{2}+c_{5}-2c_{6})+m_{\eta}^{2}(c_{1}+c_{2}-c_{5})+8c_{3}m_{\pi}^{2} \big] \nonumber   \\
&& -\frac{8\sqrt{2}F_{V}}{9\sqrt{3}FM_{V}(M_{\phi}^{2}-Q^{2})}\big[ Q^{2}(-c_{1}+c_{2}+c_{5}-2c_{6}) +m_{\eta}^{2}(c_{1}+c_{2}-c_{5})+8c_{3}(2m_{K}^{2}-m_{\pi}^{2}) \big] \,,\nonumber   \\
F^{\eta}_{c}&=&\frac{2\sqrt{2}F_{V}}{\sqrt{3}FM_{V}}\bigg(  \frac{1}{M_{\rho}^{2}}+\frac{1}{9 M_{\omega}^{2}} \bigg)\big[ Q^{2}(c_{1}-c_{2}+c_{5})+m_{\eta}^{2}(c_{1}+c_{2}-c_{5})+8c_{3}m_{\pi}^{2} \big]    \nonumber   \\
&& -\frac{8\sqrt{2}F_{V}}{9\sqrt{3}FM_{V}M_{\phi}^{2}}\big[ Q^{2}(c_{1}-c_{2}+c_{5})+m_{\eta}^{2}(c_{1}+c_{2}-c_{5})+8c_{3}(2m_{K}^{2}-m_{\pi}^{2}) \big]  \,,\nonumber   \\
F^{\eta}_{d}&=&\frac{-4F_{V}^{2}}{\sqrt{3}F} \bigg( \frac{1}{M_{\rho}^{2}(M_{\rho}^{2}-Q^{2})}+\frac{1}{9M_{\omega}^{2}(M_{\omega}^{2}-Q^{2})} \bigg) \big[ d_{3}Q^{2}+(d_{1}-d_{3})m_{\eta}^{2}+8d_{2}m_{\pi}^{2} \big] \nonumber   \\
&& +\frac{16F_{V}^{2}}{9\sqrt{3}F M_{\phi}^{2}(M_{\phi}^{2}-Q^{2})} \big[  d_{3}Q^{2}+(d_{1}-d_{3})m_{\eta}^{2}+8d_{2}(2m_{K}^{2}-m_{\pi}^{2}) \big] \,. \label{eq:Fv;P;eta}
\end{eqnarray}
The equations above are for the form factors of $e^{+}e^{-} \to P \gamma$ in the ideal mixing case. Considering the mixing angles, the form factors will be much more lengthy. We put them in the Appendix~\ref{app:F;Pgg}.

To reduce the unknown coupling constants of the RChT Lagrangians, we apply the QCD high energy ($Q^{2} \rightarrow \infty$) constraints which are established by matching the $\langle VVP \rangle$ Green functions that are calculated by RChT with those calculated from perturbative QCD with operator product expansion (OPE) at leading order.
The high energy constraints for the coupling constants appearing in our form factors are given as \cite{Ruiz-Femenia:2003jdx}:
\begin{eqnarray}
c_{1}-c_{2}+c_{5}&=&0 \,, \label{eq:c125}\\
c_{1}-c_{2}-c_{5}+2c_{6}&=&-\frac{N_{C}M_{V}}{32\sqrt{2}\pi^{2}F_{V}} \,, \label{eq:c1256} \\
c_{1}+c_{2}-c_{5}+8c_{3}&=&0\,,  \label{eq:c1253}\\
d_{1}+8d_{2}-d_{3}&=&\frac{F^{2}}{8F_{V}^{2}} \,.  \label{eq:d123}
\end{eqnarray}
Since the photon meson transition form factors have the asymptotic behavior of $1/Q^2$ in the high-energy region, it is convenient to implement that the form factors obtained in RChT vanish at $Q^2\to \infty$~\cite{Chen:2012vw}\footnote{Indeed, this is similar to what was done in Ref.~\cite{Dai:2013joa}, where the form factors of two-point Green functions of the vector current are taken to be vanished at $Q^2\to\infty$. }. One has 
\begin{eqnarray}
d_{3}&=&-\frac{N_{C}}{64\pi^{2}}\frac{M_{V}^{2}}{F_{V}^{2}} \,, \label{eq:d3}
\end{eqnarray}
Another constraint is from the requirement that the two pion vector form factors should vanish at $Q^{2}\to\infty$, 
\begin{eqnarray}
 F_{V}G_{V}=F^{2} \,. \nonumber
\end{eqnarray}
Considering these high energy constraints, the unknown couplings of RChT are $F_{V}$, $d_{2}$, $c_{3}$, and $\alpha_{V}$.
The masses of the lightest vector resonances are taken as $M_\rho=773.80$~MeV, $M_\omega=782.48$~MeV, and $M_\phi=1019.20$~MeV, and the widths of them are taken as $\Gamma_\omega=8.67$~MeV and $\Gamma_\phi=3.85$~MeV, which are compatible with those given by the PDG~\cite{ParticleDataGroup:2020ssz}\footnote{We are aware that in Refs.~\cite{Cirigliano:2003yq,Guo:2009hi}, the mass splitting and the decay widths of the resonances can be implemented by phenomenological chiral Lagrangians.}. 

In the low-energy region, the FSI is challenging to deal with in a model-independent way. Dispersive approaches to deal with FSI are possible in some cases, e.g., Refs.~\cite{Niecknig:2012sj,Schneider:2012ez,Danilkin:2014cra,Albaladejo:2017hhj,Isken:2017dkw,Colangelo:2018jxw,Yao:2020bxx} for some recent work. 
Here the situation is a bit more complicated. For example, the resonance $\rho$ appears in the intermediate states will contribute to the phase shift of $\pi\pi$ P-wave, while the Omn\'es function \cite{Omnes:1958hv}, one kind of dispersion relation, is constructed by phase shifts, too. This may cause \lq double counting' on FSI.    
In this work, We apply the method proposed in Ref.~\cite{Guerrero:1997ku} to handle the processes of $e^{+}e^{-} \to \pi\pi, K^+K^-,K_{L}^{0}K_{S}^{0}$, where the contribution of resonances has been matched with that of the Omn\'es  function, and there is no such double counting problem for the pion vector form factor. The details are provided in the Appendix \ref{pro:FSI}.

At last, the only unknown couplings in these form factors will be $F_{V}$, $d_{2}$, $c_{3}$, $\alpha_{V}$, $\beta^{'('')}_{X}$ from Eq.~(\ref{eq:BW;extend}) for the heavier resonance states, and the mixing angles between the octet and singlet states, i.e., $\theta_{P}$  for $\eta-\eta'$ mixing, $\theta_{V}$ for $\omega-\phi$ mixing, and $\delta$ for $\rho^0-\omega$ mixing. 
To extend the amplitudes to the high-energy region, we include the heavier vector resonance states $V^{'('')}$. Indeed, their masses and widths resemble well the vector resonance spectrum of PDG \cite{ParticleDataGroup:2020ssz}. In order to further determine the relevant parameters, we also calculated the decay widths related to the resonance states and fitted them together with the cross-sections of electron-positron annihilation. The specific expression of the decay widths is given in Appendix \ref{app:decay widths}. See discussions in the next section.

\section{Fit results and discussions} 
\label{Sec:III} 
Fittings to the cross sections for all processes of electron-positron annihilating into two-body final states ($\pi\pi$, $K^+K^-$,$K^{0}_{L}K^{0}_{S}$, $\pi^0\gamma$, and $\eta\gamma$) and to the decay widths of $V\to P\gamma$, $P\to V\gamma$, and $P\to\gamma\gamma$ are combined together to give an overall constraint to our solutions. 
Following the strategy of Ref.~\cite{Qin:2020udp}, we perform two fits for the analysis. 
One is to focus on the energy region below 1.1~GeV, where the ground vector states ($\rho^0$, $\omega$, $\phi$) dominate without heavier resonances $V^{'('')}$ appearing. 
It is named Fit~A, and all the $\beta_X^{'('')}$ are set to be zero. 
The other one is to extend our analysis up to $E_{CM}=2.3~GeV$, where $V^{'('')}$ are 
included, and it is named Fit~B.   
Similar to the previous work \cite{Qin:2020udp}, we take datasets after the year 2000 into counting for the $\chi^2$ and superimpose the ones before 2000 for the reader's convenience.

The fitting parameters are given in Table \ref{Tab:para}.
\begin{table}[htbp]
  \centering
{\footnotesize
 \begin{tabular}{c|ccccc}
  \hline
  & Fit~A & Fit~B   & Ref.~\cite{Qin:2020udp}   &Ref.~\cite{Dai:2013joa}   &PDG~\cite{ParticleDataGroup:2020ssz}                                                  \\ \hline
  $F_{V}$(GeV)             &0.148(1)     &0.138(1)       &0.142(1)
  &0.148(1)\\
  $d_{2}$                 &0.00271(20) &0.00753(30)    &0.0276(6)   &0.0359(7)\\
  $c_{3}$                 &0.00161(10)  &0.00211(10)    &0.00435(13) &0.00689(17)\\
  $\alpha_{V}$            &-0.0100(1) &-0.00900(35)  &-0.00113(14)  &0.0126(7)\\
  $\theta_{V}(^\circ)$    &38.60(1)    &38.62(1)  &39.56(1)      &38.94(2)\\
  $\delta(^\circ)$        &-1.80(1)   &-1.80(1)  &-1.80(1)     &-\\
  $\theta_{P}(^\circ)$    &-20.74(16)  &-20.50(30)  &-19.61(10)   &-21.37(26) \\
  $\beta'_{\pi\pi}$       &- &-0.0617(3)   &-0.0625(9)  &-\\
  $\beta''_{\pi\pi}$               &-  &0.0188(2)   &0.0118(7)   &-\\          
  $\beta'_{KK}$                    &-  &-0.195(3)&-0.0712(40)  &-\\
  $\beta''_{KK}$                   &-  &-0.139(8)  &-0.197(5)    &-\\
  $\beta'_{K_{L}^{0}K_{S}^{0}}$       &-&-0.194(15)  &-                   &-\\
  $\beta''_{K_{L}^{0}K_{S}^{0}}$       &- &-0.033(36)  &-                   &-\\
  $\beta'_{\pi^{0}\gamma}$             &-  &-0.0902(48) &-                   &-\\ 
  $\beta'_{\eta\gamma}$               &- &-0.350(50)  &-                   &- \\
  $M_{\rho'}$(GeV)              &-       &1.519(1)    &1.519(2)     &1.550(12)                &1.465(25)\\
  $\Gamma_{\rho'}$(GeV)         &-        &0.381(3)    &0.340(1)    &0.238(18)     &0.400(60)\\ 
  $M_{\omega'}$(GeV)            &-         &1.250(3)    &1.253(3)    &1.249(3)                &1.410(60)\\
  $\Gamma_{\omega'}$(GeV)       &-        &0.290(2)    &0.310(3)     &0.307(7)                &0.290(190)\\
  $M_{\phi'}$(GeV)              &-       &1.656(3)   &1.640(3)     &1.641(5)                &1.680(20)\\
  $\Gamma_{\phi'}$(GeV)         &-       &0.136(1)   &0.090(2)     &0.086(7)               &0.15(5)\\
  $M_{\rho''}$(GeV)             &-        &1.720(1)   &1.720(1)     &1.794(12)     &1.720(20)\\
  $\Gamma_{\rho''}$(GeV)        &-       &0.250(1)   &0.150(5)     &0.297(33)                &0.25(10)\\
  $M_{\omega''}$(GeV)           &-       &1.725(2)   &1.725(10)     &1.700(11)   &1.670(30)\\
  $\Gamma_{\omega''}$(GeV)      &-        & 0.400(1)  &0.400(3)     &0.400(13)     &0.315(35)\\
  $M_{\phi''}$(GeV)             &-       &2.160(1)   &2.126(25)     &2.086(22)    &2.162(70)\\
  $\Gamma_{\phi''}$(GeV)        &-       &0.105(10)  &0.100(14)    &0.108(17)               &0.100(27)\\
  \hline
  \end{tabular} 
  \caption{\label{Tab:para}Parameters of Fits A and B.  Comparison with that of Fit II of Ref.~\cite{Qin:2020udp}, Fit IV of Ref.~\cite{Dai:2013joa} and PDG \cite{ParticleDataGroup:2020ssz} are also listed. The uncertainties of the parameters are taken from MINUIT.}
}
  \end{table}
We apply the Bootstrap method \cite{Efron:1979bxm} to obtain the uncertainties of our solutions, which are calculated by varying the experimental data within its errors and multiplying a normal distribution function. Indeed, there is another source for the error, the statistical one from MINUIT \cite{James:1975dr}. However, this part is much smaller, and it can be ignored. 
Note that in Fit~B, $\beta''_{\pi^{0}\gamma}$ is set to be zero, as there are only a few data points in the high energy region ($1.8\leq E_{CM}\leq 2.3$~GeV), and they are tiny and ignorable. See discussions below. The parameters of Fit II in Ref.~\cite{Qin:2020udp} and Fit IV in Ref.~\cite{Dai:2013joa} and the masses and widths of the heavier vector resonances given by PDG \cite{ParticleDataGroup:2020ssz} are also listed in Table \ref{Tab:para} for comparison. 

In Fit~A, the parameter $d_{2}$ has a discrepancy with that of Refs.~\cite{Qin:2020udp,Dai:2013joa}\footnote{Notice that the values of $d_2$ obtained in $e^+e^-$ annihilation are much smaller than that of $\tau$ decays \cite{GomezDumm:2012dpx}.}. This is caused by fitting the cross sections of $e^+e^-\to\eta\gamma$.
The $\rho-\omega$ mixing angle $\delta$ is roughly the same as that of Ref.~\cite{Qin:2020udp}, though there is a difference in the off-diagonal matrix elements between the two mixing mechanisms.  
A large discrepancy comes from the value of the parameter $\alpha_{V}$. Here it is $-0.00900\pm0.00035$, while it is almost vanished ($-0.00113\pm0.00014$) in Ref.~\cite{Qin:2020udp} and has a positive value, $0.0126\pm0.007$ in Ref.~\cite{Dai:2013joa}. As has been checked, a negative value of $\alpha_{V}$ can give better fits to the cross section data of $e^{+}e^{-} \rightarrow K^{+}K^{-}$ from BABAR \cite{BaBar:2013jqz}. Meanwhile, the peak $\phi$(1020) will be decreased, and this is balanced by adjusting the values of $\theta_{V}$ and  $\Gamma_{\phi}$. This is why the values of $\theta_{V}$ are smaller than that of Refs.~\cite{Qin:2020udp,Dai:2013joa}. 
All the parameters of $\beta^{'}_{X}$ and $\beta{''}_{X}$ are relatively small, with magnitudes less than 0.3. This is similar to what has been found for the processes with two-body final states \cite{Qin:2020udp}, and it is compatible with empirical assumptions of lowest resonance meson dominance \cite{Knecht:2001xc,Dai:2013joa,Nugent:2013hxa}.  

The decay widths predicted by our model are given in Table \ref{Tab:width}. 
\begin{table}[htbp]
\centering\renewcommand\arraystretch{1.1}
{\footnotesize
 \begin{tabular}{c|ccccc}
 \hline
    Width   &Fit~A  & Fit~B   &Ref.~\cite{Qin:2020udp}    &Ref.~\cite{Dai:2013joa}     &PDG~\cite{ParticleDataGroup:2020ssz} \\
    \hline
    $\Gamma_{\rho \rightarrow ee} (10^{-6}$ GeV)  &6.26$\pm$0.65 &5.45$\pm$0.73  &5.81$\pm$0.52  &6.54  & 6.98$\pm$0.07 \\
    $\Gamma_{\omega \rightarrow ee}(10^{-7}$GeV)  &7.81$\pm$0.78 &6.81$\pm$0.88 &7.60$\pm$0.65  &6.69  & 6.25$\pm$0.13 \\   
    $\Gamma_{\phi \rightarrow ee}(10^{-6}$GeV)    &0.81$\pm$0.10 &0.72$\pm$0.10   &0.86$\pm$0.08  &1.20  & 1.26$\pm$0.01    \\
    $\Gamma_{\rho \rightarrow \pi\pi}(10^{-1}$GeV)   &1.14$\pm$0.14 &1.31$\pm$0.15 &1.24$\pm$0.11  &1.14  & 1.48$\pm$0.01 \\
    $\Gamma_{\omega \rightarrow \pi\pi}(10^{-4}$GeV)  &1.15$\pm$0.10 &1.32$\pm$0.09 &1.23$\pm$0.11  &1.61  & 1.30$\pm$0.05 \\
    $\Gamma_{\phi \rightarrow \pi\pi}(10^{-7}$GeV)    &1.80$\pm$0.22 &1.60$\pm$0.22  &1.91$\pm$0.18  &2.66  & 3.10$\pm$0.55 \\
    $\Gamma_{\rho^{0} \rightarrow \pi^{0}\gamma}(10^{-5}$GeV) &6.61$\pm$0.81 &7.60$\pm$0.90   &5.38$\pm$0.64 & 5.96 & 6.95$\pm$0.89    \\
    $\Gamma_{\rho^{+} \rightarrow \pi^{+}\gamma}(10^{-5}$GeV) &6.56$\pm$0.80  &7.54$\pm$0.90 &4.53$\pm$0.37  &4.81 &6.65$\pm$0.74  \\
    $\Gamma_{\omega \rightarrow \pi^{0}\gamma}(10^{-4}$GeV) &6.13$\pm$0.74  &7.05$\pm$0.83 &4.07$\pm$0.35 &4.43  &7.13$\pm$0.19    \\
    $\Gamma_{\phi \rightarrow \pi^{0}\gamma}(10^{-6}$GeV)   &8.20$\pm$1.49 &9.54$\pm$1.72 &9.17$\pm$1.30    &7.34   &5.52$\pm$0.21 \\
    $\Gamma_{\rho \rightarrow \eta\gamma}(10^{-5}$GeV)    &4.10$\pm$0.46 &5.09$\pm$0.75  &4.32$\pm$0.38    &4.85   &4.43$\pm$0.31 \\
    $\Gamma_{\omega \rightarrow \eta\gamma}(10^{-6}$GeV)  &4.47$\pm$0.47 &5.58$\pm$0.83   &3.77$\pm$0.48  &4.13 &3.82$\pm$0.34  \\
    $\Gamma_{\phi \rightarrow \eta\gamma}(10^{-5}$GeV)     &8.40$\pm$0.99 &9.73$\pm$1.48  &6.10$\pm$0.48    &6.57  &5.54$\pm$0.11  \\
    $\Gamma_{\eta' \rightarrow \rho \gamma}(10^{-5}$GeV)   &1.54$\pm$0.29 &2.60$\pm$1.28   &5.10$\pm$1.10   &5.37  &5.66$\pm$0.10\\
    $\Gamma_{\eta' \rightarrow \omega \gamma}(10^{-6}$GeV)   &2.12$\pm$0.39 &3.43$\pm$1.58   &5.52$\pm$0.94   &5.12   &4.74$\pm$0.13\\
    $\Gamma_{\phi \rightarrow \eta' \gamma}(10^{-7}$GeV)  &4.67$\pm$0.55 &5.53$\pm$0.70  &3.36$\pm$0.44   &3.93   &2.64$\pm$0.09\\
    $\Gamma_{\eta \rightarrow \gamma \gamma}(10^{-6}$GeV)    &1.40$\pm$0.07 &1.54$\pm$0.18   &- &- &0.94$\pm$0.01\\
    $\Gamma_{\eta' \rightarrow \gamma \gamma}(10^{-6}$GeV)   &6.51$\pm$0.66  &8.33$\pm$2.69   &- &-  & 4.35$\pm$0.16\\
    \hline
  \end{tabular} 
  \caption{\label{Tab:width}Our results for decay widths, compared with that of Fit~II in Ref.~\cite{Qin:2020udp}, Fit~IV of Ref.~\cite{Dai:2013joa}, and PDG~\cite{ParticleDataGroup:2020ssz}.}
}
  \end{table}
Notice that these widths have smaller weights than cross sections, where the latter dominates in constraining the parameters given in Table \ref{Tab:para}. 
In an overall view, the predicted widths are compatible with that of PDG \cite{ParticleDataGroup:2020ssz} and those given by Refs.~\cite{Dai:2013joa,Qin:2020udp}, confirming the reliability of our model.  Since $\alpha_V$ now has a minus sign, the width of $\phi\to e^+e^-$ is a bit worse than that of the previous works \cite{Dai:2013joa,Qin:2020udp}. However, as will be discussed in the next section, a minus $\alpha_V$ gives a better description of the cross sections.

\begin{figure}[htbp]
\centering 
\includegraphics[width=1\textwidth,height=0.9\textheight]{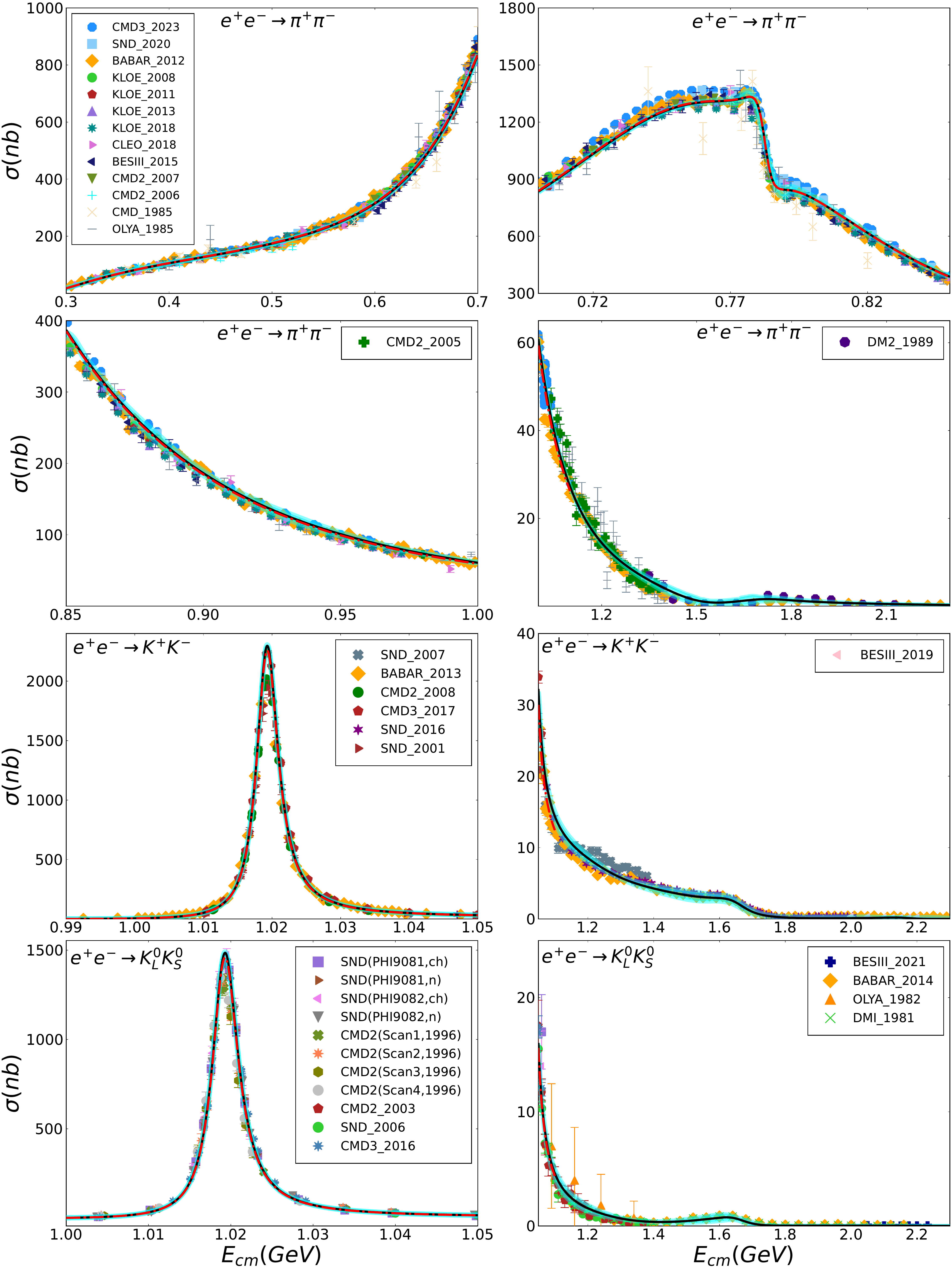}
\caption{\label{Fig:eePP} Fits to the cross sections of $e^{+}e^{-} \rightarrow \pi^{+}\pi^{-},K^{+}K^{-},K_{L}^{0}K_{S}^{0}$. The dash-doted red lines are for Fit A, and solid black lines (with the cyan bands being the uncertainty) are for Fit B. }
\end{figure}

Our fits for the cross sections of electron-positron annihilating into two pseudoscalars are shown in Fig.~\ref{Fig:eePP}, where the top four graphs are for $e^{+}e^{-} \rightarrow \pi^{+}\pi^{-}$, and the bottom four graphs are for $e^{+}e^{-} \rightarrow K^+K^-, K^0_L K^0_S$, respectively.   For $e^{+}e^{-} \rightarrow \pi^{+}\pi^{-}$, the experimental datasets are from BaBar \cite{BaBar:2012bdw}, KLOE \cite{KLOE:2008fmq,KLOE:2010qei,KLOE:2012anl,KLOE-2:2017fda}, SND \cite{SND:2020nwa}, BESIII \cite{BESIII:2015equ}, CLEO \cite{Xiao:2017dqv}, CMD-2 \cite{CMD-2:2005mvb,Aulchenko:2006dxz,CMD-2:2006gxt}, DM2 \cite{DM2:1988xqd}, and CMD \& OLYA \cite{Barkov:1985ac}\footnote{After this work is done, CMD-3 announced their new measurements on $e^+e^-\to\pi\pi$ \cite{CMD-3:2023alj}. But it would not affect our solutions much as ours are obtained through a combined analysis of various kinds of processes and datasets. Especially the $e^+e^-\to\gamma P$ processes would constrain the resonance parameters, and the most important contribution to HVP is from these resonances region.}. For $e^{+}e^{-}\rightarrow K^{+}K^{-}$, the experimental datasets are from SND \cite{Achasov:2000am,Achasov:2007kg,Achasov:2016lbc}, BaBar \cite{BaBar:2013jqz}, CMD-2 \cite{CMD-2:2008fsu}, CMD-3 \cite{Kozyrev:2017agm}, and BESIII \cite{BESIII:2018ldc}. For $e^{+}e^{-}\rightarrow K_{L}^{0}K_{S}^{0}$, the experiment datasets are from SND \cite{Achasov:2000am,Achasov:2006bv}, BaBar \cite{BaBar:2014uwz}, CMD-2 \cite{CMD-2:1999chh}, CMD-3 \cite{CMD-3:2016nhy}, BESIII \cite{BESIII:2021yam}, OLYA \cite{Ivanov:1982cr} and DM1 \cite{Mane:1980ep}.
Notice that some of the experimental datasets are given in terms of the dressed cross sections, and some others are of the bare ones. For example, the following datasets are of bare ones:  Refs.~\cite{BaBar:2013jqz,BESIII:2015equ,KLOE:2008fmq,KLOE:2010qei,KLOE:2012anl,KLOE-2:2017fda,Xiao:2017dqv} for $\pi^+\pi^-$, Refs.~\cite{BaBar:2013jqz,BESIII:2018ldc} for $K^+K^-$, and Ref.~\cite{BESIII:2021yam} for $K^0_L K^0_S$. 
We have changed all the bare cross sections into the dressed ones to give a unified description. 
Among them, some of the datasets for $e^+e^-\to \pi^+\pi^-$, e.g., Babar \cite{BaBar:2013jqz}, BESIII \cite{BESIII:2015equ}, and KLOE \cite{KLOE:2008fmq,KLOE:2010qei,KLOE:2012anl,KLOE-2:2017fda} also include the final state radiation (FSR) effects, and we will remove them by dividing the factor $1+\delta^{\pi\pi}_{add.FSR}$ \cite{BaBar:2013jqz}. Of course, for the charged kaons the FSR is ignorable. 
Some of the datasets are given by Born cross sections with the effects of initial/final states radiation removed. See e.g., Ref.~\cite{SND:2020nwa} for $\pi^+\pi^-$, Refs.~\cite{Achasov:2007kg,Achasov:2016lbc} for $K^+K^-$, and Ref.~\cite{CMD-3:2016nhy} for $K^0_L K^0_S$.
We fit our solutions to these Born cross sections. 

As can be found, our solutions fit the data rather well. 
Also, the solution of Fit~A (dash-dotted red lines) almost overlaps with that of Fit~B (solid black lines) in these processes with two final pseudoscalars. The only slight difference is for $e^+e^-\to K^+K^-$ in the energy region of $1.05-1.1$~GeV. The reason is that Fit B has to balance the data in higher energy regions. Moreover, our results of $e^+e^-\to \pi^+\pi^-$ improve a bit in the energy region of $\rho^0-\omega$ mixing, compared with the previous analysis. This is benefited from the modified mixing mechanism as given in Eq.~(\ref{eq:rho,omega}). The results of $e^+e^-\to K^0_L K^0_S$ are of high quality, and they help to refine our analyses due to coupled channels effects. 
The uncertainty bands are calculated from the Bootstrap method \cite{Efron:1979bxm} within 1~$\sigma$. It is worth pointing out that the uncertainty bands are significant in the energy region of the resonances, $\rho^0,\omega,\phi$. The reason is that it lacks high statistical measurements in these energy regions, and some of the data have apparent discrepancies with each other. It is expected that future experiments will focus on these energy regions, and this will help to fix the HVP contribution to the $(g-2)_\mu$.

\begin{figure}[htbp]
\centering 
\includegraphics[width=1\textwidth,height=0.6\textheight]{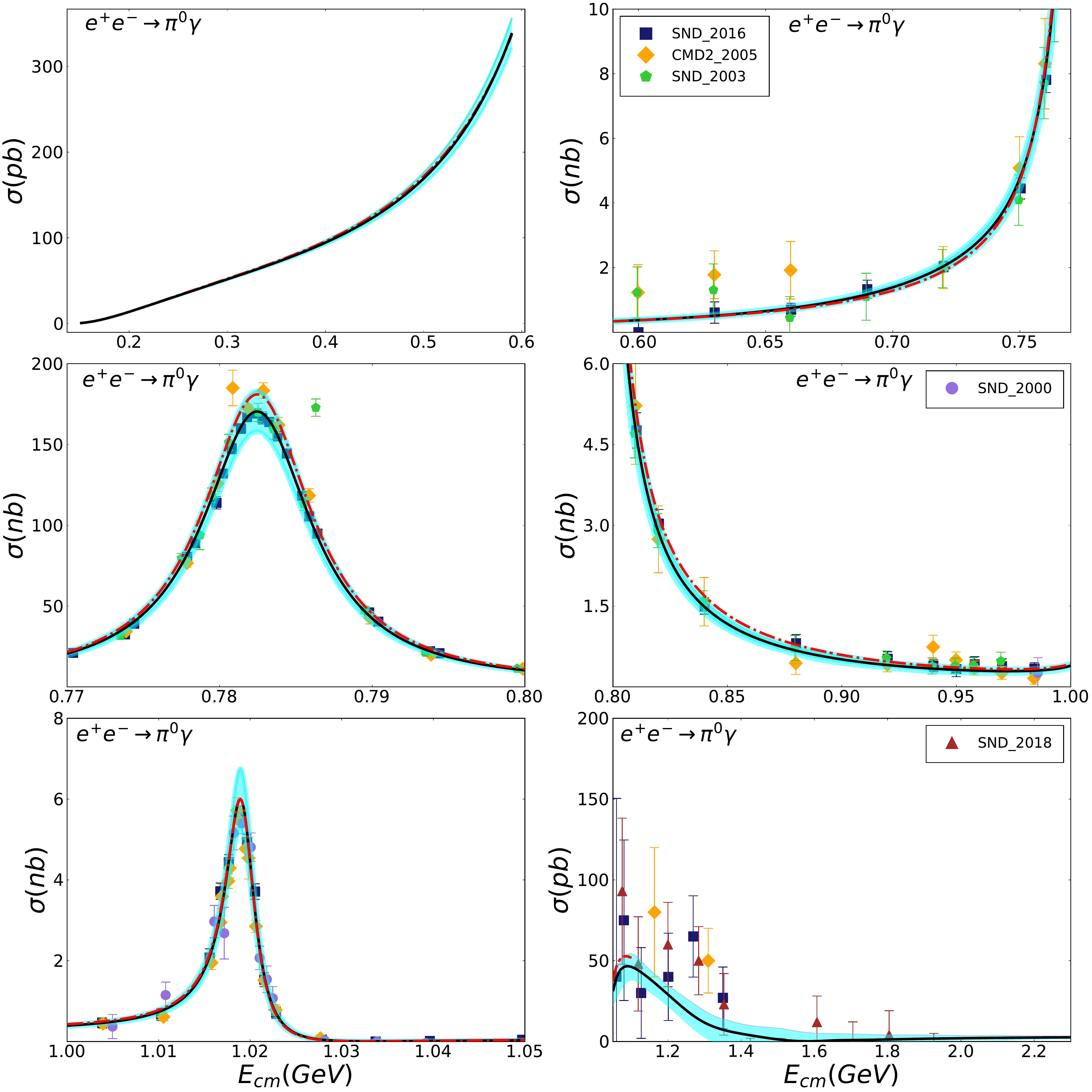}
\caption{\label{Fig:gpi} Fits for the cross sections of $e^{+}e^{-} \rightarrow \pi^{0}\gamma$. The dash-doted red lines are for Fit~A, and the solid black lines are for Fit~B, with the cyan bands the uncertainty of Fit~B. The experimental data displayed are from SND \cite{Achasov:2000zd,Achasov:2003ed,SND:2016drm,Achasov:2018ujw}, CMD-2 \cite{CMD-2:2004ahv}.}
\end{figure}

\begin{figure}[h!]
\centering 
\includegraphics[width=1\textwidth,height=0.4\textheight]{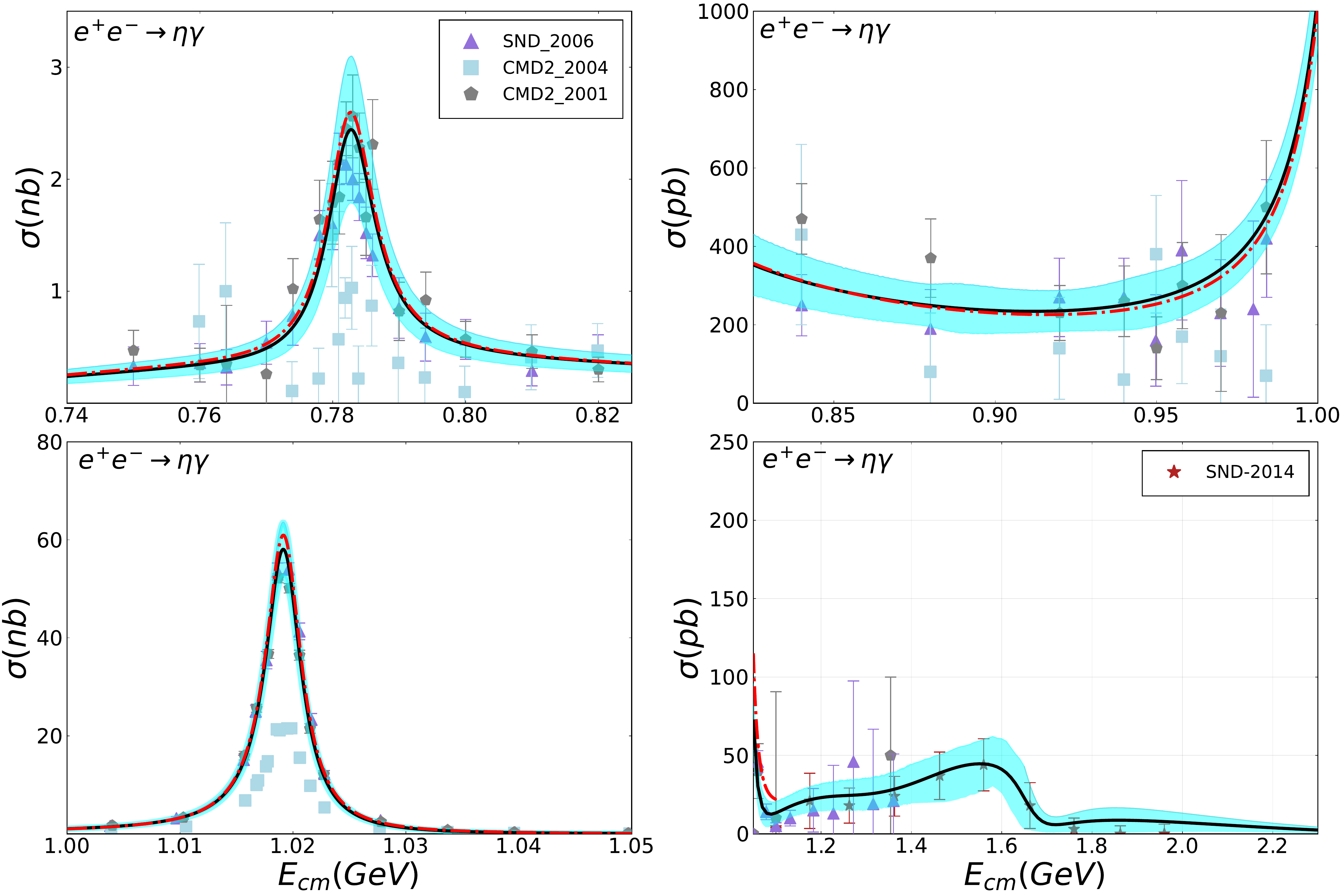}
\caption{\label{Fig:geta} Fit results for the cross sections of $e^{+}e^{-} \rightarrow\eta \gamma$ of Fit~A (dash-dotted red line) and Fit~B (solid black line), the cyan bands correspond to the uncertainty of Fit~B. The experimental data for $e^{+}e^{-}\rightarrow \eta\gamma$ are from SND~\cite{Achasov:2013eli,Achasov:2006dv}, CMD-2 in 2001~\cite{CMD-2:2001dnv}, and CMD-2 in 2004~\cite{CMD-2:2004ahv}.}
\end{figure}

The results of the processes of electron-positron annihilating into one pseudoscalar and one photon are shown in Figs.~\ref{Fig:gpi} and \ref{Fig:geta}.
For $e^{+}e^{-} \rightarrow \pi^{0}\gamma$, our results fit the datasets well. To be more careful, it is found that Fit~A does not fit better than Fit B, though Fit~A focuses on the low-energy region only. See e.g., the dash-dotted red (Fit~A) and solid black (Fit~B) lines around the $\rho^0$ peak. 
The reason is that including the heavier vector resonances would improve the amplitudes according to the interference between the contributions from ground states and heavier states. This is also presented in Ref.~\cite{Dai:2013joa}. 
Besides, the \lq peak' around $\rho,\omega$ region is sensitive to the couplings $d_2$ and $c_3$, and masses and widths of $\omega$ and $\rho$. This is helpful in fixing the amplitudes of the processes of $e^+e^-\to \pi\pi$. 
Interestingly, one would notice that the solutions on the left and right side of the $\phi(1020)$ peak are not symmetric, with the ones on the left side region larger. Conversely, the results of $e^+e^-\to K^+K^-, K^0_{L}K^0_{S}$ on the left side of $\phi$ are smaller than that on the right side. This indicates that the RChT is successful in describing the physics in the energy region where resonances appear.  

For $e^{+}e^{-} \rightarrow \eta\gamma$, the number of datasets is much less, and they are not precise enough. Especially, the CMD-2 data in 2004 \cite{CMD-2:2004ahv} are inconsistent with the other two datasets, and we do not include them in the fit. Correspondingly, the solutions here have the largest uncertainty bands in all these processes. Nevertheless, ours fit the latest data of SND \cite{Achasov:2013eli,Achasov:2006dv} well. 
In addition, the decay widths of $\eta\to\gamma\gamma$ and $\eta'\to\gamma\gamma,\rho^0\gamma, \omega\gamma$ are also fitted to constrain the unknown couplings, e.g., $\theta_{P}$, $d_{2}$, $c_{3}$.  
Besides, the \lq peak' around $\phi$ region is sensitive to the couplings $\theta_V$, $d_2$, $c_3$, and $\alpha_V$, and mass and width of $\phi$, which will be helpful in fixing the amplitudes of the processes of $e^+e^-\to \bar{K}K$. 
Similar to that discussed in $e^{+}e^{-} \rightarrow \pi^0\gamma$, the solutions on the left side of $\phi$ are larger than that on the right side. This can be checked by further experiments. 


\section{Leading-order HVP contributions to \texorpdfstring{$a_{\mu}$}{}}
\label{Sec:IV}
About the formalism of the leading order (LO) HVP correction to the muon anomalous magnetic moment, $a_\mu \equiv(g-2)_\mu/2$, it has been given in Ref.~\cite{Qin:2020udp}. It can be obtained by the cross
sections of electron-positon annihilation into hadrons, through the optical theorem and analyticity \cite{Gourdin:1969dm,Jegerlehner:2017gek}
\begin{equation}
a_{\mu}^{\mbox{\tiny HVP, LO}}=\frac{\alpha^2_{e}(0)}{3\pi^2} \int_{s_{\mbox{\tiny th}}}^{\infty}\mathrm{d}s\frac{\hat{K}(s)}{s}R_{\rm h}(s) \, ,\label{eq:amu}
\end{equation}
where $\alpha_{e}(0)=e^{2}/(4\pi)$ is the electromagnetic fine-structure constant, \lq th' represents the threshold, and $\hat{K}(s)$ is the kernel function,
\begin{equation} \label{eq:amuker}
\hat{K}(s)=\left.\left[\frac{\left(1+x^{2}\right)(1+x)^{2}}{x^{2}}\left(\ln(1+x)-x+\frac{x^{2}}{2}\right)\right.\left.+\frac{x^{2}}{2}\left(2-x^{2}\right)+\frac{1+x}{1-x}x^{2}\ln x\right]\right.
\end{equation}
with
\begin{eqnarray} \label{eq:amusup2}
x=\frac{1-\rho_{\mu}(s)}{1+\rho_{\mu}(s)}, & \qquad & \qquad \rho_{\mu}(s)=\sqrt{1-\frac{4m_{\mu}^{2}}{s}} \, .
\end{eqnarray}
The (hadronic) $R$-ratio is estimated from 
\begin{eqnarray} \label{eq:amusup}
R_{\rm h}(s)=\frac{3s}{4\pi\alpha_{e}^{2}(s)} \, \sigma\left(e^{+}e^{-}\rightarrow\text{ hadrons }\right) \, ,
\end{eqnarray}
where one has \cite{Jegerlehner:2009ry,Aoyama:2020ynm}
\begin{eqnarray}
\alpha_e(s)=\frac{\alpha_e(0)}{1-\Delta \alpha(s)}\,, \qquad\qquad \Delta \alpha(s)= \Pi_\gamma^{'}(0)-\Pi_\gamma^{'}(s) \,,
\end{eqnarray}
with $\Pi_\gamma(s)$ the vacuum polarization operator. 
Its QED part up to one loop is given as \cite{Sturm:2013uka}
\begin{eqnarray}
{\rm Re}\Pi^{'}_{\gamma~ {l}}(s)=-\frac{\alpha_e(0)}{4\pi} \left[\frac{20}{9}+\frac{16m_l^2}{3s}-\frac{2[3-\rho^2_l(s)]\rho_l(s)}{3}\ln\frac{\rho_l(s)+1}{\rho_l(s)-1}\right]\,. \label{eq:Pi;l}
\end{eqnarray}
Its hadronic part can be obtained by dispersion relations:
\begin{eqnarray}
{\rm Re} \Pi^{'}_{\gamma~ {\rm had}}(s)=\frac{\alpha_e(0)s}{3\pi } {\rm P}\int_{s_{ \rm th } }^{\infty} \frac{R(s')}{s'(s'-s)}ds'\,, \label{eq:Pi;had}
\end{eqnarray}
where \lq$l$' is for different types of leptons, and \lq P' is for the principal value. The $R$-ratio in Eq.~(\ref{eq:Pi;had}) is from \cite{ParticleDataGroup:2020ssz}, where the linear interpolation method has been used to give results between the nearby two data points. Notice that here the $s_{\rm th}$ is started from $m_\pi^2$, where the $R$ value between $m_\pi^2$-$4m_\pi^2$ has been estimated from our analysis on $e^+e^-\to\pi^0\gamma$. 
Notice that we also add the contributions of the narrow resonances in terms of Breit-Wigner forms \cite{BESIII:2021wib} for the $R$ value
\begin{eqnarray}
\Pi^{'}_{\gamma~ {\rm res}}(s)=-\frac{3s}{\alpha_e(0) } \sum_r\frac{\Gamma_{0,r}^{ee}}{M_r} \frac{1}{M_r^2-s-i M_r \Gamma_r}\,, \label{eq:Pi;res}
\end{eqnarray}
where $r$ is for $J/\psi$, $\psi(2S)$, $\Upsilon(1S)$,  $\Upsilon(2S)$, and $\Upsilon(3S)$, respectively.

\begin{table}[tp]
  \centering{\scalebox{.8}{\small\renewcommand\arraystretch{1.5}
  \begin{tabular}{cccccc}
  \hline
  $a_{\mu}^{C}\times 10^{-10}$  & Ref.~\cite{Colangelo:2018mtw}  & Ref.~\cite{Davier:2019can}  & Ref.~\cite{Qin:2020udp}     & Fit~A                           & Fit~B                                                  \\ \hline
  $a_{\mu}^{\pi\pi} \vert \leq  0.63$ GeV      & 132.8(0.4)(1.0)    &    &130.10 $\pm$0.67         & 130.70 $\pm$0.02 &129.91$\pm$0.19\\
  $a_{\mu}^{\pi\pi} \vert \leq  1$ GeV         & 495.0(1.5)(2.1)    &    &496.45 $\pm$2.33     & 496.50$\pm$0.93   &495.28$\pm$2.14\\
  $a_{\mu}^{\pi\pi} \vert \leq  1.8$ GeV       & -              &$507.85\pm0.83\pm3.23\pm0.55$                        &506.69 $\pm$2.45      &-  &505.64$\pm$2.33 \\
  $a_{\mu}^{\pi\pi} \vert \leq  2.3$ GeV       &-                   &    &506.93 $\pm$2.48   &-    &505.89$\pm$2.34\\
  \hline
  $a_{\mu}^{KK} \vert \leq  1.1$ GeV           & -   &              &19.88$\pm$0.88             &18.92 $\pm$0.41   &19.42$\pm$0.15\\
  $a_{\mu}^{KK} \vert \leq  1.8$ GeV           & -   & $23.08\pm0.20\pm0.33\pm0.21$  &23.44 $\pm$0.97   &-    &22.99$\pm$0.37\\
  $a_{\mu}^{KK} \vert \leq  2.3$ GeV           & -   &              &23.52 $\pm$1.01       &-  &23.03$\pm$0.43\\ 
   \hline
  $a_{\mu}^{K_{L}^{0}K_{S}^{0}} \vert \leq  1.1$ GeV  & -      &         &     &11.79$\pm$0.27   &12.02$\pm$0.11        \\
  $a_{\mu}^{K_{L}^{0}K_{S}^{0}} \vert \leq  1.8$ GeV  & -      & $12.82\pm0.06\pm0.18\pm0.15$  &          &-  &12.67$\pm$0.24\\
  $a_{\mu}^{K_{L}^{0}K_{S}^{0}} \vert \leq  2.3$ GeV  & -      &       &     &-   &12.67$\pm$0.24 \\
  \hline
  $a_{\mu}^{\pi^{0}\gamma} \vert \leq  0.63$ GeV      & -      &      &     &0.15$\pm$0.01  &0.15$\pm$0.01 \\
  $a_{\mu}^{\pi^{0}\gamma} \vert \leq  1$ GeV         & -      &      &    &4.66$\pm$0.04   &4.55$\pm$0.08  \\
  $a_{\mu}^{\pi^{0}\gamma} \vert \leq  1.8$ GeV       & -      & $4.41\pm0.06\pm0.04\pm0.07$      &         &-   &4.61$\pm$0.09\\
  $a_{\mu}^{\pi^{0}\gamma} \vert \leq  2.3$ GeV       & -      &    &         &-  &4.61$\pm$0.10\\
  \hline 
  $a_{\mu}^{\eta\gamma} \vert \leq  1$ GeV            & -      &    &       &0.20$\pm$0.02       &0.19$\pm$0.02    \\
  $a_{\mu}^{\eta\gamma} \vert \leq  1.8$ GeV          & -      & $0.65\pm0.02\pm0.01\pm0.01$  &         &- &0.66$\pm$0.06\\
  $a_{\mu}^{\eta\gamma} \vert \leq  2.3$ GeV          & -      &    &     &-  &0.66$\pm$0.06\\
  \hline 
  $a_{\mu}^{\pi\pi\pi} \vert \leq  1.8$ GeV          & -      & $46.21\pm0.40\pm1.10\pm0.86$  &48.46$\pm$1.42    &      &\\
  $a_{\mu}^{\pi\pi\pi} \vert \leq  2.3$ GeV          & -      &    &48.68$\pm$1.45   &    & \\
  \hline 
  $a_{\mu}^{\pi\pi\eta} \vert \leq  1.8$ GeV          & -      &$1.19\pm0.02\pm0.04\pm0.02$  & 1.28$\pm$0.10    &      &\\
  $a_{\mu}^{\pi\pi\eta} \vert \leq  2.3$ GeV          & -      &    &1.49$\pm$0.12    &    &\\
  \hline 
  $a_{\mu}^{HVP.LO}$      & -   & $694.0\pm4.0$       & 695.54$\pm$3.35      &-&694.10$\pm$3.14       \\
  $a_{\mu}^{SM}$          & -   & $11659183.1\pm4.8$  & 11659183.4$\pm$3.7          & -   & 11659182.0$\pm$3.5\\
  $\Delta a_{\mu}$        & -   & $26.0\pm7.9(3.3\sigma)$ &  22.7$\pm$5.5(4.1$\sigma$)        & -   & 24.1$\pm$5.4(4.5$\sigma$)  \\
  \hline
  \end{tabular}
  }
  \caption{\label{Tab:gm2} Our predictions of muon anomalous magnetic moment, where other contributions are from Ref.~\cite{Aoyama:2020ynm} and references therein. Notice that we replace the results of Ref.~\cite{Davier:2019can} with ours to obtain $a_\mu^{HVP,LO}$. The averaged experimental value is $a_{\mu}^{\exp}=116 592 061(41) \times 10^{-11}$ \cite{Muong-2:2021ojo}.}
  } 
  \end{table}
As is well known, due to the $1/s$ factor in Eq.~(\ref{eq:amu}), the LO HVP contributions to $a_{\mu}^{\mbox{\tiny had}}$ are dominated by the low-energy physics. 
In particular, the lowest-lying resonance $\rho (770)$ is coupled strongly to $\pi^+ \pi^-$ and the pion pair production gives roughly 73\% contribution to $a_{\mu}^{\mbox{\tiny had}}$ with 58\% errors \cite{Davier:2019can}.
However, the resonance parameters are correlated in all the processes discussed above. Especially, the processes of $e^+e^-\to\pi^0\gamma, \eta\gamma$ are helpful to fix the resonance parameters of $\rho, \omega, \phi$, resulting in strong constraints on the amplitudes of the processes of $e^+e^-\to \pi\pi, \bar{K}K$. 
That is why we need an overall analysis of all these processes.
Nevertheless, we are in the position to determine the contributions to the muon anomalous magnetic moment relevant
to the two pseudoscalar final states that we discussed above. The LO HVP contributions are shown in Table~\ref{Tab:gm2}, given as $a_{\mu}^{C}$, with $C$ the label of process, i.e., $C=\pi\pi, KK, K_{L}^{0}K_{S}^{0},\pi^{0}\gamma, \eta\gamma$ represent $e^{+}e^{-} \rightarrow \pi^+\pi^-, K^+K^-, K_{L}^{0}K_{S}^{0}, \pi^{0}\gamma, \eta\gamma$, respectively. Note that the  $a_{\mu}^{C}$ cut at different energies is also given for comparison.

Summing over all the contributions from the five processes calculated in this paper, the contributions from $e^{+}e^{-}\rightarrow \pi\pi\pi, \pi\pi\eta$ given by Ref.~\cite{Qin:2020udp}\footnote{It should be noticed that in Ref.~\cite{Qin:2020udp}, the correction of the fine structure constant is ignored, here we correct it and the contribution to HVP is smaller.},  and the contributions of the other channels given by Ref.~\cite{Davier:2019can}, we get the total contributions to the LO HVP as $a_{\mu}^{HVP,LO}=(694.10\pm3.14)\times 10^{-10}$ for Fit B. The uncertainties are estimated by the bootstrap method within 1~$\sigma$. At the end of the day, combining ours with the other contributions such as QED, EW, NLO HVP, NNLO HVP, and HLBL within the SM \cite{Aoyama:2020ynm}, we have a discrepancy of 4.5~$\sigma$ for the anomalous magnetic moment of the muon between theoretical prediction and the latest average value of experimental measurements \cite{Muong-2:2021ojo}\footnote{Recently, Lattice QCD announced the latest results on hadronic light-by-light contribution to the muon anomaly \cite{Blum:2023vlm}. If we replace the result of HLBL in the standard model \cite{Aoyama:2020ynm} with that of lattice QCD \cite{Blum:2023vlm}, the final result of deviation of muon $g-2$ is 3.9$\sigma$ for Fit B.}.  
In addition, ours is indeed an improved data-driven method where RChT is applied, and final state interactions are considered. It has an apparent discrepancy with that of the lattice QCD \cite{Borsanyi:2020mff}, which needs further study.  

\section{Conclusion}
\label{Sec:V}
In the framework of resonance chiral theory, we have carried out a combined comprehensive analysis of the processes with two final states: $e^{+}e^{-}\rightarrow \pi\pi, KK, K_{L}^{0}K_{S}^{0}, \pi^{0}\gamma, \eta\gamma$. The final state interactions are considered.
Two solutions are obtained: In Fit~A, the focused energy region is up to 1.1 GeV, and in Fit~B, it is up to 2.3 GeV. Both solutions are of high quality. With the obtained form factors, we evaluate the LO HVP contribution to the muon magnetic anomaly, as shown in Table \ref{Tab:gm2}. Although the contributions to the LO HVP are tiny, the processes of $e^+e^-\to\pi^0\gamma, \eta\gamma$ are helpful to fix the resonance parameters, resulting in a refined analysis of the processes of electron-positron annihilating into two pseudoscalars, and further for the LO HVP estimation. 
Combining ours with the results of electron-positron annihilation into three pseudoscalars, $\pi\pi\pi$ and $\eta\pi\pi$, and other processes predicted by Ref.~\cite{Davier:2019can}, we get $a_\mu^{HVP,LO}=694.10\pm3.14$. This gives $\Delta a_{\mu}=24.1 \pm 5.4$ (4.5$\sigma$) with the other contributions taken from the SM prediction\cite{Aoyama:2020ynm},  compared with the latest results from experimental measurements \cite{Muong-2:2021ojo}. This large discrepancy between theoretical predictions and experimental measurements implies new physics beyond the standard model. Further measurements of the electron-positron annihilation  in the resonance region would be essential to check the HVP contributions and the muon $g-2$.

\section*{Acknowledgements}
\label{Sec:VI}
We thank helpful discussions with Professors  W. Qin, W.-B. Yan, J. Portoles, and H.-Q. Zheng, and Mr. Y.-L. Ye. This project is supported by Joint Large Scale Scientific Facility Funds of the National Natural Science Foundation of China (NSFC) and Chinese Academy of Sciences (CAS) under Contract No.U1932110, NSFC Grant with No. 11905055 and 12061141006, the Natural Science Foundation of Hunan Province of China under Grant No. 2023JJ30115, and Fundamental Research Funds for the Central Universities. 

\appendix
\section{The vector form factors for \texorpdfstring{$e^+e^-\to PP$}{} }\label{app:Fv}
The relevant vector form factors for the processes of $e^{+}e^{-} \rightarrow \pi^{+}\pi^{-},K^{+}K^{-},K_{L}^{0}K_{S}^{0}$, as defined in Eq.~(\ref{eq:Fv}), are given as follows
\begin{eqnarray*}
F^{\pi^+\pi^-}_{V} &=& \bigg( 1+\frac{F_{V}G_{V}}{F^{2}} Q^2 BW_{R}^{\pi^+\pi^-}[\rho, Q^{2}]~  (\frac{1}{\sqrt{3}} \sin\theta_{V} \sin\delta^{\rho} + \cos\delta)\cos\delta \nonumber\\  &&-\frac{F_{V}G_{V}}{F^{2}}Q^{2}BW_{R}^{\pi^+\pi^-}[\omega, Q^{2}]~ (\frac{1}{\sqrt{3}}\sin\theta_{V} \cos\delta -\sin\delta^{\omega})   \sin\delta^{\omega} \bigg)  \nonumber\\
&& \exp\bigg[ \frac{-Q^2}{96\pi^{2}F^{2}}{\rm Re}\big[ A(m_{\pi},M_{\rho},Q^{2})+\frac{1}{2}A(m_{K}, M_{\rho}, Q^{2}) \big] \bigg]   \,, \label{Eq,A:FF;pipi}\\[3mm]
F^{K^+K^-}_{V}&=& \bigg( \frac{F_{V}G_{V}}{24F^{2}}(1+8\sqrt{2}\alpha_{V}\frac{m_{\pi}^{2}}{M_{V}^{2}})M_{\rho}^{2} BW_{R}^{K^+K^-}[\rho, Q^{2}]~(16\sqrt{3} \cos\delta \sin\delta^{\rho} \sin\theta_{V} + 6\cos 2\delta \nonumber\\
&& -6\sin^{2}\delta^{\rho} \cos 2\theta_{V} +6\sin^{2}\delta^{\rho} +6) \bigg)  \exp \bigg[ \frac{-Q^{2}}{96\pi^{2}F^{2}} {\rm Re} \big[ A(m_{\pi},M_{\rho}, Q^{2})+\frac{1}{2}A(m_{K}, M_{\rho}, Q^{2}) \big] \bigg] \nonumber\\
&& +\bigg( \frac{F_{V}G_{V}}{24F^{2}} (1+8\sqrt{2} \alpha_{V} \frac{m_{\pi}^{2}}{M_{V}^{2}})M_{\omega}^{2} BW_{R}^{K^+K^-}[\omega, Q^{2}]~(-16\sqrt{3}\cos\delta\sin\delta^{\omega} \sin\theta_{V} \nonumber\\
&& -6\cos^{2}\delta \cos 2\theta_{V}  +3\cos 2\delta +12 \sin^{2}\delta^{\omega}+3)+\frac{\cos^{2}\theta_{V}}{2} \frac{F_{V}G_{V}}{F^{2}}(1+8\sqrt{2}\alpha_{V}\frac{2m_{K}^{2}-m_{\pi}^{2}}{M_{V}^{2}})  \nonumber\\
&& M_{\phi}^{2}  BW_{R}^{K^+K^-}[\phi, Q^{2}] \bigg) \exp \bigg[ \frac{-Q^{2}}{64\pi^{2}F^{2}} {\rm Re} [A(m_{K}, M_{\rho},Q^{2})] \bigg]  \,, \label{Eq,A:FF;KK}  \\[3mm]
F^{K_{L}^{0}K_{S}^{0}}_{V} & = & \bigg( -\frac{F_{V}G_{V}}{24F^{2}} (1+8\sqrt{2} \alpha_{V}\frac{m_{\pi}^{2}}{M_{V}^{2}})M_{\rho}^{2} BW_R^{K_{L}^{0}K_{S}^{0}}[\rho,Q^{2}]~(-8\sqrt{3} \cos \delta \sin\delta^{\rho} \sin\theta_{V} +6\cos2\delta\nonumber\\
&& +6\sin^{2}\delta^{\rho} \cos2\theta_{V} -6\sin^{2}\delta^{\rho} +6) \bigg)\exp\bigg[\frac{-Q^{2}}{96\pi^{2}F^{2}}{\rm Re}\big[ A(m_{\pi},M_{\rho},Q^{2})+\frac{1}{2}A(m_{K},M_{\rho}, Q^{2})\big]   \bigg] \nonumber\\
&& +\bigg( \frac{F_{V}G_{V}}{24F^{2}}(1+8\sqrt{2}\alpha_{V}\frac{m_{\pi^{2}}}{M_{V}^{2}})M_{\omega}^{2} BW_R^{K_{L}^{0}K_{S}^{0}}[\omega, Q^{2}] ~(-8\sqrt{3}\cos\delta \sin\delta^{\omega} \sin\theta_{V}  \nonumber\\
&& -6\cos^{2}\delta \cos2\theta_{V} +3 \cos 2\delta-6\sin^{2}\delta^{\omega} +3)+\frac{\cos^{2}\theta_{V}}{2}\frac{F_{V}G_{V}}{F^{2}}(1+8\sqrt{2}\alpha_{V}\frac{2m_{K}^{2}-m_{\pi}^{2}}{M_{V}^{2}})  \nonumber\\
&&  M_{\phi}^{2} BW_R^{K_{L}^{0}K_{S}^{0}}[\phi,Q^{2}] \bigg) 
\exp \bigg[\frac{-Q^{2}}{64\pi^{2}F^{2}} {\rm Re} [A(m_{K},M_{\rho}, Q^{2}) ]  \bigg]   \,.  \label{Eq,A:FF;KLKS}
\end{eqnarray*}
The loop functions in the above expressions are given by:
\begin{eqnarray}
A(m_{P}, \mu , Q^{2})  &=& \ln(\frac{m_{P}^{2}}{\mu^{2}}) + \frac{8m_{P}^{2}}{Q^{2}} -\frac{5}{3} +\sigma_{P}^{3}\ln \bigg(\frac{\sigma_{P}+1}{\sigma_{P}-1} \bigg) \,,\nonumber 
\end{eqnarray}
with the phase space factor $\sigma_{P} \equiv \sqrt{1-4m_{P}^{2}/Q^{2}}$.
\lq BW' is the generalized Breit-Wigner propagators of the resonances \cite{Dai:2013joa}
\begin{eqnarray} \label{eq:func2}
  BW(M_{V},\Gamma_{V},Q^{2})& = & \left[ M_{V}^{2}-iM_{V}\Gamma_{V}(Q^{2})-Q^{2} \right]^{-1}  \, , \nonumber \\
  BW_R^P[V,x]&=& BW[V,x]+ \beta_P' BW[V',x] + \beta_P'' BW[V'',x] \, , \nonumber \\
  BW_{RR}^P[V_1,V_2,x,y] &=& BW_R^P[V_1,x] \, BW_R^P[V_2,y] \, ,
\end{eqnarray} 
with the off-shell widths of the vector resonances taken from Ref.~\cite{Dai:2013joa}:
\begin{eqnarray}
 \Gamma_\rho(q^2)&=&\frac{M_\rho \,  q^2}{96\pi F^2}\left[\sigma^{3}_\pi(q^2)\ \theta(q^2-4m_\pi^2)+\frac{1}{2}\sigma^{3}_K(q^2)\ \theta(q^2-4m_K^2)\right] \, , \nonumber \\
 \Gamma_{\rho'}(q^2)&=&\Gamma_{\rho_0'}(M_{\rho '}^2) \frac{\sqrt{q^2}}{M_{\rho'}}
 \left( \frac{\sigma_\pi(q^2)}{\sigma_\pi(M_{\rho'}^2)}\right)^{3} \theta(q^2-4m_\pi^2),\nonumber\\
 \Gamma_{\rho''}(q^2)&=&\Gamma_{\rho_0''}(M_{\rho ''}^2)\frac{\sqrt{q^2}}{M_{\rho''}}
 \left( \frac{\sigma_\pi(q^2)}{\sigma_\pi(M_{\rho''}^2)}\right)^{3} \theta(q^2-4m_\pi^2) \,, \nonumber 
\end{eqnarray}
where $\theta(x)$ is the step function.

\section{The transition form factors for \texorpdfstring{$e^{+}e^{-}\rightarrow P\gamma$}{}} \label{app:F;Pgg}
\subsection{\texorpdfstring{$P =\pi^{0}$}{}}
The transition form factors relevant to the process of $e^{+} e^{-} \rightarrow \pi^{0} \gamma$, as defined by Eq.~(\ref{eq:EM;transit}), are given as 
\begin{eqnarray*}
F_{V}^{\pi} = F_{a}^{\pi} + F_{b}^{\pi} + F_{c}^{\pi} + F_{d}^{\pi} \,, 
\end{eqnarray*}
with each part given as
\begin{eqnarray*}
F_{a}^{\pi}&=&\frac{N_{C}}{12\pi^{2}F} \,, \label{Eq,A:FF;V}\\[3mm]
F_{b}^{\pi}&=&\bigg(\frac{F_{V}(1+8\sqrt{2}\alpha_{V}\frac{m_{\pi}^{2}}{M_{V}^{2}})}{18F M_{V}} BW_{R}^{\pi}[\rho,s] \big(24\sqrt{3}\cos\delta  \sin\delta^{\rho}(s) \cos\theta_{V}+16\sqrt{6}\cos\delta \sin\delta^{\rho}(s)  \sin\theta_{V} \nonumber\\
&&+3\sqrt{2}(2\cos 2\delta -(1-2\sin^{2}\delta^{\rho}(s))+3)+12\sin^{2}\delta^{\rho} \sin 2\theta_{V}-6\sqrt{2}\sin^{2}\delta^{\rho}(s)\cos 2\theta_{V}\big) \nonumber\\
&& +\frac{F_{V}(1+8\sqrt{2}\alpha_{V}\frac{m_{\pi}^{2}}{M_{V}^{2}})}{18FM_{V}}BW_{R}^{\pi}[\omega,s] \big(-24\sqrt{3} \cos\delta \sin\delta^{\omega}(s) \cos\theta_{V} -16\sqrt{6}\cos\delta \sin\delta^{\omega}(s) \sin\theta_{V} \nonumber\\ 
&& +3\sqrt{2}(\cos 2\delta -2(1-2\sin^{2}\delta^{\omega}(s))+3)-6\sqrt{2}\cos^{2}\delta \cos 2\theta_{V}  +12\cos^{2}\delta \sin 2\theta_{V}\big) \nonumber\\ && +\frac{2F_{V}(1+8\sqrt{2}\alpha_{V}\frac{2m_{K}^{2}-m_{\pi}^{2}}{M_{V}^{2}})}{3FM_{V}}BW_{R}^{\pi}[\phi,s]\cos\theta_{V}(\sqrt{2}\cos\theta_{V}-2\sin\theta_{V})\bigg)C_{R\pi}(0,s)   \,, \nonumber\\[3mm]
F_{c}^{\pi}&=&\bigg(\frac{F_{V}(1+8\sqrt{2}\alpha_{V}\frac{m_{\pi}^{2}}{M_{V}^{2}})}{18F M_{V}} BW_{R}^{\pi}[\rho,0] \big(24\sqrt{3}\cos\delta \sin\delta^{\rho}(0)  \cos\theta_{V}+16\sqrt{6}\cos\delta \sin\delta^{\rho}(0) \sin\theta_{V}  \nonumber\\
&&+3\sqrt{2}(2\cos 2\delta-(1-2\sin^{2}\delta^{\rho}(0))+3)+12\sin^{2}\delta^{\rho}(0) \sin 2\theta_{V}-6\sqrt{2}\sin^{2}\delta^{\rho}(0) \cos 2\theta_{V}\big)   \nonumber\\
&&+\frac{F_{V}(1+8\sqrt{2}\alpha_{V}\frac{m_{\pi}^{2}}{M_{V}^{2}})}{18FM_{V}} BW_{R}^{\pi}[\omega,0] \big(-24\sqrt{3}\cos\delta \sin\delta^{\omega}(0) \cos\theta_{V}-16\sqrt{6}\cos\delta \sin\delta^{\omega}(0) \sin\theta_{V}   \nonumber\\
&&+3\sqrt{2}(\cos 2\delta-2(1-2\sin^{2}\delta^{\omega}(0))+3)-6\sqrt{2}\cos^{2}\delta\cos 2\theta_{V}+12\cos^{2}\delta \sin 2\theta_{V}\big)   \nonumber\\
&&+\frac{2F_{V}(1+8\sqrt{2}\alpha_{V}\frac{2m_{K}^{2}-m_{\pi}^{2}}{M_{V}^{2}})}{3FM_{V}} BW_{R}^{\pi}[\phi,0] \cos\theta_{V}(\sqrt{2}\cos\theta_{V}-2\sin\theta_{V})\bigg) C_{R\pi}(s,0)   \,,\nonumber\\[3mm]
F_{d}^{\pi}&=&\bigg(-\frac{F_{V}^{2}(1+8\sqrt{2}\alpha_{V}\frac{m_{\pi}^{2}}{M_{V}^{2}})^{2}}{9F} BW_{RR}^{\pi}[\rho,\rho,0,s]  \big(\cos\delta (\sin \delta^{\rho}(s)+\sin \delta^{\rho}(0))\\
&&(\sin\theta_{V}+\sqrt{2}\cos\theta_{V}) (12\cos\delta (\sin \delta^{\rho}(s)+\sin \delta^{\rho}(0)) \sin \theta_{V} +\sqrt{3}(6\cos 2\delta\\
&&-(1-2\sin\delta^{\rho}(0)\sin \delta^{\rho}(s))+7)-2\sqrt{3}\sin\delta^{\rho}(0)\sin \delta^{\rho}(s)\cos 2\theta_{V})\big) \nonumber\\ 
&& +\frac{F_{V}^{2}(1+8\sqrt{2}\alpha_{V}\frac{m_{\pi}^{2}}{M_{V}^{2}})^{2}}{18F} BW_{RR}^{\pi}[\omega,\omega,0,s]\big(\cos\delta (\sin \delta^{\omega}(0)+\sin \delta^{\omega}(s))\\
&&(\sin\theta_{V}+\sqrt{2}\cos\theta_{V})  (-24\cos\delta (\sin\delta^{\omega}(s)+\sin\delta^{\omega}(0)) \sin\theta_{V}+2\sqrt{3}(\cos 2\delta\\
&&-6(1-2\sin\delta^{\omega}(s)\sin\delta^{\omega}(0))+7)-4\sqrt{3}\cos^{2}\delta \cos 2\theta_{V})\big) \nonumber\\ 
&& -\frac{2F_{V}^{2}(1+8\sqrt{2}\alpha_{V}\frac{m_{\pi}^{2}}{M_{V}^{2}})^{2}}{9F}BW_{RR}^{\pi}[\rho,\omega,0,s] (\sin\theta_{V} +\sqrt{2}\cos\theta_{V}) \nonumber\\ 
&& (\cos^{2}\delta-\sin\delta^{\omega}(s) \sin\delta^{\rho}(0))(-3\sin\theta_{V}(\sin^{2}\delta+2\sin\delta^{\rho}(0) \sin\delta^{\omega}(s)-1)\\
&&+2\sqrt{3}\cos\delta (-3\sin\delta^{\omega}(s) +\sin\delta^{\rho}(s) \sin^{2}\theta_{V})+3\cos^{ 2}\delta \sin\theta_{V})\\ 
&& -\frac{2F_{V}^{2}(1+8\sqrt{2}\alpha_{V}\frac{m_{\pi}^{2}}{M_{V}^{2}})^{2}}{9F}BW_{RR}^{\pi}[\omega,\rho,0,s](\sin\theta_{V} +\sqrt{2}\cos\theta_{V}) \nonumber\\ 
&& (\cos^{2}\delta-\sin\delta^{\omega}(0) \sin\delta^{\rho}(s))(-3\sin\theta_{V}(\sin^{2}\delta+2\sin\delta^{\rho}(s) \sin\delta^{\omega}(0)-1)\\
&&+2\sqrt{3}\cos\delta (-3\sin\delta^{\omega}(0) +\sin\delta^{\rho}(s) \sin^{2}\theta_{V})+3\cos^{ 2}\delta \sin\theta_{V})\\ 
&&-\frac{4F_{V}^{2}(1+8\sqrt{2}\alpha_{V}\frac{m_{\pi}^{2}}{M_{V}^{2}})(1+8\sqrt{2}\alpha_{V}\frac{2m_{K}^{2}-m_{\pi}^{2}}{M_{V}^{2}})}{9F} BW_{RR}^{\pi}[\rho,\phi,0,s]\cos\delta \cos\theta_{V}\\
&&(\cos\theta_{V}-\sqrt{2} \sin\theta_{V}) (3\cos\delta+\sqrt{3}\sin\delta^{\rho}(0) \sin\theta_{V})\\
&&-\frac{4F_{V}^{2}(1+8\sqrt{2}\alpha_{V}\frac{m_{\pi}^{2}}{M_{V}^{2}})(1+8\sqrt{2}\alpha_{V}\frac{2m_{K}^{2}-m_{\pi}^{2}}{M_{V}^{2}})}{9F}BW_{RR}^{\pi}[\phi,\rho,0,s]\cos\delta \cos\theta_{V} \\
&&(\cos\theta_{V}-\sqrt{2} \sin\theta_{V}) (3\cos\delta+\sqrt{3}\sin\delta^{\rho}(s) \sin\theta_{V})\\
&&+\frac{4F_{V}^{2}(1+8\sqrt{2}\alpha_{V}\frac{m_{\pi}^{2}}{M_{V}^{2}})(1+8\sqrt{2}\alpha_{V}\frac{2m_{K}^{2}-m_{\pi}^{2}}{M_{V}^{2}})}{9F} BW_{RR}^{\pi}[\omega,\phi,0,s]\sin\delta^{\omega}(0) \cos\theta_{V}\\
&&(\cos\theta_{V}-\sqrt{2}\sin\theta_{V})  (-3\sin\delta^{\omega}(0)+\sqrt{3}\cos\delta \sin\theta_{V})\\
&&+\frac{4F_{V}^{2}(1+8\sqrt{2}\alpha_{V}\frac{m_{\pi}^{2}}{M_{V}^{2}})(1+8\sqrt{2}\alpha_{V}\frac{2m_{K}^{2}-m_{\pi}^{2}}{M_{V}^{2}})}{9F}BW_{RR}^{\pi}[\phi,\omega,0,s]\sin\delta^{\omega}(s) \cos\theta_{V}\\
&&(\cos\theta_{V}-\sqrt{2}\sin\theta_{V}) (-3\sin\delta^{\omega}(s)+\sqrt{3}\cos\delta \sin\theta_{V})\bigg)D_{R\pi}(0,s)   \, .  \label{Eq,A:FF;V;abcd}
\end{eqnarray*}
Notice that the functions such as $C_{R}$ and $D_{R}$ functions are defined in Ref.~\cite{Dai:2013joa}
\begin{eqnarray}
  C_{R\pi}(Q^2,x)&=&(c_1-c_2+c_5) \, Q^2- (c_1-c_2-c_5+2 c_6) \,  x+ (c_1+c_2+8 c_3-c_5) \,  m_\pi^2 \; ,  \nonumber \\
  D_{R\pi}(Q^2,x)&=& d_3 \, (Q^2+x)+(d_1+8d_2-d_3) \, m_\pi^2 \; ,  \nonumber \\
  C_{R\eta1}(Q^2,x,m^2)&=&  (c_1-c_2+c_5) Q^2-(c_1-c_2-c_5+2 c_6) x+(c_1+c_2-c_5)m^2 \; , \nonumber \\
  C_{R\eta2}&=& 8 \, c_3 \;  , \nonumber \\
  D_{R\eta 1}(Q^2,x,m^2)&=&d_3(Q^2+x)+(d_1-d_3) \, m^2 \; , \nonumber \\
  D_{R\eta 2}&=& 8 \, d_2 \; . \nonumber 
  \end{eqnarray}
The equations with superscript $\eta$ will be applied in the process with final states $\eta\gamma$. 

\subsection{\texorpdfstring{$ P = \eta $}{}}
The transition form factors relevant to the processes of $e^{+} e^{-} \rightarrow \eta \gamma$, as defined by Eq.~(\ref{eq:EM;transit}), are given by 
\begin{eqnarray*}
F_{V}^{\eta} = F_{a}^{\eta} + F_{b}^{\eta} + F_{c}^{\eta} + F_{d}^{\eta} \,, \label{Eq,A:FF;t}
\end{eqnarray*}
with each part of them given as 
\begin{eqnarray*}
F_{a}^{\eta}&=&\frac{N_{C} (\cos \theta_{P}-2 \sqrt{2} \sin\theta_{P})}{12 \sqrt{3} \pi ^2 F}     \,, \nonumber\\[3mm]
F_{b}^{\eta}&=&-\frac{2 F_{V}(1+8\sqrt{2}\alpha_{V}\frac{m_{\pi}^{2}}{M_{V}^{2}})}{27 F M_{V}} BW_{R}^{\eta}[\rho,s](3 \cos \delta+\sqrt{3} \sin\delta^{\rho}(s) \sin \theta_{V}) \nonumber\\
&&\bigg( (-3C_{R\eta1}(0,s,m_{\eta}^{2}))[\sqrt{3} \cos \delta (\sqrt{2} \cos \theta_{P}-2 \sin \theta_{P})\\
&&+2 \sin \delta^{\rho}(s) \cos \theta_{V} \cos \theta_{P}-\sin\delta^{\rho}(s) \sin \theta_{V} (2 \sin \theta_{P}+\sqrt{2} \cos \theta_{P})]\\
&&+2C_{R\eta2} \sin \delta^{\rho}(s) m_{K}^2 (\sqrt{2} (3 \sin (\theta_{V} -\theta_{P} )+\sin (\theta_{V} +\theta_{P} ))-4 \cos (\theta_{V} +\theta_{P} ))\nonumber\\
&&-\frac{C_{R\eta2}}{2} m_{\pi }^2 [6 \sqrt{3} \cos \delta (\sqrt{2} \cos \theta_{P}-2 \sin \theta_{P})\\
&&+\sin\delta^{\rho}(s) (\sqrt{2} (9 \sin (\theta_{V} -\theta_{P} )+\sin (\theta_{V} +\theta_{P} ))-4 \cos (\theta_{V} +\theta_{P} ))]\bigg) \nonumber\\
&&-\frac{2 F_{V}(1+8\sqrt{2}\alpha_{V}\frac{m_{\pi}^{2}}{M_{V}^{2}})}{27 F M_{V}}BW_{R}^{\eta}[\omega,s]  (\sqrt{3} \cos \delta \sin\theta_{V}-3 \sin \delta^{\omega}(s)) \bigg(3 \big(\cos \delta (2 \cos \theta_{V} \cos \theta_{P}\\
&&-\sin \theta_{V} (2 \sin \theta_{P}+\sqrt{2} \cos \theta_{P}))-\sqrt{3} \sin\delta^{\omega}(s) (\sqrt{2} \cos\theta_{P}-2 \sin\theta_{P})\big)(-C_{R\eta1}(0,s,m_{\eta}^{2})) \nonumber\\
&&+2C_{R\eta2} \cos \delta  m_{K}^{2} (\sqrt{2} (3 \sin (\theta_{V} -\theta_{P} )+\sin (\theta_{V} +\theta_{P} ))-4 \cos (\theta_{V} +\theta_{P} ))+C_{R\eta2} m_{\pi }^2 (\cos \delta\nonumber\\
&&(2 \cos (\theta_{V} +\theta_{P} )+\sqrt{2} (4 \cos \theta_{V} \sin \theta_{P}-5 \sin \theta_{V} \cos \theta_{P} ))\\
&&-6 \sqrt{3} \sin \delta^{\omega}(s) \sin \theta_{P}+3 \sqrt{6} \sin \delta^{\omega}(s) \cos \theta_{P})\bigg) \nonumber\\
&&+\frac{F_{V}(1+8\sqrt{2}\alpha_{V}\frac{2m_{K}^{2}-m_{\pi}^{2}}{M_{V}^{2}})}{9 \sqrt{3} F M_{V}}BW_{R}^{\eta}[\phi,s]\cos \theta_{V}  \bigg(6 \big(\cos\theta_{V} (2 \sin \theta_{P}+\sqrt{2} \cos \theta_{P})\\
&&+2 \sin \theta_{V} \cos \theta_{P}\big)(-C_{R\eta1}(0,s,m_{\eta}^{2}))-C_{R\eta2} \big(3 \sqrt{2} (4 m_{K}^{2}-3 m_{\pi }^{2}) \cos (\theta_{V} -\theta_{P})\nonumber\\
&&+(4 m_{K}^{2}-m_{\pi}^{2}) (4 \sin (\theta_{V} +\theta_{P} )+\sqrt{2} \cos (\theta_{V}+\theta_{P}))\big)\bigg)  \,, \nonumber \\[3mm]
F_{c}^{\eta}&=&\frac{2 F_{V}(1+8\sqrt{2}\alpha_{V}\frac{m_{\pi}^{2}}{M_{V}^{2}})}{27 F M_{V} }BW_{R}^{\eta}[\rho,0]   (3 \cos \delta+\sqrt{3} \sin \delta^{\rho}(0) \sin \theta_{V}) \bigg(3C_{R\eta1}(s,0,m_{\eta}^{2})\nonumber\\
&&(\sqrt{3} \cos \delta (\sqrt{2} \cos \theta_{P}-2 \sin \theta_{P})+2 \sin \delta^{\rho}(0) \cos \theta_{V} \cos \theta_{P}-\sin\delta^{\rho}(0) \sin \theta_{V} (2 \sin \theta_{P}\\
&&+\sqrt{2} \cos \theta_{P}))-2C_{R\eta2} \sin\delta^{\rho}(0) m_{K}^{2} (\sqrt{2} (3 \sin (\theta_{V} -\theta_{P})+\sin (\theta_{V} +\theta_{P}))-4 \cos (\theta_{V} +\theta_{P}))\\
&&+\frac{C_{R\eta2}}{2} m_{\pi }^2 \big(6 \sqrt{3} \cos \delta (\sqrt{2} \cos \theta_{P}-2 \sin \theta_{P})+\sin \delta^{\rho}(0) (\sqrt{2} (9 \sin (\theta_{V} -\theta_{P})\\
&&+\sin (\theta_{V} +\theta_{P}))-4 \cos (\theta_{V} +\theta_{P}))\big)\bigg)\nonumber\\
&&+\frac{2 F_{V}(1+8\sqrt{2}\alpha_{V}\frac{m_{\pi}^{2}}{M_{V}^{2}})}{27 F M_{V}}BW_R^{\eta}[\omega,0]  (\sqrt{3} \cos \delta \sin\theta_{V}-3 \sin \delta^{\omega}(0)) \bigg(3C_{R\eta1}(s,0,m_{\eta}^{2}) \\
&&\big(\cos \theta_{P}(2 \cos \delta  \cos\theta_{V}-\sqrt{2} \cos \delta  \sin\theta_{V}-\sqrt{6} \sin\delta^{\omega}(0))-2 \sin \theta_{P} (\cos \delta \sin\theta_{V}\\
&&-\sqrt{3} \sin \delta^{\omega}(0))\big)-2C_{R\eta2} \cos \delta m_{K}^{2} (\sqrt{2} (3 \sin (\theta_{V} -\theta_{P})+\sin (\theta_{V} +\theta_{P}))-4 \cos (\theta_{V} +\theta_{P}))\\
&&+\frac{C_{R\eta2}}{2} m_{\pi }^2\big(\cos \delta (\sqrt{2} (9 \sin (\theta_{V} -\theta_{P})+\sin (\theta_{V} +\theta_{P}))-4 \cos (\theta_{V} +\theta_{P}))-6 \sqrt{3} \sin\delta^{\omega}(0)\\
&&(\sqrt{2} \cos \theta_{P}-2 \sin \theta_{P})\big)\bigg)\nonumber\\
&& -\frac{F_{V}(1+8\sqrt{2}\alpha_{V}\frac{2m_{K}^{2}-m_{\pi}^{2}}{M_{V}^{2}})}{9 \sqrt{3} F M_{V}}BW_R^{\eta}[\phi,0] \cos\theta_{V} \bigg(6 \big(\cos\theta_{V} (2 \sin \theta_{P}+\sqrt{2} \cos \theta_{P})\\
&&+2 \sin\theta_{V} \cos \theta_{P}\big)C_{R\eta1}(s,0,m_{\eta}^{2})+C_{R\eta2} \big(3 \sqrt{2} (4 m_{K}^2-3 m_{\pi }^2) \cos (\theta_{V} -\theta_{P})\\
&&+(4 m_{K}^{2}-m_{\pi }^{2})(4 \sin (\theta_{V} +\theta_{P})+\sqrt{2} \cos (\theta_{V} +\theta_{P}))\big)\bigg) \,.  \nonumber\\[3mm]
F_{d}^{\eta} &=&-\frac{4 F_{V}^{2}(1+8\sqrt{2}\alpha_{V}\frac{m_{\pi}^{2}}{M_{V}^{2}})^{2}}{27 \sqrt{3} F }BW_{RR}^{\eta}[\rho,\rho,0,s]  (3 \cos \delta +\sqrt{3} \sin \delta^{\rho}(0)\sin\theta_{V})(3 \cos \delta \\
&&+\sqrt{3} \sin \delta^{\rho}(s)\sin\theta_{V})\bigg(\frac{3}{4} D_{R\eta1}(0,s,m_{\eta}^{2}) \big(\cos \theta_{P} (2 \cos 2 \delta + 2 \sin\delta^{\rho}(0)\sin\delta^{\rho}(s)\\
&&(2 \sqrt{2} \sin 2 \theta_{V}+\cos 2 \theta_{V})+(1-2\sin\delta^{\rho}(s)\sin\delta^{\rho}(0))+1)-2 \sqrt{2} \sin \theta_{P} (\cos 2 \delta\\
&&-(1-2\sin\delta^{\rho}(0)\sin\delta^{\rho}(s))+2)\big)+D_{R\eta2} \sin\delta^{\rho}(s) \sin\delta^{\rho}(0) m_{K}^{2} (2 \cos \theta_{P} (2 \sqrt{2} \sin 2 \theta_{V}\\
&&+\cos 2 \theta_{V}-3)+\sin \theta_{P} (4 \sin 2 \theta_{V}+\sqrt{2} \cos 2 \theta_{V}-3 \sqrt{2}))\\
&&+\frac{D_{R\eta2}}{4} m_{\pi }^2 [~12 \cos ^2\delta (\cos \theta_{P}-\sqrt{2} \sin \theta_{P})
+\sin \delta^{\rho}(0)\sin\delta^{\rho}(s) (-4 \sqrt{2} \sin (2 \theta_{V}+\theta_{P})\\
&&-9 \cos (2 \theta_{V}-\theta_{P})+7 \cos (2 \theta_{V}+\theta_{P})+18 \cos \theta_{P})~]\bigg)\\
&&-\frac{4 F_{V}^{2}(1+8\sqrt{2}\alpha_{V}\frac{m_{\pi}^{2}}{M_{V}^{2}})^{2}}{27 \sqrt{3} F }BW_{RR}^{\eta}[\omega,\omega,0,s](\sqrt{3} \cos \delta  \sin \theta_{V} -3 \sin \delta^{\omega}(s))(\sqrt{3} \cos \delta  \sin \theta_{V}\\
&&-3 \sin \delta^{\omega}(0))\bigg(-\frac{3}{4} D_{R\eta1}(0,s,m_{\eta}^{2}) (\cos \theta_{P} (-2 \cos ^2\delta  (2 \sqrt{2} \sin 2 \theta_{V}+\cos 2 \theta_{V})+\cos 2 \delta \\
&&+2 (1-2\sin\delta^{\omega}(0)\sin\delta^{\omega}(s))-1)+2 \sqrt{2} \sin \theta_{P} (\cos 2 \delta-(1-2\sin\delta^{\omega}(0)\sin\delta^{\omega}(s))+2))\\
&&+D_{R\eta2} \cos ^2\delta m_{K}^{2} (2 \cos \theta_{P} (2 \sqrt{2} \sin 2 \theta_{V}+\cos 2 \theta_{V}-3)+\sin \theta_{P} (4 \sin 2 \theta_{V}+\sqrt{2} \cos 2 \theta_{V}\\
&&-3 \sqrt{2}))+D_{R\eta2} m_{\pi }^2 \big(\frac{1}{4} \cos ^2\delta (-4 \sqrt{2} \sin (2 \theta_{V}+\theta_{P})-9 \cos (2 \theta_{V}-\theta_{P})\\
&&+7 \cos (2 \theta_{V}+\theta_{P})+18 \cos \theta_{P})+3 \sin\delta^{\omega}(0)\sin\delta^{\omega}(s) (\cos \theta_{P}-\sqrt{2} \sin \theta_{P})\big)\bigg)\\
&&+\frac{4 F_{V}^2(1+8\sqrt{2}\alpha_{V}\frac{2m_{K}^{2}-m_{\pi}^{2}}{M_{V}^{2}})^{2}}{9 \sqrt{3} F }BW_{RR}^{\eta}[\phi,\phi,0,s] \cos ^2\theta_{V} \bigg(3 (\cos \theta_{V} \cos \theta_{P} \\
&&(2 \sqrt{2} \sin \theta_{V}+\cos \theta_{V})+\sqrt{2} \sin \theta_{P})D_{R\eta1}(0,s,m_{\eta}^{2})+D_{R\eta2} m_{K}^{2} (2 \cos \theta_{P} (2 \sqrt{2} \sin 2 \theta_{V}\\
&&+\cos 2 \theta_{V}+3)+\sin\theta_{P} (4 \sin 2 \theta_{V}+\sqrt{2} \cos 2 \theta_{V}+3 \sqrt{2}))-\frac{D_{R\eta2}}{4}m_{\pi }^2 (4 \sqrt{2} \sin (2 \theta_{V}+\theta_{P})\\
&&+9 \cos (2 \theta_{V}-\theta_{P})-7 \cos (2 \theta_{V}+\theta_{P})+18 \cos \theta_{P})\bigg)\\
&&-\frac{4 F_{V}^{2}(1+8\sqrt{2}\alpha_{V}\frac{m_{\pi}^{2}}{M_{V}^{2}})^{2}}{27 \sqrt{3} F }BW_{RR}^{\eta}[\omega,\rho,0,s] \cos \delta (\sqrt{3} \cos \delta  \sin \theta_{V}-3 \sin \delta^{\omega}(0)) (3 \cos \delta\\
&&+\sqrt{3} \sin \delta^{\rho}(s) \sin \theta_{V})\bigg(3 D_{R\eta1}(0,s,m_{\eta}^{2}) [\cos \theta_{P} (-\sin \delta^{\omega}(0) \\
&&+\sin \delta^{\rho}(s) \sin \theta_{V} (2 \sqrt{2} \cos \theta_{V}-\sin \theta_{V}))-\sqrt{2} \sin \theta_{P} (-\sin\delta^{\omega}(0)+\sin \delta^{\rho}(s))~]\\
&&+D_{R\eta2} \sin\delta^{\rho}(s) m_{K}^{2} (2 \cos \theta_{P} (2 \sqrt{2} \sin 2 \theta_{V}+\cos 2 \theta_{V}-3)\\
&&+\sin \theta_{P} (4 \sin 2 \theta_{V}+\sqrt{2} \cos 2 \theta_{V}-3 \sqrt{2}))+\frac{D_{R\eta2}}{4} m_{\pi }^2 \big(-12 \sin\delta^{\omega}(0) (\cos \theta_{P}-\sqrt{2} \sin \theta_{P})\\
&&+\sin \delta^{\rho}(s)(-4 \sqrt{2} \sin (2 \theta_{V}+\theta_{P})-9 \cos (2 \theta_{V}-\theta_{P})+7 \cos (2 \theta_{V}+\theta_{P})+18 \cos \theta_{P})\big)\bigg)   \\
&&-\frac{4 F_{V}^2(1+8\sqrt{2}\alpha_{V}\frac{m_{\pi}^{2}}{M_{V}^{2}})^{2}}{27 \sqrt{3} F}BW_{RR}^{\eta}[\rho,\omega,0,s] \cos \delta  (\sqrt{3} \cos\delta  \sin \theta_{V}-3 \sin \delta^{\omega}(s)) \\
&&(3 \cos \delta+\sqrt{3} \sin \delta^{\rho}(0) \sin \theta_{V}) \bigg(3 D_{R\eta1}(0,s,m_{\eta}^{2}) (\cos \theta_{P} (-\sin \delta^{\omega}(s)+\sin \delta^{\rho}(0) \sin \theta_{V}\\
&&(2 \sqrt{2} \cos \theta_{V}-\sin \theta_{V}))-\sqrt{2} \sin \theta_{P} (-\sin\delta^{\omega}(s)+\sin \delta^{\rho}(0)))+D_{R\eta2}\sin\delta^{\rho}(0) m_{K}^2 (2 \cos \theta_{P}\\
&&(2 \sqrt{2} \sin 2 \theta_{V}+\cos 2 \theta_{V}-3)+\sin \theta_{P} (4 \sin 2 \theta_{V}+\sqrt{2} \cos 2 \theta_{V}-3 \sqrt{2}))\\
&&+\frac{D_{R\eta2}}{4} m_{\pi }^2 \big(-12 \sin \delta^{\omega}(s) (\cos \theta_{P}-\sqrt{2} \sin \theta_{P})+\sin \delta^{\rho}(0) (-4 \sqrt{2} \sin (2 \theta_{V}+\theta_{P})\\
&&-9 \cos (2 \theta_{V}-\theta_{P})+7 \cos (2 \theta_{V}+\theta_{P})+18 \cos \theta_{P})\big)\bigg)\\
&&+\frac{2 F_{V}^2(1+8\sqrt{2}\alpha_{V}\frac{m_{\pi}^{2}}{M_{V}^{2}})(1+8\sqrt{2}\alpha_{V}\frac{2m_{K}^{2}-m_{\pi}^{2}}{M_{V}^{2}})}{27 F}BW_{RR}^{\eta}[\phi,\rho,0,s] \sin\delta^{\rho}(s) \cos \theta_{V} \\
&&(3 \cos \delta +\sqrt{3} \sin \delta^{\rho}(s) \sin \theta_{V}) \bigg(2D_{R\eta2} (m_{K}^2-m_{\pi }^2) \sin \theta_{P} (\sqrt{2} \sin 2 \theta_{V}-4 \cos 2 \theta_{V})\\
&&-\cos \theta_{P} (2 \sqrt{2} \cos 2 \theta_{V}-\sin 2 \theta_{V})\big(3 D_{R\eta1}(0,s,m_{\eta}^{2})+D_{R\eta2}(4m_{K}^2-m_{\pi }^2)\big)\bigg)\\
&&+\frac{2 F_{V}^2(1+8\sqrt{2}\alpha_{V}\frac{m_{\pi}^{2}}{M_{V}^{2}})(1+8\sqrt{2}\alpha_{V}\frac{2m_{K}^{2}-m_{\pi}^{2}}{M_{V}^{2}})}{27 F}BW_{RR}^{\eta}[\rho,\phi,0,s] \sin \delta^{\rho}(0) \cos\theta_{V}\\
&&(3 \cos \delta +\sqrt{3} \sin \delta^{\rho}(0) \sin \theta_{V}) \bigg(2D_{R\eta1} (m_{K}^2-m_{\pi }^2) \sin \theta_{P} (\sqrt{2} \sin 2\theta_{V} -4 \cos 2\theta_{V} )\\
&&-\cos \theta_{P} (2 \sqrt{2} \cos 2\theta_{V}-\sin 2\theta_{V} )\big(3D_{R\eta1}(0,s,m_{\eta}^2)+D_{R\eta2}(4m_{K}^2- m_{\pi }^2)\big)\bigg)\\
&&-\frac{F_{V}^2(1+8\sqrt{2}\alpha_{V}\frac{m_{\pi}^{2}}{M_{V}^{2}})(1+8\sqrt{2}\alpha_{V}\frac{2m_{K}^{2}-m_{\pi}^{2}}{M_{V}^{2}})}{27 F} BW_{RR}^{\eta}[\phi,\omega,0,s]\cos \delta  \cos \theta_{V} \\
&&(2 \sqrt{3} \cos \delta  \sin \theta_{V}-6 \sin\delta^{\omega}(s))\bigg(\cos \theta_{P}(2 \sqrt{2} \cos 2 \theta_{V}-\sin 2 \theta_{V}) (3 D_{R\eta 1}(0,s,m_{\eta}^{2})\\
&&+D_{R\eta2} (4m_{K}^2-m_{\pi }^2))-2D_{R\eta2} (m_{K}^2-m_{\pi }^2) \sin \theta_{P} (\sqrt{2} \sin 2 \theta_{V}-4 \cos 2 \theta_{V})\bigg)\\
&&-\frac{F_{V}^2(1+8\sqrt{2}\alpha_{V}\frac{m_{\pi}^{2}}{M_{V}^{2}})(1+8\sqrt{2}\alpha_{V}\frac{2m_{K}^{2}-m_{\pi}^{2}}{M_{V}^{2}})}{27 F}BW_{RR}^{\eta}[\omega,\phi,0,s] \cos \delta  \cos \theta_{V} \\
&&(2 \sqrt{3} \cos \delta  \sin \theta_{V}-6 \sin \delta^{\omega}(0)) \bigg(\cos \theta_{P} (2 \sqrt{2} \cos 2 \theta_{V}-\sin 2 \theta_{V}) (3D_{\eta1}(0,s,m_{\eta}^{2})\\
&&+D_{R\eta2} (4m_{K}^2-m_{\pi }^2))-2D_{R\eta 2} (m_{K}^2-m_{\pi }^2) \sin \theta_{P}  (\sqrt{2} \sin 2 \theta_{V}-4 \cos 2 \theta_{V})\bigg)\,. \label{Eq,A:FF;eta;abcd}
\end{eqnarray*}

\section{Discussions on final state interactions}
\label{pro:FSI}
\begin{figure}[h]
\centering 
\includegraphics[width=1\textwidth,]{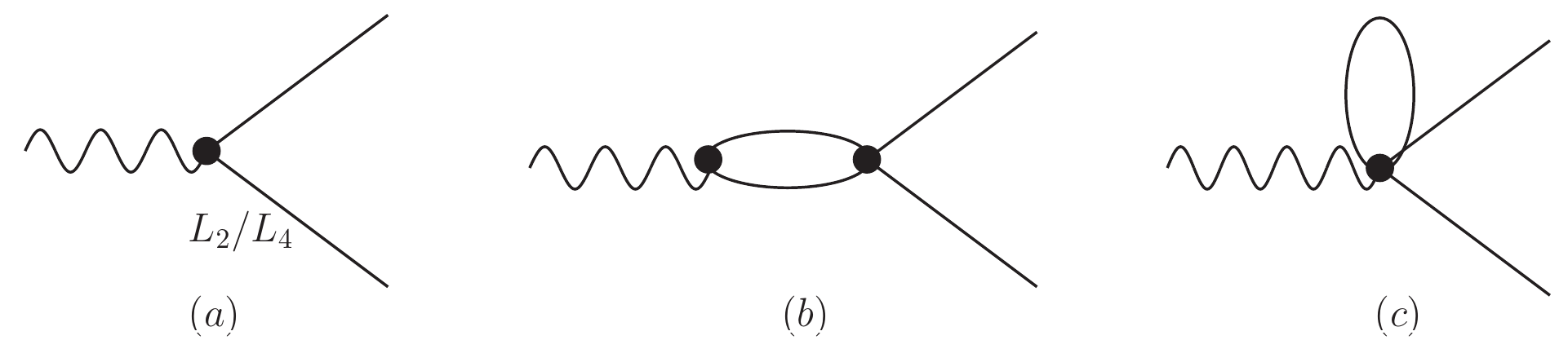}
\caption{\label{Fig:FD;KK}Feynman diagrams contributing to the hadronization of the vector current of $\gamma^*\to K^{+} K^{-}, K^{0}_{L} K^{0}_{S}$ in ChPT. }
\end{figure}

In this section, we discuss how to consider the final state interactions (FSI) in our form factors. In phenomenological analyses of scatterings and decays of hadrons, FSI should be taken into account to describe the physics appropriately \cite{Dai:2014zta,Dai:2014lza,Danilkin:2014cra,Yao:2020bxx}. The critical thought of this work is following Ref.~\cite{Guerrero:1997ku}, where the matching is performed between the two form factors of ChPT and RChT in the low energy region. 
For $\gamma^*\to\pi^+\pi^-$, it has been discussed in Ref.~\cite{Guerrero:1997ku}. For the processes of $\gamma^*\rightarrow K^{+} K^{-}, K^{0}_{L} K^{0}_{S}$, one has such Feynman diagrams given by ChPT. See Fig.~\ref{Fig:FD;KK}.
The form factors for each kind of diagram are given as
\begin{eqnarray}
F^{a}&=&1+\frac{1}{F^{2}}\big( 2 L_{9}Q^{2}+8L_{4}m_{K}^{2}+4m_{K}^{2}L_{5}+4m_{\pi}^{2}L_{4}\big) \,,\nonumber\\[3mm]
F^{b}&=&\frac{1}{576\pi^{2}F^{2}}(3(Q^{2}-4m_{K}^{2})B_{0}^{r}(Q^{2},m_{\pi}^{2},m_{\pi}^{2})-6A_{0}^{r}(m_{\pi}^{2})-12m_{\pi}^{2}+2Q^{2})  \nonumber\\
&&+\frac{1}{288\pi^{2}F^{2}}(3(Q^{2}-4m_{K}^{2})B_{0}^{r}(Q^{2},m_{K}^{2},m_{K}^{2})-6A_{0}^{r}(m_{K}^{2})-12m_{K}^{2}+2Q^{2})  \,,\nonumber\\[3mm]
F^{c}&=&\frac{A_{0}^{r}(m_{\pi}^{2})}{192\pi^{2}F^{2}}+\frac{A_{0}^{r}(m_{\pi}^{2})}{24\pi^{2}F^{2}}+\frac{A_{0}^{r}(m_{K}^{2})}{96\pi^{2}F^{2}}+\frac{A_{0}^{r}(m_{K}^{2})}{12\pi^{2}F^{2}}+\frac{A_{0}^{r}(m_{\eta}^{2})}{64\pi^{2}F^{2}} \,,\nonumber
\end{eqnarray}
where the function $B^r_0(Q^2,m_a^2,m_b^2)$ is defined as \cite{Passarino:1978jh}
 \begin{eqnarray}
    B^r_0(Q^2,m_a^2,m_b^2)&=&\frac{\lambda^{1/2}(Q^2,m_a^2,m_b^2) }{Q^2}\ln \left(\frac{\sqrt{Q^2-({m_a}+{m_b})^2}-\sqrt{Q^2-({m_a}-{m_b})^2}}{\sqrt{Q^2-({m_a}-{m_b})^2}+\sqrt{Q^2-({m_a}+{m_b})^2}}\right)\nonumber\\ 
    &&-\ln \left(\frac{{m_b}^2}{\mu ^2}\right)+2+\frac{\left({m_a}^2-{m_b}^2+s\right) }{2 Q^2}\ln \left(\frac{{m_b}^2}{{m_a}^2}\right)\,, \nonumber
 \end{eqnarray}
where $\lambda(a,b,c)=(a+b-c)^2-4ab$ is the triangle function, and the renormalization scale is fixed as $\mu=M_V$. 
Also, the wave function renormalization up to $\mathcal{O}(p^4)$ is given as 
\begin{eqnarray*}
Z_{K} = 1-\frac{A_{0}^{r}(m_{\pi}^{2})}{64\pi^{2}F^{2}}-\frac{A_{0}^{r}(m_{\eta}^{2})}{64\pi^{2}F^{2}}-\frac{A_{0}^{r}(m_{K}^{2})}{32\pi^{2}F^{2}}-\frac{8}{F^{2}}(2L_{4}m_{K}^{2}+L_{4}m_{\pi}^{2}+L_{5}m_{K}^{2}) \,.\nonumber
\end{eqnarray*}
Combining all of these form factors, one can obtain the vector form factors of $\gamma^*\to K^+K^-$ within ChPT up to $\mathcal{O}(p^4)$  
\begin{eqnarray}
  F^{\rm ChPT}_{K^+K^-}&=&1+\frac{2L_{9}}{F^{2}}Q^2+\frac{-Q^2}{96\pi^{2}F^{2}}(A[m_{K},M_{\rho},Q^2]+\frac{1}{2}A[m_{\pi},M_{\rho},Q^2])\,. \label{Eq,A:FKK;ChPT}
\end{eqnarray}
Note that $L_9=F^2/(2M_V^2)$ \cite{Passarino:1978jh}.  
In contrast, the vector form factors of $\gamma^*\to K^+K^-$ within RChT are given as
\begin{eqnarray}
F^{\rm RChT}_{K^+K^-}&=&\frac{M_{\rho}^{2}}{2(M_{\rho}^{2}-Q^{2})}+\frac{M_{\omega}^{2}}{6(M_{\omega}^{2}-Q^{2})}+\frac{M_{\phi}^{2}}{3(M_{\phi}^{2}-Q^{2})}
   \,. 
\end{eqnarray}
Notice that the form factor of RChT is calculated in the ideal mixing case through 
Eq.~(\ref{Eq:FPP;ideal}), setting the mixing angles $\delta=0$, ignoring $\eta-\eta'$ mixing, taking all the high energy constraints on the couplings and setting $\alpha_V=0$. Also, the total widths of the resonances in the propagators have been ignored. 
However, in the \lq physical' world, the width of $\rho$ is too large to be ignored, and only $\Gamma_\omega$ and $\Gamma_\phi$ can be safely set to be zero. Hence, we consider the FSI in the same way as Ref.~\cite{Guerrero:1997ku} 
\begin{eqnarray*}
F^{\rm RChT,phy}_{K^+K^-}&=&\frac{M_{\rho}^{2}}{2(M_{\rho}^{2}-Q^{2}-i M_V \Gamma_\rho(Q^2))}\exp\bigg[\frac{-Q^2}{96\pi^{2}F^{2}}Re(A[m_{\pi},M_{\rho},Q^2]+\frac{1}{2}A[m_{K},M_{\rho},Q^2])\bigg]\\
  &&+\big[\frac{M_{\omega}^{2}}{2(M_{\omega}^{2}-Q^{2})}+\frac{M_{\phi}^{2}}{3(M_{\phi}^{2}-Q^{2})}\big]\exp\bigg[\frac{-Q^2}{96\pi^{2}F^{2}} {\rm Re}\frac{3}{2}A[m_{K},M_{\rho},Q^2] \bigg] \,.\\
\end{eqnarray*}
Performing an expansion on the $Q$ with the limit of $Q\to 0$ and taking the $SU(3)$ limit $M_\rho=M_\omega=M_\phi=M_V$, we have 
\begin{eqnarray*}
  F^{\rm RChT,phy}_{K^+K^-}&=&(\frac{1}{2}+\frac{Q^{2}}{2M_{\rho}^{2}}+i\frac{\Gamma_{\rho}}{2M_{\rho}})\big(1+\frac{-Q^2}{96\pi^{2}F^{2}}Re(A[m_{\pi},M_{\rho},Q^2]+\frac{1}{2}A[m_{K},M_{\rho},Q^2])\big)\\
  &\quad+&(\frac{1}{6}+\frac{Q^{2}}{6M_{\omega}^{2}})(1+\frac{-Q^2}{96\pi^{2}F^{2}}Re\frac{3}{2}A[m_{K},M_{\rho},Q^2])+(\frac{1}{3}+\frac{Q^{2}}{3M_{\phi}^{2}}) \nonumber\\ && (1+\frac{-Q^2}{96\pi^{2}F^{2}}\frac{3}{2}Re A[m_{K},M_{\rho},Q^2])\\
    &=&1+\frac{Q^2}{M_V^2}-\frac{Q^2}{96\pi^{2}F^{2}}(\frac{1}{2}A[m_{\pi},M_{\rho},Q^2]+A[m_{K},M_{\rho},Q^2])\,.
  \end{eqnarray*}
It is the same as that of ChPT up to $\mathcal{O}(p^4)$.
Similar discussions can be made for the vector form factors of $\gamma*\to K^{0}_{L} K^{0}_{S}$. In ChPT, one has    
\begin{eqnarray*}
F^{\rm ChPT}_{K^{0}_{L} K^{0}_{S}} &=& \frac{-Q^2}{96\pi^{2}F^{2}}(\frac{1}{2}A[m_{K},M_{\rho},Q^2]-\frac{1}{2}A[m_{\pi},M_{\rho},Q^2]) \,. 
\end{eqnarray*}
In the ideal mixing case of RChT and taking the same conditions as discussed in the $K^+K^-$ form factors to simplify the model, one has 
\begin{eqnarray*}
F^{\rm RChT}_{K_{L}^{0}K_{S}^{0}}&=&-\frac{M_\rho^{2}}{2(M_{\rho}^{2}-Q^{2})}+\frac{M_\omega^{2}}{6(M_{\omega}^{2}-Q^{2})}+\frac{M_\phi^{2}}{3(M_{\phi}^{2}-Q^{2})}  \,. 
\end{eqnarray*}
The final state interactions are taken into account as 
\begin{eqnarray*}
F^{\rm RChT, phy}_{K_{L}^{0}K_{S}^{0}}&=&
-\frac{M_{\rho}^{2}}{2(M_{\rho}^{2}-Q^{2}-i M_\rho \Gamma_\rho(Q^2))}\exp\bigg[\frac{-Q^{2}}{96\pi^{2}F^{2}}Re(A[m_{\pi},M_{\rho},Q^{2}]+\frac{1}{2}A[m_{K},M_{\rho},Q^{2}])\bigg]\\
&&+(\frac{M_\omega^{2}}{6(M_{\omega}^{2}-Q^{2})}+\frac{M_\phi^{2}}{3(M_{\phi}^{2}-Q^{2})})\exp\bigg[\frac{-Q^{2}}{96\pi^{2}F^{2}}Re(\frac{3}{2}A[m_{K},M_{\rho},Q^{2}])\bigg] \,. \nonumber
\end{eqnarray*}
In the low energy expansion and taking $SU(3)$ limit, one has
\begin{eqnarray*}
  F^{\rm RChT, phy}_{K_{L}^{0}K_{S}^{0}}&=&
-\frac{1}{2}-\frac{Q^{2}}{2M_{V}^{2}}-\frac{i\Gamma_{V}}{2M_{V}}+\frac{1}{2}\frac{Q^{2}}{96\pi^{2}F^{2}}Re[A[m_{\pi},M_{\rho},Q^{2}]+\frac{1}{2}A[m_{K},M_{\rho},Q^2]]\\
  &&+\frac{1}{6}+\frac{Q^{2}}{6M_{V}^{2}}+\frac{1}{6}\frac{-Q^2}{96\pi^{2}F^{2}}Re[\frac{3}{2}A[m_{K},M_{\rho},Q^2]]\\
  &&+\frac{1}{3}+\frac{Q^{2}}{3M_{V}^{2}}+\frac{1}{3}\frac{-Q^2}{96\pi^{2}F^{2}}Re[\frac{3}{2}A[m_{K},M_{\rho},Q^2]]\\
  &=&-\frac{Q^2}{96\pi^{2}F^{2}}[\frac{1}{2}A[m_{K},M_{\rho},Q^2]-\frac{1}{2}A[m_{\pi},M_{\rho},Q^2]] \,.
  \end{eqnarray*}
Again, it is the same as that of ChPT up to $\mathcal{O}(p^4)$. 

For the transition form factor of $\gamma^*\to \pi^0\gamma$, it is found that there is no need to include the FSI. This is partly because the high energy constraints on the coupling constants are obtained in the chiral limit, and the form factors will vanish at $Q\to\infty$. The terms violating these constraints are those involved with the pion mass, which is still suppressed by $m_\pi^2$ and can be ignored. 
However, for the transition form factor of $\gamma^*\to \eta\gamma$, the high energy behavior of the form factors is not so well as the high energy constraints on the coupling constants will be violated by the terms with the mass of ${m_\eta}^2$, which are large and obvious. Here we use an exponential regulator as applied in Refs.~\cite{Dai:2017ont,Yang:2022qoy}, $f(Q)=\exp\left[ -Q^6/\Lambda^6\right]$. $\Lambda$ is the cut-off, and we set it to be 2.3~GeV. The merit of this regulator is that it behaves as a step function with almost no effects in the low energy region ($Q\leq 1.7GeV$) but suppresses the form factors strongly in the high energy region. Notice that the behavior is compatible with the data \cite{Achasov:2013eli}, where the cross sections are tiny and ignorable above 1.7~GeV. 
\section{Decay widths involving vector resonances}\label{app:decay widths}
The decay widths involving the lightest vector resonances are given below. 
Note that some of them are only slightly different from the ones given in Ref.~\cite{Qin:2020udp}, as only the $\rho-\omega$ mixing mechanism has been changed. 
\begin{eqnarray*}
\Gamma_{\rho \rightarrow \pi\pi}=\frac{G_{V}^{2}M_{\rho}^{3}}{48\pi F^{4}}\cos^{2}\delta (1-\frac{4m_{\pi}^{2}}{M_{\rho}^{2}})^{\frac{3}{2}} \,,
\end{eqnarray*}

\begin{eqnarray*}
    \Gamma_{\omega \rightarrow \pi\pi}=\frac{G_{V}^{2}M_{\omega}^{3}}{48\pi F^{4}} \sin^{2}\delta^{\omega}(M_{\omega}^{2}) (1-\frac{4m_{\pi}^{2}}{M_{\omega}^{2}})^{\frac{3}{2}} \,,
\end{eqnarray*}

\begin{eqnarray*}
    \Gamma_{\phi \rightarrow \pi\pi} = \frac{\alpha^{2}\pi F_{V}^{2}}{9M_{\phi}}(1+8\sqrt{2}\alpha_{V}\frac{2m_{K}^{2}-m_{\pi}^{2}}{M_{V}^{2}})^{2} \cos^{2}\theta_{V}(1-\frac{4m_{\pi}^{2}}{M_{\phi}^{2}})^{\frac{3}{2}}\,, 
\end{eqnarray*}

\begin{eqnarray*}
    \Gamma_{\rho \rightarrow l^{+}l^{-}} = \frac{4\alpha^{2}\pi F_{V}^{2}}{3M_{\rho}}(1+8\sqrt{2}\alpha_{V}\frac{m_{\pi}^{2}}{M_{V}^{2}})^{2}(\cos\delta +\frac{1}{\sqrt{3}}\sin\theta_{V}\sin\delta^{\rho}(M_{\rho}^{2}))^{2}(1+\frac{2m_{l}^{2}}{M_{\rho}^{2}})(1-\frac{4m_{l}^{2}}{M_{\rho}^{2}})^{\frac{1}{2}} \,,
\end{eqnarray*}

\begin{eqnarray*}
    \Gamma_{\omega \rightarrow l^{+}l^{-}} = \frac{4\alpha^{2}\pi F_{V}^{2}}{27 M_{\omega}}(1+8\sqrt{2}\alpha_{V}\frac{m_{\pi}^{2}}{M_{V}^{2}})^{2}(\sqrt{3} \sin\theta_{V} \cos\delta-3\sin\delta^{\omega}(M_{\omega}^{2}))^{2}(1+\frac{2m_{l}^{2}}{M_{\omega}^{2}})(1-\frac{4m_{l}^{2}}{M_{\omega}^{2}})^{\frac{1}{2}}\,,
\end{eqnarray*}

\begin{eqnarray*}
    \Gamma_{\phi \rightarrow l^{+}l^{-}} = \frac{4\alpha^{2}\pi F_{V}^{2}}{9 M_{\phi}}(1+8\sqrt{2}\alpha_{V}\frac{2m_{K}^{2}-m_{\pi}^{2}}{M_{V}^{2}})^{2}\cos^{2}\theta_{V}(1+\frac{2m_{l}^{2}}{M_{\phi}^{2}})(1-\frac{4m_{l}^{2}}{M_{\phi}^{2}})^{\frac{1}{2}}\,,
\end{eqnarray*}

\begin{eqnarray*}
    F_{\rho^{0} \rightarrow \pi^{0}\gamma} &=& \frac{2\sqrt{2}}{3 M_{V} F}C_{R\pi}(0,M_{\rho}^{2})\big(\cos\delta+\sqrt{3} \sin\delta^{\rho}(M_{\rho}^{2}) (\sin \theta_{V}+\sqrt{2} \cos\theta_{V})\big)\\
    &&-\frac{4F_{V}(1+8\sqrt{2}\alpha_{V}\frac{m_{\pi}^{2}}{M_{V}^{2}})}{3 F M_{\rho}^{2}} \cos\delta [\sin\delta^{\rho}(M_{\rho}^{2})+\sin\delta^{\rho}(0)] (\sin\theta_{V}+\sqrt{2} \cos\theta_{V})\\
    &&(\sqrt{3} \cos\delta+\sin\delta^{\rho}(0) \sin\theta_{V}) D_{R\pi}(0,M_{\rho}^{2})-\frac{4F_{V}(1+8\sqrt{2}\alpha_{V}\frac{m_{\pi}^{2}}{M_{V}^{2}})}{3 F M_{\omega }^2}D_{R\pi}(0,M_{\rho}^{2})\\
    &&(\sin\theta_{V}+\sqrt{2} \cos\theta_{V})[\cos^{2}\delta-\sin\delta^{\rho}(M_{\rho}^{2})\sin\delta^{\omega}(0)] (\cos\delta \sin\theta_{V}-\sqrt{3}\sin\delta^{\omega}(0))\\
    &&-\frac{4F_{V}(1+8\sqrt{2}\alpha_{V}\frac{2m_{K}^{2}-m_{\pi}^{2}}{M_{V}^{2}})} {3 F M_{\phi }^2}D_{R\pi}(0,M_{\rho}^{2})\cos\delta \cos\theta_{V}(\cos\theta_{V}-\sqrt{2} \sin\theta_{V})\,,
\end{eqnarray*}

\begin{eqnarray*}
    F_{\omega \rightarrow \pi^{0}\gamma} &=& \frac{2\sqrt{2}}{3 F M_{V}}C_{R\pi}(0,M_\omega^{2})\big(\sqrt{3} \cos\delta (\sin\theta_{V}+\sqrt{2} \cos\theta_{V})-\sin\delta^{\omega}(M_{\omega}^{2})\big) \\
    &&-\frac{4 F_{V}(1+8\sqrt{2}\alpha_{V}\frac{m_{\pi}^{2}}{M_{V}^{2}})}{3 F M_{\rho }^2}(\sin\theta_{V}+\sqrt{2}\cos\theta_{V})(\cos^{2}\delta-\sin\delta^{\rho}(0) \sin \delta^{\omega}(M_{\omega}^{2}))\\
    &&(\sqrt{3}  \cos\delta+\sin\delta^{\rho}(0) \sin\theta_{V}) D_{R\pi}(0,M_{\omega}^{2})\\
    &&+\frac{4 F_{V}(1+8\sqrt{2}\alpha_{V}\frac{m_{\pi}^{2}}{M_{V}^{2}})}{3 F M_{\omega }^2} D_{R\pi}(0,M_\omega^{2})\cos\delta (\sin\delta^{\omega}(0)+\sin\delta^{\omega}(M_{\omega}^{2}))(\sin \theta_{V}+\sqrt{2} \cos\theta_{V}) \\
    &&(\cos\delta \sin\theta_{V}-\sqrt{3} \sin\delta^{\rho}(0))+\frac{4 F_{V}(1+8\sqrt{2}\alpha_{V}\frac{2m_{K}^{2}-m_{\pi}^{2}}{M_{V}^{2}})}{3 F M_{\phi }^2}D_{R\pi}(0,M_{\omega}^{2})\sin\delta^{\omega}(M_{\omega}^{2})\\
    &&\cos\theta_{V} (\cos\theta_{V}-\sqrt{2} \sin\theta_{V})\,,
\end{eqnarray*}

\begin{eqnarray*}
    F_{\phi \rightarrow \pi^{0}\gamma} &=& \frac{2\sqrt{2}}{\sqrt{3} F M_{V}} (\cos\theta_{V}-\sqrt{2} \sin\theta_{V}) C_{R\pi}(0,M_{\phi}^{2})-\frac{4 F_{V}(1+8\sqrt{2}\alpha_{V}\frac{m_{\pi}^{2}}{M_{V}^{2}})}{3 F M_{\rho }^2}D_{R\pi}(0,M_{\phi}^{2})\\
    &&\cos\delta(\cos\theta_{V}-\sqrt{2} \sin\theta_{V})(\sqrt{3} \cos\delta+\sin\delta^{\rho}(0) \sin\theta_{V})+\frac{4F_{V}(1+8\sqrt{2}\alpha_{V}\frac{m_{\pi}^{2}}{M_{V}^{2}})}{3 F M_{\omega }^2}\\
    &&D_{R\pi}(0,M_{\phi}^{2})\sin\delta^{\omega}(0)(\cos\theta_{V}-\sqrt{2} \sin\theta_{V})  (\cos\delta \sin\theta_{V}-\sqrt{3} \sin\delta^{\omega}(0))\,,
\end{eqnarray*}

\begin{eqnarray*}
    F_{\rho^{+} \rightarrow \pi^{+}\gamma} &=&\frac{2\sqrt{2}}{3M_{V}F}C_{R\pi}(0,M_{\rho}^{2})-\frac{4F_{V}(1+8\sqrt{2}\alpha_{V}\frac{m_{\pi}^{2}}{M_{V}^{2}}) }{3 F M_{\rho }^2}D_{R\pi}(0,M_{\rho}^{2})(\sin\theta_{V}+\sqrt{2}\cos\theta_{V}) \nonumber\\ 
    && \sin\delta^{\rho}(0)(\sqrt{3} \cos \delta +\sin\delta^{\rho}(0)\sin \theta_{V})-\frac{4 F_{V}(1+8\sqrt{2}\alpha_{V}\frac{m_{\pi}^{2}}{M_{V}^{2}}) }{3 F M_{\omega }^2}D_{R\pi}(0,M_{\rho}^{2})\cos\delta\\
    && (\sin \theta_{V}+\sqrt{2} \cos\theta_{V})
    (\cos\delta \sin\theta_{V}-\sqrt{3} \sin\delta^{\omega}(0))\\
    &&-\frac{4 F_{V}(1+8\sqrt{2}\alpha_{V}\frac{2m_{K}^{2}-m_{\pi}^{2}}{M_{V}^{2}})}{3 F M_{\phi }^2}D_{R\pi}(0,M_{\rho}^{2}) \cos\theta_{V}(\cos\theta_{V}-\sqrt{2} \sin\theta_{V})  \,,
\end{eqnarray*}

\begin{eqnarray*}
F_{\omega\to\eta\gamma} &=&\frac{2\sqrt{2}}{3M_{V}F}C_{R\eta1}(0,M_{\omega}^{2},m_{\eta}^{2}) \Big\{ \sqrt{3}\sin\delta^{\omega}(M_{\omega}^{2})(-\cos\theta_{P}+\sqrt{2}\sin\theta_{P})+\cos\delta[\sqrt{2}\cos\theta_{V}\cos\theta_{P}\\
&&-\sin\theta_{V}(\cos\theta_{P}+\sqrt{2}\sin\theta_{P})]\Big\}  +\frac{2\sqrt{2}}{9M_{V}F}C_{R\eta2}\bigg\{ 4\cos\delta(\sqrt{2}\cos(\theta_{V}+\theta_{P})-2\cos\theta_{P}\sin\theta_{V}\\
&&+\cos\theta_{V}\sin\theta_{P})m_{K}^{2} -(3\sqrt{3}\sin\delta^{\omega}(M_{\omega}^{2})(\cos\theta_{P}-\sqrt{2}\sin\theta_{P})+\cos\delta[\sqrt{2}\cos(\theta_{V}+\theta_{P})\\
&&-5\cos\theta_{P}\sin\theta_{V}+4\cos\theta_{V}\sin\theta_{P}])m_{\pi}^{2}\bigg\} \\
&& -\frac{F_{V}\left(1+8\sqrt{2}\alpha_{V}\frac{m_{\pi}^{2}}{M_{V}^{2}}\right)}{3\sqrt{2}M_{\rho}^{2}F}D_{R\eta1}(0,M_{\omega}^{2},m_{\eta}^{2})(\sin\theta_{V}\sin\delta^{\rho}(0)+\sqrt{3}\cos\delta)\bigg\{(-4\sqrt{2}\cos\delta)\\
&&\bigg(-\frac{1}{2}\cos^{2}\theta_{V}\cos\theta_{p}\sin\delta^{\rho}(0)+\frac{1}{2} \sin^{2}\theta_{V} \cos\theta_{p}\sin\delta^{\rho}(0)-2\sqrt{2} \sin\theta_{V} \cos\theta_{V} \cos\theta_{P}\\
&&\sin\delta^{\rho}(0)+\sqrt{2}\sin\theta_{P}\sin\delta^{\rho}(0)+\frac{1}{2}\cos\theta_{P}\sin\delta^{\rho}(0)-\sqrt{2}\sin\theta_{P}\sin\delta^{\omega}(M_{\omega}^{2})+\cos\theta_{P}\\
&&\sin\delta^{\omega}(M_{\omega}^{2})\bigg)\bigg\} -\frac{F_{V}\left(1+8\sqrt{2}\alpha_{V}\frac{m_{\pi}^{2}}{M_{V}^{2}}\right)}{9\sqrt{2}M_{\rho}^{2}F}D_{R\eta2}(\sin\theta_{V}\sin\delta^{\rho}(0)+\sqrt{3}\cos\delta)\\
&&\bigg\{ -\sqrt{2}\cos\delta\ [m_{\pi}^{2}(\sin\delta^{\rho}(0)(4\sqrt{2}\sin(2\theta_{V}+\theta_{P})+9\cos(2\theta_{V}-\theta_{P})-7\cos(2\theta_{V}+\theta_{P})\\
 &&-18\cos\theta_{P}) +12\sin\delta^{\omega}(M_{\omega}^{2})(\cos\theta_{P}-\sqrt{2}\sin\theta_{P}))-4m_{K}^{2}\sin\delta^{\rho}(0)(2\cos\theta_{P}\\
 &&(2\sqrt{2}\sin2\theta_{V}+\cos2\theta_{V}-3)+\sin\theta_{P}(4\sin2\theta_{V}+\sqrt{2}\cos2\theta_{V}-3\sqrt{2}))] \bigg\}\\
 && -\frac{2\sqrt{2}F_{V}\left(1+8\sqrt{2}\alpha_{V}\frac{m_{\pi}^{2}}{M_{V}^{2}}\right)}{3M_{\omega}^{2}F}D_{R\eta1}(0,M_{\omega}^{2},m_{\eta}^{2})(\sin\theta_{V}\cos\delta-\sqrt{3}\sin\delta^{\omega}(0))\\
 &&\bigg\{\cos\theta_{P}[\sqrt{2}\sin\delta^{\omega}(0)\sin\delta^{\omega}(M_{\omega}^{2})+\cos^{2}\delta\sin\theta_{V}(4\cos\theta_{V}-\sqrt{2}\sin\theta_{V})]\\
 && -2[\sin\delta^{\omega}(0)\sin\delta^{\omega}(M_{\omega}^{2})+\cos^{2}\delta]\sin\theta_{P}\bigg\}\\
 &&-\frac{F_{V}\left(1+8\sqrt{2}\alpha_{V}\frac{m_{\pi}^{2}}{M_{V}^{2}}\right)}{9\sqrt{2}M_{\omega}^{2}F}D_{R\eta2}(\sin\theta_{V}\cos\delta-\sqrt{3}\sin\delta^{\omega}(0))\bigg\{ 8\cos^{2}\delta(\cos\theta_{P}(-3\sqrt{2}\\
 &&+\sqrt{2}\cos2\theta_{V}+4\sin2\theta_{V}) +(-3+\cos2\theta_{V}+2\sqrt{2}\sin2\theta_{V})\sin\theta_{P})m_{K}^{2}\\
&&+(12\sin\delta^{\omega}(0)\sin\delta^{\omega}(M_{\omega}^{2})(\sqrt{2}\cos\theta_{P}-2\sin\theta_{P})+\cos^{2}\delta[-9\sqrt{2}\cos(2\theta_{V}-\theta_{P})\\
&&+18\sqrt{2}\cos\theta_{P}+7\sqrt{2}\cos(2\theta_{V}+\theta_{P})-8\sin(2\theta_{V}+\theta_{P})])m_{\pi}^{2}\bigg\}  \\
&&+\frac{\sqrt{2}F_{V}\left(1+8\sqrt{2}\alpha_{V}\frac{2m_{K}^{2}-m_{\pi}^{2}}{M_{V}^{2}}\right)}{3M_{\phi}^{2}F}D_{R\eta1}(0,M_{\omega}^{2},m_{\eta}^{2})[\cos\theta_{V}\cos\delta\cos\theta_{P}(-4\cos2\theta_{V} \\
 &&+\sqrt{2}\sin2\theta_{V})]-\frac{\sqrt{2}F_{V}\left(1+8\sqrt{2}\alpha_{V}\frac{2m_{K}^{2}-m_{\pi}^{2}}{M_{V}^{2}}\right)}{9M_{\phi}^{2}F}D_{R\eta2}\cos\theta_{V}\cos\delta\bigg\{ 4(2\sqrt{2}\cos2\theta_{V}\\
 && -\sin2\theta_{V})\sin\theta_{P}(m_{K}^{2}-m_{\pi}^{2})+\cos\theta_{P}(4\cos2\theta_{V}-\sqrt{2}\sin2\theta_{V})(4m_{K}^{2}-m_{\pi}^{2})\bigg\} \,,
\\
\end{eqnarray*}

\begin{eqnarray*}
F_{\rho^{0}\to\eta\gamma}  &=&\frac{2\sqrt{2}}{3M_{V}F}C_{R\eta1}(0,M_{\rho}^{2},m_{\eta}^{2})\bigg\{ \sqrt{3}\cos\delta(\cos\theta_{P}-\sqrt{2}\sin\theta_{P})+\sin\delta^{\rho}(M_{\rho}^{2})[\sqrt{2}\cos\theta_{V}\cos\theta_{P}\\
&&-\sin\theta_{V}(\cos\theta_{P}+\sqrt{2}\sin\theta_{P})]\bigg\} +\frac{2\sqrt{2}}{9M_{V}F}C_{R\eta2}\Big\{ 4\sin\delta^{\rho}(M_{\rho}^{2})\Big(\sqrt{2}\cos(\theta_{V}+\theta_{P})\\
&&-2\cos\theta_{P}\sin\theta_{V}+\cos\theta_{V}\sin\theta_{P}\Big)m_{K}^{2}+\left(3\sqrt{3}\cos\delta(\cos\theta_{P}-\sqrt{2}\sin\theta_{P})\right.\\
&& \left.-\sin\delta^{\rho}(M_{\rho}^{2})[\sqrt{2}\cos(\theta_{V}+\theta_{P})-5\cos\theta_{P}\sin\theta_{V}+4\cos\theta_{V}\sin\theta_{P}]\right)m_{\pi}^{2}\Big\} \\
&& -\frac{2\sqrt{2}F_{V}\left(1+8\sqrt{2}\alpha_{V}\frac{m_{\pi}^{2}}{M_{V}^{2}}\right)}{3M_{\rho}^{2}F}D_{R\eta1}(0,M_{\rho}^{2},m_{\eta}^{2})(\sin\theta_{V}\sin\delta^{\rho}(0)+\sqrt{3}\cos\delta) \bigg\{ \cos^{2}\delta\\
&&(\sqrt{2}\cos\theta_{P}-2\sin\theta_{P})
+\sin\delta^{\rho}(M_{\rho}^{2})\sin\delta^{\rho}(0)[\cos\theta_{P}\sin\theta_{V}(4\cos\theta_{V}-\sqrt{2}\sin\theta_{V}) \\
&&-2\sin\theta_{P}]\bigg\}\\
&&-\frac{F_{V}\left(1+8\sqrt{2}\alpha_{V}\frac{m_{\pi}^{2}}{M_{V}^{2}}\right)}{9\sqrt{2}M_{\rho}^{2}F}D_{R\eta2}(\sin\theta_{V}\sin\delta^{\rho}(0)+\sqrt{3}\cos\delta)\bigg\{8\sin\delta^{\rho}(0)\sin\delta^{\rho}(M_{\rho}^{2})\\
&&\bigg(\cos\theta_{P}(-3\sqrt{2}+\sqrt{2}\cos2\theta_{V}+4\sin2\theta_{V})+(-3+\cos2\theta_{V}+2\sqrt{2}\sin2\theta_{V})\sin\theta_{P}\bigg)\\
&&m_{K}^{2}+(12\cos^{2}\delta(\sqrt{2}\cos\theta_{P}-2\sin\theta_{P})+\sin\delta^{\rho}(M_{\rho}^{2})\sin\delta^{\rho}(0)[-9\sqrt{2}\cos(2\theta_{V}-\theta_{P})\\
&&+18\sqrt{2}\cos\theta_{P}+7\sqrt{2}\cos(2\theta_{V}+\theta_{P})-8\sin(2\theta_{V}+\theta_{P})])m_{\pi}^{2}\bigg\} \\
 && -\frac{F_{V}\left(1+8\sqrt{2}\alpha_{V}\frac{m_{\pi}^{2}}{M_{V}^{2}}\right)}{3\sqrt{2}M_{\omega}^{2}F}D_{R\eta1}(0,M_{\rho}^{2},m_{\eta}^{2})(\sin\theta_{V}\cos\delta-\sqrt{3}\sin\delta^{\omega}(0))\bigg\{(-4\sqrt{2}\cos\delta)\\
 &&\big[-\frac{1}{2}\cos^{2}\theta_{V}\cos\theta_{p}\sin\delta^{\rho}(M_{\rho}^{2})+\frac{1}{2}\sin^{2}\theta_{V}\cos\theta_{p}\sin\delta^{\rho}(M_{\rho}^{2})-2\sqrt{2}\sin\theta_{V}\cos\theta_{V}\\
 && \cos\theta_{P}\sin\delta^{\rho}(M_{\rho}^{2})+\sqrt{2}\sin\theta_{P}\sin\delta^{\rho}(M_{\rho}^{2}) +\frac{1}{2}\cos\theta_{P}\sin\delta^{\rho}(M_{\rho}^{2})\\
 &&-\sqrt{2}\sin\theta_{P}\sin\delta^{\omega}(0)+\cos\theta_{P}\sin\delta^{\omega}(0)\big]\bigg\}\\
 && -\frac{F_{V}\left(1+8\sqrt{2}\alpha_{V}\frac{m_{\pi}^{2}}{M_{V}^{2}}\right)}{9\sqrt{2}M_{\omega}^{2}F}D_{R\eta2}(\sin\theta_{V}\cos\delta-\sqrt{3}\sin\delta^{\omega}(0))\bigg\{ -\sqrt{2}\cos\delta\ \big[m_{\pi}^{2}\\
 &&\big(\sin\delta^{\rho}(M_{\rho}^{2})(4\sqrt{2}\sin(2\theta_{V}+\theta_{P})+9\cos(2\theta_{V}-\theta_{P})-7\cos(2\theta_{V}+\theta_{P})-18\cos\theta_{P})\\
 && +12\sin\delta^{\omega}(0)(\cos\theta_{P}-\sqrt{2}\sin\theta_{P})\big)-4m_{K}^{2}\sin\delta^{\rho}(M_{\rho}^{2})(2\cos\theta_{P}(2\sqrt{2}\sin2\theta_{V}+\cos2\theta_{V}\\
 && -3)+\sin\theta_{P}(4\sin2\theta_{V}+\sqrt{2}\cos2\theta_{V}-3\sqrt{2}))\big]\bigg\}+\frac{\sqrt{2}F_{V}\left(1+8\sqrt{2}\alpha_{V}\frac{2m_{K}^{2}-m_{\pi}^{2}}{M_{V}^{2}}\right)}{3M_{\phi}^{2}F} \\
 && D_{R\eta1}(0,M_{\rho}^{2},m_{\eta}^{2})\cos\theta_{V}\cos\theta_{P}\sin\delta^{\rho}(M_{\rho}^{2})(-4\cos2\theta_{V}+\sqrt{2}\sin2\theta_{V})\\
 &&-\frac{\sqrt{2}F_{V}\left(1+8\sqrt{2}\alpha_{V}\frac{2m_{K}^{2}-m_{\pi}^{2}}{M_{V}^{2}}\right)}{9M_{\phi}^{2}F}D_{R\eta2}\cos\theta_{V}\sin\delta^{\rho}(M_{\rho}^{2})\bigg\{ 4(2\sqrt{2}\cos2\theta_{V}-\sin2\theta_{V})\\
 &&\sin\theta_{P}(m_{K}^{2}-m_{\pi}^{2})+\cos\theta_{P}(4\cos2\theta_{V}-\sqrt{2}\sin2\theta_{V})(4m_{K}^{2}-m_{\pi}^{2})\bigg\} ,
\\
\end{eqnarray*}
\begin{eqnarray*}
F_{\phi\to\eta\gamma}  &=& \frac{2\sqrt{2}}{3M_{V}F}C_{R\eta1}(0,M_{\phi}^{2},m_{\eta}^{2})\left\{ -\sqrt{2}\cos\theta_{P}\sin\theta_{V}-\cos\theta_{V}(\cos\theta_{P}+\sqrt{2}\sin\theta_{P})\right\} \\
 && +\frac{\sqrt{2}}{9M_{V}F}C_{R\eta2}\left\{ -4\left(3\cos(\theta_{V}-\theta_{P})+\cos(\theta_{V}+\theta_{P})+2\sqrt{2}\sin(\theta_{V}+\theta_{P})\right)m_{K}^{2}\right.\\
 && \left.+\left(9\cos(\theta_{V}-\theta_{P})+\cos(\theta_{V}+\theta_{P})+2\sqrt{2}\sin(\theta_{V}+\theta_{P})\right)m_{\pi}^{2}\right\} \\
 && +\frac{\sqrt{2}F_{V}\left(1+8\sqrt{2}\alpha_{V}\frac{m_{\pi}^{2}}{M_{V}^{2}}\right)}{3M_{\rho}^{2}F}D_{R\eta1}(0,M_{\phi}^{2},m_{\eta}^{2})(\sin\theta_{V}\sin\delta^{\rho}(0)+\sqrt{3}\cos\delta) \cos\theta_{P}\\
 &&\sin\delta^{\rho}(0)(-4\cos2\theta_{V}+\sqrt{2}\sin2\theta_{V}) \\
 &&-\frac{\sqrt{2}F_{V}\left(1+8\sqrt{2}\alpha_{V}\frac{m_{\pi}^{2}}{M_{V}^{2}}\right)}{9M_{\rho}^{2}F}D_{R\eta2}(\sin\theta_{V}\sin\delta^{\rho}(0)+\sqrt{3}\cos\delta)\sin\delta^{\rho}(0)\Big\{4(2\sqrt{2}\cos2\theta_{V}\\
 &&-\sin2\theta_{V}) \sin\theta_{P}(m_{K}^{2}-m_{\pi}^{2})+\cos\theta_{P}(4\cos2\theta_{V}-\sqrt{2}\sin2\theta_{V})(4m_{K}^{2}-m_{\pi}^{2})\Big\} \\
 && +\frac{\sqrt{2}F_{V}\left(1+8\sqrt{2}\alpha_{V}\frac{m_{\pi}^{2}}{M_{V}^{2}}\right)}{3M_{\omega}^{2}F}D_{R\eta1}(0,M_{\phi}^{2},m_{\eta}^{2})(\sin\theta_{V}\cos\delta-\sqrt{3}\sin\delta^{\omega}(0))\cos\delta\cos\theta_{P}\\
 &&(-4\cos2\theta_{V}+\sqrt{2}\sin2\theta_{V}) \\
 &&-\frac{\sqrt{2}F_{V}\left(1+8\sqrt{2}\alpha_{V}\frac{m_{\pi}^{2}}{M_{V}^{2}}\right)}{9M_{\omega}^{2}F}D_{R\eta2}(\sin\theta_{V}\cos\delta-\sqrt{3}\sin\delta^{\omega}(0))\cos\delta\big\{ 4(2\sqrt{2}\cos2\theta_{V}\\
 &&-\sin2\theta_{V}) \sin\theta_{P}(m_{K}^{2}-m_{\pi}^{2})+\cos\theta_{P}(4\cos2\theta_{V}-\sqrt{2}\sin2\theta_{V})(4m_{K}^{2}-m_{\pi}^{2})\big\}\\
 &&-\frac{2\sqrt{2}F_{V}\left(1+8\sqrt{2}\alpha_{V}\frac{2m_{K}^{2}-m_{\pi}^{2}}{M_{V}^{2}}\right)}{3M_{\phi}^{2}F}D_{R\eta1}(0,M_{\phi}^{2},m_{\eta}^{2}) \\
 &&\cos\theta_{V}\bigg\{ -\cos\theta_{V}\cos\theta_{P}(\sqrt{2}\cos\theta_{V}+4\sin\theta_{V})-2\sin\theta_{P}\bigg\}
 \\&&-\frac{\sqrt{2}F_{V}\left(1+8\sqrt{2}\alpha_{V}\frac{2m_{K}^{2}-m_{\pi}^{2}}{M_{V}^{2}}\right)}{9M_{\phi}^{2}F}D_{R\eta2} \cos\theta_{V}\bigg\{ (\sqrt{2}\cos\theta_{V}-2\sin\theta_{V})^{2}(\sqrt{2}\cos\theta_{P}
 \\
 &&-2\sin\theta_{P})m_{\pi}^{2}-4(\sqrt{2}\cos\theta_{V}+\sin\theta_{V})^{2}(\sqrt{2}\cos\theta_{P}+\sin\theta_{P})(2m_{K}^{2}-m_{\pi}^{2})\bigg\} \,,
\\
\end{eqnarray*}
\begin{eqnarray*}
F_{\eta'\to\omega\gamma}  &=& \frac{2\sqrt{2}}{3M_{V}F}C_{R\eta1}(0,M_{\omega}^{2},m_{\eta'}^{2})\bigg\{ \cos\delta\sin\theta_{V}(\sqrt{2}\cos\theta_{P}-\sin\theta_{P})+\sqrt{2}\cos\delta\cos\theta_{V}\sin\theta_{P}\\
&&-\sqrt{3}\sin\delta^{\omega}(M_{\omega}^{2})(\sqrt{2}\cos\theta_{P}+\sin\theta_{P})\bigg\} +\frac{\sqrt{2}}{9M_{V}F}C_{R\eta2}\bigg\{ 4\cos\delta(-3\cos(\theta_{V}\!-\!\theta_{P})\!\\
&&+\!\cos(\theta_{V}\!+\!\theta_{P})\!+\!2\sqrt{2}\sin(\theta_{V}\!+\!\theta_{P}))m_{K}^{2} +(-6\sqrt{3}\sin\delta^{\omega}(M_{\omega}^{2})(\sqrt{2}\cos\theta_{P}+\sin\theta_{P})\\
 && -\cos\delta[-9\cos(\theta_{V}-\theta_{P})+\cos(\theta_{V}+\theta_{P})+2\sqrt{2}\sin(\theta_{V}+\theta_{P})])m_{\pi}^{2}\bigg\} \\
 && -\frac{F_{V}\left(1+8\sqrt{2}\alpha_{V}\frac{m_{\pi}^{2}}{M_{V}^{2}}\right)}{3\sqrt{2}M_{\rho}^{2}F}D_{R\eta1}(0,M_{\omega}^{2},m_{\eta'}^{2})(\sin\theta_{V}\sin\delta^{\rho}(0)+\sqrt{3}\cos\delta)\bigg\{(-4\sqrt{2}\cos\delta)\\
 &&\{\sin\theta_{P}[\sin\theta_{V}\sin\delta^{\rho}(0)(\sin\theta_{V}-2\sqrt{2}\cos\theta_{V})+\sin\delta^{\omega}(M_{\omega}^{2})]+\sqrt{2}\cos\theta_{P}(\sin\delta^{\omega}(M_{\omega}^{2})\\
 &&-\sin\delta^{\rho}(0))\}\bigg\} -\frac{F_{V}\left(1+8\sqrt{2}\alpha_{V}\frac{m_{\pi}^{2}}{M_{V}^{2}}\right)}{9\sqrt{2}M_{\rho}^{2}F}D_{R\eta2}(\sin\theta_{V}\sin\delta^{\rho}(0)+\sqrt{3}\cos\delta)\\
 && \bigg\{(-2\sqrt{2}\cos\delta)\bigg[\sin\delta^{\rho}(0)[2m_{K}^{2}\ (\cos\theta_{P}\ (4\sin2\theta_{V}+\sqrt{2}\cos2\theta_{V}-3\sqrt{2})\\
 && -2\sin\theta_{P}(2\sqrt{2}\sin2\theta_{V}+\cos2\theta_{V}-3))+m_{\pi}^{2}\ (-2\sqrt{2}\cos(2\theta_{V}+\theta_{P})\\
 && -8\sin2\theta_{V}\cos\theta_{P}\!+\!(\cos2\theta_{V}\!-\!9)\sin\theta_{P})]\!+\!6m_{\pi}^{2}\sin\delta^{\omega}(M_{\omega}^{2})(\sin\theta_{P}\!+\!\sqrt{2}\cos\theta_{P})\bigg] \!\bigg\}\\
 && -\frac{2\sqrt{2}F_{V}\left(1+8\sqrt{2}\alpha_{V}\frac{m_{\pi}^{2}}{M_{V}^{2}}\right)}{3M_{\omega}^{2}F}D_{R\eta1}(0,M_{\omega}^{2},m_{\eta'}^{2})(\sin\theta_{V}\cos\delta-\sqrt{3}\sin\delta^{\omega}(0))\\
&&\bigg\{\sin\delta^{\omega}(M_{\omega}^{2}) \times\sin\delta^{\omega}(0)(2\cos\theta_{P}+\sqrt{2}\sin\theta_{P})\\
&&+\cos^{2}\delta[2\cos\theta_{P}+\sin\theta_{V}(4\cos\theta_{V}-\sqrt{2}\sin\theta_{V})\sin\theta_{P}]\bigg\}\\
 && -\frac{\sqrt{2}F_{V}\left(1+8\sqrt{2}\alpha_{V}\frac{m_{\pi}^{2}}{M_{V}^{2}}\right)}{9M_{\omega}^{2}F}D_{R\eta2}(\sin\theta_{V}\cos\delta-\sqrt{3}\sin\delta^{\omega}(0))\bigg\{ -4\cos^{2}\delta(\cos\theta_{P}\\
 &&(-3+\cos2\theta_{V}+2\sqrt{2}\sin2\theta_{V})-(-3\sqrt{2}+\sqrt{2}\cos2\theta_{V}+4\sin2\theta_{V})\sin\theta_{P})m_{K}^{2}\\
 &&+(6\sin\delta^{\omega}(M_{\omega}^{2})\sin\delta^{\omega}(0)(2\cos\theta_{P}+\sqrt{2}\sin\theta_{P})+\cos^{2}\delta(4\cos(2\theta_{V}\!+\!\theta_{P})\!\\
 &&+\!\sqrt{2}[8\cos\theta_{P}\sin2\theta_{V}-(-9\!+\!\cos2\theta_{V})\sin\theta_{P}]))m_{\pi}^{2}\bigg\} 
\\
&&+\frac{\sqrt{2}F_{V}\left(1+8\sqrt{2}\alpha_{V}\frac{2m_{K}^{2}-m_{\pi}^{2}}{M_{V}^{2}}\right)}{3M_{\phi}^{2}F}D_{R\eta1}(0,M_{\omega}^{2},m_{\eta'}^{2})
 \left\{ \cos\theta_{V}\cos\delta\sin\theta_{P}\right.\\
 &&\left.(-4\cos2\theta_{V}+\sqrt{2}\sin2\theta_{V})\right\}\\
 &&-\frac{\sqrt{2}F_{V}\left(1+8\sqrt{2}\alpha_{V}\frac{2m_{K}^{2}-m_{\pi}^{2}}{M_{V}^{2}}\right)}{9M_{\phi}^{2}F}D_{R\eta2}\cos\theta_{V}\cos\delta\bigg\{ -4\cos\theta_{P}(2\sqrt{2}\cos2\theta_{V}\\
 &&-\sin2\theta_{V})(m_{K}^{2}-m_{\pi}^{2})+(4\cos2\theta_{V}-\sqrt{2}\sin2\theta_{V})\sin\theta_{P}(4m_{K}^{2}-m_{\pi}^{2})\bigg\} \,,
\\
\end{eqnarray*}
\begin{eqnarray*}
F_{\eta'\to\rho\gamma}  &=& \frac{2\sqrt{2}}{3M_{V}F}C_{R\eta1}(0,M_{\rho}^{2},m_{\eta'}^{2})\bigg\{ \sqrt{3}\cos\delta(\sqrt{2}\cos\theta_{P}+\sin\theta_{P})+\sin\delta^{\rho}(M_{\rho}^{2})[\sqrt{2}\cos\theta_{P}\sin\theta_{V}\\
&&+(\sqrt{2}\cos\theta_{V}-\sin\theta_{V})\sin\theta_{P}]\bigg\} +\frac{\sqrt{2}}{9M_{V}F}C_{R\eta2}\Big\{ 4\sin\delta^{\rho}(M_{\rho}^{2})\Big(-3\cos(\theta_{V}-\theta_{P})\\
 && +\cos(\theta_{V}+\theta_{P})+2\sqrt{2}\sin(\theta_{V}+\theta_{P})\Big)m_{K}^{2}+\left(6\sqrt{3}\cos\delta(\sqrt{2}\cos\theta_{P}+\sin\theta_{P})\right.\\
 && \left.-\sin\delta^{\rho}(M_{\rho}^{2})[-9\cos(\theta_{V}-\theta_{P})+\cos(\theta_{V}+\theta_{P})+2\sqrt{2}\sin(\theta_{V}+\theta_{P})]\right)m_{\pi}^{2}\Big\} \\
 && -\frac{2\sqrt{2}F_{V}\left(1+8\sqrt{2}\alpha_{V}\frac{m_{\pi}^{2}}{M_{V}^{2}}\right)}{3M_{\rho}^{2}F}D_{R\eta1}(0,M_{\rho}^{2},m_{\eta'}^{2})(\sin\theta_{V}\sin\delta^{\rho}(0)+\sqrt{3}\cos\delta)\\
 && \bigg\{ \cos^{2}\delta(2\cos\theta_{P}+\sqrt{2}\sin\theta_{P})+\sin\delta^{\rho}(0)\sin\delta^{\rho}(M_{\rho}^{2})[2\cos\theta_{P} \\
 &&+\sin\theta_{V}(4\cos\theta_{V}-\sqrt{2}\sin\theta_{V})\sin\theta_{P}]\bigg\}\\
 && -\frac{\sqrt{2}F_{V}\left(1+8\sqrt{2}\alpha_{V}\frac{m_{\pi}^{2}}{M_{V}^{2}}\right)}{9M_{\rho}^{2}F}D_{R\eta2}(\sin\theta_{V}\sin\delta^{\rho}(0)+\sqrt{3}\cos\delta)\bigg\{ -4\sin\delta^{\rho}(0)\\
 && \sin\delta^{\rho}(M_{\rho}^{2})(\cos\theta_{P}(-3+\cos2\theta_{V}+2\sqrt{2}\sin2\theta_{V})-(-3\sqrt{2}+\sqrt{2}\cos2\theta_{V}\\
 && +4\sin2\theta_{V})\sin\theta_{P}\Big)m_{K}^{2}\!+\!\Big(6\cos^{2}\delta(2\cos\theta_{P}\!+\!\sqrt{2}\sin\theta_{P})+\sin\delta^{\rho}(0)\sin\delta^{\rho}(M_{\rho}^{2})\\
 && \Big(4\cos(2\theta_{V}+\theta_{P})+\sqrt{2}[8\cos\theta_{P}\sin2\theta_{V}-(-9+\cos2\theta_{V})\sin\theta_{P}]\Big)\Big)m_{\pi}^{2}\bigg\} \\
 && -\frac{\sqrt{2}F_{V}\left(1+8\sqrt{2}\alpha_{V}\frac{m_{\pi}^{2}}{M_{V}^{2}}\right)}{6M_{\omega}^{2}F}D_{R\eta1}(0,M_{\rho}^{2},m_{\eta'}^{2})(\sin\theta_{V}\cos\delta-\sqrt{3}\sin\delta^{\omega}(0))\\
 && \times\bigg\{(-4\sqrt{2}\cos\delta)\{\sin\theta_{P}[\sin\theta_{V}\sin\delta^{\rho}(M_{\rho}^{2})(\sin\theta_{V}-2\sqrt{2}\cos\theta_{V})\\
 && +\sin\delta^{\omega}(0)]+\sqrt{2}\cos\theta_{P}(\sin\delta^{\omega}(0)-\sin\delta^{\rho}(M_{\rho}^{2}))\}\bigg\}\\
 && -\frac{\sqrt{2}F_{V}\left(1+8\sqrt{2}\alpha_{V}\frac{m_{\pi}^{2}}{M_{V}^{2}}\right)}{18M_{\omega}^{2}F}D_{R\eta2}(\sin\theta_{V}\cos\delta-\sqrt{3}\sin\delta^{\omega}(0))\\
 && \times\bigg\{(-2\sqrt{2}\cos\delta)\bigg(\sin\delta^{\rho}(M_{\rho}^{2})[2m_{K}^{2}\ (\cos\theta_{P}\ (4\sin2\theta_{V}+\sqrt{2}\cos2\theta_{V}-3\sqrt{2})\\
 && -2\sin\theta_{P}(2\sqrt{2}\sin2\theta_{V}+\cos2\theta_{V}-3))+m_{\pi}^{2}\ (-2\sqrt{2}\cos(2\theta_{V}+\theta_{P})\\
 && -8\sin2\theta_{V}\cos\theta_{P}\!+\!(\cos2\theta_{V}\!-\!9)\sin\theta_{P})]+\!6m_{\pi}^{2}\sin\delta^{\omega}(0)(\sin\theta_{P}\!+\!\sqrt{2}\cos\theta_{P})\bigg) \bigg\}\\
 && +\frac{\sqrt{2}F_{V}\left(1+8\sqrt{2}\alpha_{V}\frac{2m_{K}^{2}-m_{\pi}^{2}}{M_{V}^{2}}\right)}{3M_{\phi}^{2}F}D_{R\eta1}(0,M_{\rho}^{2},m_{\eta'}^{2}) \bigg\{ \cos\theta_{V}\sin\delta^{\rho}(M_{\rho}^{2})\\
 &&(-4\cos2\theta_{V}+\sqrt{2}\sin2\theta_{V})\sin\theta_{P}\bigg\} \\
 &&-\frac{\sqrt{2}F_{V}\left(1+8\sqrt{2}\alpha_{V}\frac{2m_{K}^{2}-m_{\pi}^{2}}{M_{V}^{2}}\right)}{9M_{\phi}^{2}F}D_{R\eta2}\cos\theta_{V}\sin\delta^{\rho}(M_{\rho}^{2})\bigg\{ -4\cos\theta_{P}(2\sqrt{2}\cos2\theta_{V}\\
 &&-\sin2\theta_{V})(m_{K}^{2}-m_{\pi}^{2})+(4\cos2\theta_{V}-\sqrt{2}\sin2\theta_{V})\sin\theta_{P}(4m_{K}^{2}-m_{\pi}^{2})\bigg\} \,,
\\
\end{eqnarray*}
\begin{eqnarray*}
F_{\phi\to\eta'\gamma}  &=& \frac{2\sqrt{2}}{3M_{V}F}C_{R\eta1}(0,M_{\phi}^{2},m_{\eta'}^{2})( \sqrt{2}\cos(\theta_{V}+\theta_{P})-\cos\theta_{V}\sin\theta_{P}) \\
 && +\frac{\sqrt{2}}{9M_{V}F}C_{R\eta2}\left\{ 8\left(\sqrt{2}\cos(\theta_{V}+\theta_{P})+\cos\theta_{P}\sin\theta_{V}-2\cos\theta_{V}\sin\theta_{P}\right)m_{K}^{2}\right.\\
 && \left.+\left(-2\sqrt{2}\cos(\theta_{V}+\theta_{P})-9\sin(\theta_{V}-\theta_{P})+\sin(\theta_{V}+\theta_{P})\right)m_{\pi}^{2}\right\} \phantom{\frac{1}{1}}\\
 && +\frac{\sqrt{2}F_{V}\left(1+8\sqrt{2}\alpha_{V}\frac{m_{\pi}^{2}}{M_{V}^{2}}\right)}{3M_{\rho}^{2}F}D_{R\eta1}(0,M_{\phi}^{2},m_{\eta'}^{2})(\sin\theta_{V}\sin\delta^{\rho}(0)+\sqrt{3}\cos\delta)\\
 && \times\left\{ \sin\delta^{\rho}(0)(-4\cos2\theta_{V}+\sqrt{2}\sin2\theta_{V})\sin\theta_{P}\right\} \phantom{\frac{1}{1}} \\
 && -\frac{\sqrt{2}F_{V}\left(1+8\sqrt{2}\alpha_{V}\frac{m_{\pi}^{2}}{M_{V}^{2}}\right)}{9M_{\rho}^{2}F}D_{R\eta2}(\sin\theta_{V}\sin\delta^{\rho}(0)+\sqrt{3}\cos\delta)\sin\delta^{\rho}(0) \phantom{\frac{1}{1}}\\
 && \{ -4\cos\theta_{P}(2\sqrt{2}\cos2\theta_{V}-\sin2\theta_{V})(m_{K}^{2}-m_{\pi}^{2})+(4\cos2\theta_{V}-\sqrt{2}\sin2\theta_{V})\phantom{\frac{1}{1}} \\
 &&\sin\theta_{P} (4m_{K}^{2}-m_{\pi}^{2})\} \\
 &&+\frac{\sqrt{2}F_{V}\left(1+8\sqrt{2}\alpha_{V}\frac{m_{\pi}^{2}}{M_{V}^{2}}\right)}{3M_{\omega}^{2}F}D_{R\eta1}(0,M_{\phi}^{2},m_{\eta'}^{2})(\sin\theta_{V}\cos\delta-\sqrt{3}\sin\delta^{\omega}(0)) \phantom{\frac{1}{1}}\\
 && \left\{ \cos\delta(-4\cos2\theta_{V}+\sqrt{2}\sin2\theta_{V})\sin\theta_{P}\right\}  \phantom{\frac{1}{1}}\\
 && -\frac{\sqrt{2}F_{V}\left(1+8\sqrt{2}\alpha_{V}\frac{m_{\pi}^{2}}{M_{V}^{2}}\right)}{9M_{\omega}^{2}F}D_{R\eta2}(\sin\theta_{V}\cos\delta-\sqrt{3}\sin\delta^{\omega}(0))\cos\delta\Big\{ -4\cos\theta_{P} \phantom{\frac{1}{1}}\\
 && (2\sqrt{2}\cos2\theta_{V}-\sin2\theta_{V})(m_{K}^{2}-m_{\pi}^{2})+(4\cos2\theta_{V}-\sqrt{2}\sin2\theta_{V})\sin\theta_{P}(4m_{K}^{2}-m_{\pi}^{2})\Big\}  \phantom{\frac{1}{1}}\\
 && -\frac{2\sqrt{2}F_{V}\left(1+8\sqrt{2}\alpha_{V}\frac{2m_{K}^{2}-m_{\pi}^{2}}{M_{V}^{2}}\right)}{3M_{\phi}^{2}F}D_{R\eta1}(0,M_{\phi}^{2},m_{\eta'}^{2})\cos\theta_{V}\\
 && \times\left\{ 2\cos\theta_{P}-\cos\theta_{V}(\sqrt{2}\cos\theta_{V}+4\sin\theta_{V})\sin\theta_{P}\right\} \\
 && -\frac{\sqrt{2}F_{V}\left(1+8\sqrt{2}\alpha_{V}\frac{2m_{K}^{2}-m_{\pi}^{2}}{M_{V}^{2}}\right)}{9M_{\phi}^{2}F}D_{R\eta2}\cos\theta_{V}\left\{ (\sqrt{2}\cos\theta_{V}-2\sin\theta_{V})^{2}(2\cos\theta_{P}\right.\\
 && \left.+\sqrt{2}\sin\theta_{P})m_{\pi}^{2}-4(\sqrt{2}\cos\theta_{V}+\sin\theta_{V})^{2}(-\cos\theta_{P}+\sqrt{2}\sin\theta_{P})(2m_{K}^{2}-m_{\pi}^{2})\right\} \,,
\end{eqnarray*}

\begin{eqnarray*}
    F_{\eta \rightarrow \gamma\gamma}&=&-\frac{N_{C} (\cos \theta_{P}-2 \sqrt{2} \sin \theta_{P}) }{12 \sqrt{3} \pi ^2 F}\\
    &&-\frac{2 F_{V}(1+8\sqrt{2}\alpha_{V}\frac{m_{\pi}^{2}}{M_{V}^{2}})}{27 FM_{V} M_{\rho }^2} (3 \cos \delta+\sqrt{3} \sin \delta^{\rho}(0) \sin \theta_{V})\bigg\{3 C_{R\eta1}(0,0,m_{\eta}^{2}) [\sqrt{3} \cos \delta\\
    &&(\sqrt{2} \cos \theta_{P}-2 \sin \theta_{P})+2 \sin \delta^{\rho}(0) \cos \theta_{V} \cos \theta_{P}-\sin \delta^{\rho}(0) \sin \theta_{V} (2 \sin \theta_{P}\\
    &&+\sqrt{2} \cos \theta_{P})]-2C_{R\eta2} \sin \delta^{\rho}(0) m_{K}^2 (\sqrt{2} (3 \sin (\theta_{V}-\theta_{P})+\sin (\theta_{V}+\theta_{P}))\\
    &&-4 \cos (\theta_{V}+\theta_{P}))+\frac{C_{R\eta2}}{2} m_{\pi }^2 [6 \sqrt{3} \cos \delta (\sqrt{2} \cos \theta_{P}-2 \sin \theta_{P})+\sin \delta^{\rho}(0)\\
    &&(\sqrt{2} (9 \sin (\theta_{V}-\theta_{P})+\sin (\theta_{V}+\theta_{P}))-4 \cos (\theta_{V}+\theta_{P}))]\bigg\}\\
     &&-\frac{4 F_{V}(1+8\sqrt{2}\alpha_{V}\frac{m_{\pi}^{2}}{M_{V}^{2}})}{27 FM_{V} M_{\omega }^2} (\sqrt{3} \cos \delta \sin \theta_{V}-3 \sin\delta^{\omega}(0)) \bigg\{\frac{3}{2} C_{R\eta1}(0,0,m_{\eta}^{2}) [\cos \theta_{P}\\
    &&(2 \cos \delta \cos \theta_{V}-\sqrt{2} \cos \delta \sin \theta_{V}-\sqrt{6} \sin\delta^{\omega}(0))-2 \sin \theta_{P} (\cos \delta \sin \theta_{V}\\
    &&-\sqrt{3} \sin \delta^{\omega}(0))]-C_{\eta2} \cos \delta m_{K}^2 (\sqrt{2} (3 \sin (\theta_{V}-\theta_{P})+\sin (\theta_{V}+\theta_{P}))\\
    &&-4 \cos (\theta_{V}+\theta_{P}))+\frac{C_{R\eta2}}{4} m_{\pi }^2 [\cos \delta (\sqrt{2}(9 \sin (\theta_{V}-\theta_{P})+\sin (\theta_{V}+\theta_{P}))\\
    &&-4 \cos (\theta_{V}+\theta_{P}))-6 \sqrt{3} \sin \delta^{\omega}(0) (\sqrt{2} \cos \theta_{P}-2 \sin \theta_{P})]\bigg\}\\
    &&+\frac{2 F_{V}(1+8\sqrt{2}\alpha_{V}\frac{2m_{K}^{2}-m_{\pi}^{2}}{M_{V}^{2}})}{9 \sqrt{3} FM_{V} M_{\phi }^2} \cos \theta_{V}  \bigg\{3C_{R\eta1}(0,0,m_{\eta}^{2}) (\cos \theta_{V} (2 \sin \theta_{P}+\sqrt{2} \cos \theta_{P})\\
    &&+2 \sin \theta_{V} \cos \theta_{P})+\frac{C_{R\eta2}}{2} [3 \sqrt{2} (4 m_{K}^2-3 m_{\pi }^2) \cos (\theta_{V}-\theta_{P})+(4 m_{K}^2-m_{\pi }^2)\\
    &&(4 \sin (\theta_{V}+\theta_{P})+\sqrt{2} \cos (\theta_{V}+\theta_{P}))]\bigg\}\\
    &&+\frac{4 F_{V}^2(1+8\sqrt{2}\alpha_{V}\frac{m_{\pi}^{2}}{M_{V}^{2}})^{2}}{27 \sqrt{3} F M_{\rho }^4} (3 \cos \delta+\sqrt{3} \sin \delta^{\rho}(0) \sin \theta_{V})^2 \bigg\{\sin ^2\delta^{\rho}(0) [-3D_{R\eta1}(0,0, m_{\eta }^2) \\
    &&(\sin ^2\theta_{V}  \cos \theta_{P}+\sqrt{2} (\sin \theta_{P}-\sin 2\theta_{V} \cos \theta_{P}))+D_{R\eta2} m_{K}^2\\
    &&(2 \cos \theta_{P} (2 \sqrt{2} \sin 2\theta_{V}+\cos 2\theta_{V}-3)+\sin \theta_{P} (4 \sin 2\theta_{V}+\sqrt{2} \cos 2\theta_{V}-3 \sqrt{2}))\\
    &&+\frac{D_{\eta2}}{4} m_{\pi }^2(-4 \sqrt{2} \sin (2 \theta_{V} +\theta_{P})-9 \cos (2 \theta_{V} -\theta_{P})+7 \cos (2 \theta_{V} +\theta_{P})\\
    &&+18 \cos \theta_{P})]+3 \cos ^2\delta (\cos \theta_{P}-\sqrt{2} \sin \theta_{P}) (D_{R\eta1}(0,0,m_{\eta }^2)+D_{R\eta2} m_{\pi }^2)\bigg\}\\
     &&+\frac{2 F_{V}^2(1+8\sqrt{2}\alpha_{V}\frac{m_{\pi}^{2}}{M_{V}^{2}})^{2}}{27 \sqrt{3} F M_{\omega }^4}(\sqrt{3} \cos \delta \sin \theta_{V}-3 \sin \delta^{\omega}(0))^2 \bigg\{3 D_{R\eta1}(0,0,m_{\eta}^{2}) \\
     &&[2 \cos \theta_{P} (\cos ^2 \delta \sin \theta_{V} (2 \sqrt{2} \cos \theta_{V}-\sin \theta_{V})+\sin ^2\delta^{\omega}(0))+\sqrt{2} \sin \theta_{P} \\
     &&(-\cos 2 \delta +(1-2\sin^{2}\delta^{\omega}(0))-2)]+2D_{R\eta2} \cos ^2\delta m_{K}^2 [2 \cos \theta_{P} (2 \sqrt{2} \sin 2 \theta_{V}\\
     &&+\cos 2 \theta_{V}-3)+\sin \theta_{P} (4 \sin 2 \theta_{V}+\sqrt{2} \cos 2 \theta_{V}-3 \sqrt{2})]\\
     &&+\frac{D_{R\eta2}}{2} m_{\pi }^2 [\cos ^2(\delta ) (-4 \sqrt{2}\sin (2 \theta_{V}+\theta_{P})-9 \cos (2 \theta_{V}-\theta_{P})\\
     &&+7 \cos (2 \theta_{V}+\theta_{P})+18 \cos \theta_{P})+12 \sin ^2\delta^{\omega}(0) (\cos \theta_{P}-\sqrt{2} \sin \theta_{P})]\bigg\}\\
    &&-\frac{4 F_{V}^2(1+8\sqrt{2}\alpha_{V}\frac{2m_{K}^{2}-m_{\pi}^{2}}{M_{V}^{2}})^{2}}{9 \sqrt{3} F M_{\phi }^4}\cos^2 \theta_{V}\bigg \{3 D_{R\eta1}(0,0,m_{\eta}^{2}) (\cos ^2 \theta_{V} \cos \theta_{P}+\sqrt{2}\\
    &&(\sin 2 \theta_{V} \cos \theta_{P}+\sin \theta_{P}))+D_{R\eta2} m_{K}^2 (2 \cos \theta_{P} (2 \sqrt{2} \sin 2\theta_{V}+\cos 2\theta_{V}+3)+\sin \theta_{P} \\
    &&(4 \sin 2\theta_{V}+\sqrt{2} \cos 2\theta_{V}+3 \sqrt{2}))-\frac{D_{\eta2}}{4} m_{\pi }^2 (4 \sqrt{2} \sin (2 \theta_{V}+\theta_{P})+9 \cos (2 \theta_{V}-\theta_{P})\\
    &&-7 \cos (2 \theta_{V}+\theta_{P})+18 \cos \theta_{P})\bigg\}\\ 
    &&+\frac{4 F_{V}^2(1+8\sqrt{2}\alpha_{V}\frac{m_{\pi}^{2}}{M_{V}^{2}})^{2}}{27 \sqrt{3} F M_{\rho }^2 M_{\omega }^2} \cos \delta  (\sqrt{3} \cos \delta \sin \theta_{V}-3 \sin \delta^{\omega}(0)) (3 \cos \delta+\sqrt{3} \sin \delta^{\rho}(0)  \\
    &&\sin \theta_{V})\bigg\{3D_{R\eta1}(0,0,m_{\eta}^{2}) [\cos \theta_{P} (-\sin \delta^{\omega}(0)+\sin\delta^{\rho}(0)\sin \theta_{V} (2 \sqrt{2} \cos \theta_{V}-\sin \theta_{V}))\\
    &&-\sqrt{2} \sin \theta_{P} (-\sin \delta^{\omega}(0)+\sin \delta^{\rho}(0))]+D_{R\eta2} \sin \delta^{\rho}(0) m_{K}^2 [2 \cos \theta_{P} (2 \sqrt{2} \sin 2\theta_{V}\\
    &&+\cos 2\theta_{V}-3)+\sin \theta_{P} (4 \sin 2\theta_{V}+\sqrt{2} \cos 2\theta_{V}-3 \sqrt{2})]+\frac{D_{R\eta2}}{4} m_{\pi }^2 [-12 \sin \delta^{\omega}(0) \\
    &&(\cos \theta_{P}-\sqrt{2} \sin \theta_{P})+\sin \delta^{\rho}(0)(-4 \sqrt{2} \sin (2 \theta_{V}+\theta_{P})-9 \cos (2 \theta_{V}-\theta_{P})\\
    &&+7 \cos (2 \theta_{V}+\theta_{P})+18 \cos \theta_{P})]\bigg\}\\
   &&-\frac{2 F_{V}^2(1+8\sqrt{2}\alpha_{V}\frac{m_{\pi}^{2}}{M_{V}^{2}})(1+8\sqrt{2}\alpha_{V}\frac{2m_{K}^{2}-m_{\pi}^{2}}{M_{V}^{2}})}{27 F M_{\rho }^2 M_{\phi }^2}\sin\delta^{\rho}(0) \cos \theta_{V}(3 \cos \delta +\sqrt{3} \sin \delta^{\rho}(0)  \\
   &&\sin \theta_{V})\bigg\{2D_{R\eta2} (m_{K}^2-m_{\pi }^2) \sin \theta_{P} (\sqrt{2} \sin 2\theta_{V}-4 \cos 2\theta_{V})-\cos \theta_{P} (2 \sqrt{2} \cos 2\theta_{V} \\
   &&-\sin 2\theta_{V})(3 D_{R\eta1}(0,0,m_{\eta}^{2})+D_{R\eta2} (4m_{K}^2-m_{\pi }^2))\bigg\}\\
    &&+\frac{F_{V}^2(1+8\sqrt{2}\alpha_{V}\frac{m_{\pi}^{2}}{M_{V}^{2}})(1+8\sqrt{2}\alpha_{V}\frac{2m_{K}^{2}-m_{\pi}^{2}}{M_{V}^{2}})}{27 F M_{\omega }^2 M_{\phi }^2}\cos \delta \cos \theta_{V}  (2 \sqrt{3} \cos \delta \sin \theta_{V}\\
    &&-6 \sin \delta^{\omega}(0))\bigg\{\cos \theta_{P}(2 \sqrt{2} \cos 2\theta_{V}-\sin 2\theta_{V})(3 D_{R\eta1}(0,0,m_{\eta}^{2})+D_{R\eta2} (4m_{K}^2-m_{\pi }^2))\\
    &&-2D_{R\eta2} (m_{K}^2-m_{\pi }^2) \sin \theta_{P} (\sqrt{2} \sin 2\theta_{V}-4 \cos 2\theta_{V})\bigg\}\,,
\end{eqnarray*}

 \begin{eqnarray*}
   F_{\eta' \rightarrow \gamma\gamma}&=&-\frac{N_{C}(\sin \theta_{P}+2 \sqrt{2} \cos \theta_{P}) }{12 \sqrt{3} \pi ^2 F}-\frac{2 F_{V}(1+8\sqrt{2}\alpha_{V}\frac{m_{\pi}^{2}}{M_{V}^{2}})}{27 FM_{V} M_{\rho }^2} (3 \cos \delta+\sqrt{3} \sin \delta^{\rho}(0) \cos \theta_{V})\\
   &&\bigg\{3 C_{\eta1}(0,0,m_{\eta'}^{2})[\sqrt{3} \cos \delta(\sqrt{2} \sin \theta_{P}+2 \cos \theta_{P})+\sin \delta^{\rho}(0)(-\sqrt{2} \cos \theta_{V} \sin \theta_{P}\\
   &&+2 \cos \theta_{V} \cos \theta_{P}+2 \cos \theta_{V} \sin \theta_{P})]+2C_{R\eta2} \sin \delta^{\rho}(0) m_{K}^2(4 \sin (\theta_{V} +\theta_{P})\\
   &&-3 \sqrt{2} \cos (\theta_{V} -\theta_{P})+\sqrt{2} \cos (\theta_{V} +\theta_{P}))+\frac{C_{R\eta2}}{2} m_{\pi }^2[6 \sqrt{3} \cos \delta(\sqrt{2} \sin \theta_{P}+2 \cos \theta_{P})\\
   &&-\sin \delta^{\rho}(0) (4 \sin (\theta_{V} +\theta_{P})-9 \sqrt{2} \cos (\theta_{V} -\theta_{P})+\sqrt{2} \cos (\theta_{V} +\theta_{P}))]\bigg\}\\
   &&-\frac{2F_{V}(1+8\sqrt{2}\alpha_{V}\frac{m_{\pi}^{2}}{M_{V}^{2}})}{27 FM_{V} M_{\omega }^2}(\sqrt{3} \cos \delta \cos \theta_{V}-3 \sin \delta^{\omega}(0))\bigg\{3C_{R\eta1}(0,0,m_{\eta'}^{2}) [2 \cos \theta_{P}\\
   &&(\cos \delta \cos \theta_{V}-\sqrt{3} \sin \delta^{\omega}(0))+\sin \theta_{P}(2 \cos \delta \cos \theta_{V}-\sqrt{2} \cos \delta \cos \theta_{V}-\sqrt{6} \sin \delta^{\omega}(0))]\\
   &&+2C_{R\eta2} \cos \delta m_{K}^2(4 \sin (\theta_{V} +\theta_{P})-3 \sqrt{2} \cos (\theta_{V} -\theta_{P})+\sqrt{2} \cos (\theta_{V} +\theta_{P}))\\
   &&+\frac{C_{R\eta2}}{2} m_{\pi }^2[-6 \sqrt{3} \sin \delta^{\omega}(0) (\sqrt{2} \sin \theta_{P}+2 \cos \theta_{P})-\cos \delta(4 \sin (\theta_{V} +\theta_{P})\\
   &&-9 \sqrt{2} \cos (\theta_{V} -\theta_{P})+\sqrt{2} \cos (\theta_{V} +\theta_{P}))]\bigg\}\\
   &&+\frac{2F_{V}(1+8\sqrt{2}\alpha_{V}\frac{2m_{K}^{2}-m_{\pi}^{2}}{M_{V}^{2}})}{9 \sqrt{3} FM_{V} M_{\phi }^2} \cos \theta_{V} \bigg\{3C_{R\eta1}(0,0,m_{\eta'}^{2})(2 \cos \theta_{V} \sin \theta_{P}+\cos \theta_{V}\\
   &&(\sqrt{2} \sin \theta_{P}-2 \cos \theta_{P}))-4C_{R\eta2} m_{K}^2(2 \cos (\theta_{V} +\theta_{P})+\sqrt{2} (\cos \theta_{V} \cos \theta_{P}-2 \cos \theta_{V}\\
   && \sin \theta_{P}))+C_{R\eta2} m_{\pi }^2(2 \cos (\theta_{V} +\theta_{P})+\sqrt{2} (4 \cos \theta_{V} \cos \theta_{P}-5 \cos \theta_{V} \sin \theta_{P}))\bigg\}\\
   &&+\frac{4 F_{V}^2(1+8\sqrt{2}\alpha_{V}\frac{m_{\pi}^{2}}{M_{V}^{2}})^{2}}{27 \sqrt{3} F M_{\rho }^4} (3 \cos \delta+\sqrt{3} \sin \delta^{\rho}(0) \cos \theta_{V})^2\bigg\{3 D_{R\eta1}(0,0,m_{\eta'}^{2})\\
   &&[\cos ^2 \delta (\sin \theta_{P}+\sqrt{2} \cos \theta_{P})+\sin ^2 \delta^{\rho}(0)(\cos \theta_{V} \sin \theta_{P}(2 \sqrt{2} \cos \theta_{V}-\cos \theta_{V})\\
   &&+\sqrt{2} \cos \theta_{P})]-D_{R\eta2} \sin ^2\delta^{\rho}(0) m_{K}^2[\cos \theta_{P}(4 \sin 2\theta_{V}+\sqrt{2} \cos 2\theta_{V}-3 \sqrt{2})\\
   &&-2 \sin \theta_{P}(2 \sqrt{2} \sin 2\theta_{V}+\cos 2\theta_{V}-3)]+\frac{D_{R\eta2}}{2} m_{\pi }^2[6 \cos ^2 \delta (\sin \theta_{P}+\sqrt{2} \cos \theta_{P})\\
   &&+\sin ^2\delta^{\rho}(0)(2 \sqrt{2} \cos (2 \theta_{V} +\theta_{P})+8 \sin 2\theta_{V} \cos \theta_{P}-(\cos 2\theta_{V}-9) \sin \theta_{P})]\bigg\}\\
   &&+\frac{4 F_{V}^2(1+8\sqrt{2}\alpha_{V}\frac{m_{\pi}^{2}}{M_{V}^{2}})^{2}}{27 \sqrt{3} F M_{\omega }^4} (\sqrt{3} \cos \delta \cos \theta_{V}-3 \sin\delta^{\omega}(0))^2\bigg\{3D_{R\eta1}(0,0,m_{\eta'}^{2})[\cos ^2\delta \\
   &&(\cos \theta_{V} \sin \theta_{P}(2 \sqrt{2} \cos \theta_{V}-\cos \theta_{V})+\sqrt{2} \cos \theta_{P})+\sin ^2\delta^{\omega}(0) (\sin \theta_{P}+\sqrt{2} \cos \theta_{P})]\\
   &&-D_{R\eta2} \cos ^2 \delta m_{K}^2[\cos \theta_{P}(4 \sin 2\theta_{V}+\sqrt{2} \cos 2\theta_{V}-3 \sqrt{2})\\
   &&-2 \sin \theta_{P}(2 \sqrt{2} \sin 2\theta_{V}+\cos 2\theta_{V}-3)]+\frac{D_{R\eta2}}{2} m_{\pi }^2[\cos ^2 \delta (2 \sqrt{2} \cos (2 \theta_{V} \\
   &&+\theta_{P})+8 \sin 2\theta_{V} \cos \theta_{P}-(\cos 2\theta_{V}-9) \sin \theta_{P})+6 \sin ^2\delta^{\omega}(0)(\sin \theta_{P}+\sqrt{2} \cos \theta_{P})]\bigg\}\\
   &&+\frac{4 F_{V}^2(1+8\sqrt{2}\alpha_{V}\frac{2m_{K}^{2}-m_{\pi}^{2}}{M_{V}^{2}})^{2}}{9 \sqrt{3} F M_{\phi }^4} \cos ^2\theta_{V} \bigg\{3 D_{R\eta1}(0,0,m_{\eta'}^{2})(\sqrt{2} \cos \theta_{P}-\cos \theta_{V} \sin \theta_{P}\\
   &&(2 \sqrt{2} \cos \theta_{V}+\cos \theta_{V}))+D_{R\eta 2} m_{K}^2(\cos \theta_{P}(4 \sin 2\theta_{V}+\sqrt{2} \cos 2\theta_{V}+3 \sqrt{2})\\
   &&-2 \sin \theta_{P}(2 \sqrt{2} \sin 2\theta_{V}+\cos 2\theta_{V}+3))+\frac{D_{R\eta2}}{2} m_{\pi }^2(-2 \sqrt{2} \cos (2 \theta_{V} +\theta_{P})\\
   &&-8 \sin 2\theta_{V} \cos \theta_{P}+(\cos 2\theta_{V}+9) \sin \theta_{P})\bigg\}\\
   &&+\frac{4 F_{V}^2(1+8\sqrt{2}\alpha_{V}\frac{m_{\pi}^{2}}{M_{V}^{2}})^{2}}{27 \sqrt{3} F M_{\rho }^2 M_{\omega }^2} \cos \delta (\sqrt{3} \cos \delta \cos \theta_{V}-3 \sin \delta^{\omega}(0))(3 \cos \delta+\sqrt{3} \sin \delta^{\rho}(0) \cos \theta_{V})\\
   &&\bigg\{3 D_{R\eta1}(0,0,m_{\eta'}^{2})(\sin \theta_{P}(-\sin \delta^{\omega}(0)+\sin \delta^{\rho}(0) \cos \theta_{V}(2 \sqrt{2} \cos \theta_{V}-\cos \theta_{V}))+\sqrt{2} \cos \theta_{P}\\
   &&(-\sin \delta^{\omega}(0)+\sin\delta^{\rho}(0)))-D_{R\eta2} \sin \delta^{\rho}(0) m_{K}^2(\cos \theta_{P}(4 \sin 2\theta_{V}+\sqrt{2} \cos 2\theta_{V}-3 \sqrt{2})\\
   &&-2 \sin \theta_{P}(2 \sqrt{2} \sin 2\theta_{V}+\cos 2\theta_{V}-3))+\frac{D_{R\eta2}}{2} m_{\pi }^2(\cos \theta_{P}(-6 \sqrt{2} \sin \delta^{\omega}(0)\\
   &&+8 \sin \delta^{\rho}(0) \sin 2\theta_{V})-6 \sin\delta^{\omega}(0) \sin \theta_{P}+\sin \delta^{\rho}(0)(2 \sqrt{2} \cos (2 \theta_{V} +\theta_{P})\\
   &&-(\cos 2\theta_{V}-9) \sin \theta_{P}))\bigg\}\\
   &&+\frac{2 F_{V}^2(1+8\sqrt{2}\alpha_{V}\frac{m_{\pi}^{2}}{M_{V}^{2}})(1+8\sqrt{2}\alpha_{V}\frac{2m_{K}^{2}-m_{\pi}^{2}}{M_{V}^{2}})}{27 F M_{\rho }^2 M_{\phi }^2}\sin \delta^{\rho}(0) \cos \theta_{V} (3 \cos \delta\\
   &&+\sqrt{3} \sin \delta^{\rho}(0) \sin \theta_{V})\bigg\{\sin \theta_{P} (2 \sqrt{2} \cos 2\theta_{V}-\sin 2\theta_{V}) (3 D_{R\eta1}(0,0,m_{\eta'}^{2}) \\
   &&+D_{R\eta2}(4 m_{K}^2-m_{\pi }^2))+2D_{R\eta2}(m_{K}^2-m_{\pi }^2)\cos \theta_{P} (\sqrt{2} \sin 2\theta_{V}-4 \cos 2\theta_{V})\bigg\}\\
   &&+\frac{2F_{V}^2(1+8\sqrt{2}\alpha_{V}\frac{m_{\pi}^{2}}{M_{V}^{2}})(1+8\sqrt{2}\alpha_{V}\frac{2m_{K}^{2}-m_{\pi}^{2}}{M_{V}^{2}})}{27 F M_{\omega }^2 M_{\phi }^2} \cos \delta \cos \theta_{V} (\sqrt{3} \cos \delta \cos \theta_{V}\\
   &&-3 \sin \delta^{\omega}(0))\bigg\{\sin \theta_{P}(2 \sqrt{2} \cos 2\theta_{V}-\sin 2\theta_{V})(3 D_{R\eta1}(0,0,m_{\eta'}^{2})+D_{R\eta2}\\
   &&(4 m_{K}^2-m_{\pi }^2))+2D_{R\eta 2}(m_{K}^2-m_{\pi }^2) \cos \theta_{P}(\sqrt{2} \sin 2\theta_{V}-4 \cos 2\theta_{V})\bigg\}\,,
\end{eqnarray*}
and the two-photon decay widths of the $\eta$, $\eta'$ are
\begin{eqnarray*}
\Gamma_{P \rightarrow \gamma\gamma}=\frac{1}{4}\pi^{2}\alpha^{2}m_{P}^{3}\vert F \vert^{2} \,.
\end{eqnarray*}



\bibliographystyle{unsrt}
\bibliography{ref}

\begin{thebibliography}{100}

\bibitem{Weinberg:1978kz}
Steven Weinberg.
\newblock {Phenomenological Lagrangians}.
\newblock {\em Physica A}, 96(1-2):327--340, 1979.

\bibitem{Gasser:1983yg}
J.~Gasser and H.~Leutwyler.
\newblock {Chiral Perturbation Theory to One Loop}.
\newblock {\em Annals Phys.}, 158:142, 1984.

\bibitem{Ecker:1988te}
G.~Ecker, J.~Gasser, A.~Pich, and E.~de~Rafael.
\newblock {The Role of Resonances in Chiral Perturbation Theory}.
\newblock {\em Nucl. Phys. B}, 321:311--342, 1989.

\bibitem{Ecker:1989yg}
G.~Ecker, J.~Gasser, H.~Leutwyler, A.~Pich, and E.~de~Rafael.
\newblock {Chiral Lagrangians for Massive Spin 1 Fields}.
\newblock {\em Phys. Lett. B}, 223:425--432, 1989.

\bibitem{Cirigliano:2006hb}
V.~Cirigliano, G.~Ecker, M.~Eidemuller, Roland Kaiser, A.~Pich, and
  J.~Portoles.
\newblock {Towards a consistent estimate of the chiral low-energy constants}.
\newblock {\em Nucl. Phys. B}, 753:139--177, 2006.

\bibitem{Kampf:2006yf}
Karol Kampf, Jiri Novotny, and Jaroslav Trnka.
\newblock {On different lagrangian formalisms for vector resonances within
  chiral perturbation theory}.
\newblock {\em Eur. Phys. J. C}, 50:385--403, 2007.

\bibitem{Portoles:2010yt}
J.~Portoles.
\newblock {Basics of Resonance Chiral Theory}.
\newblock {\em AIP Conf. Proc.}, 1322(1):178--187, 2010.

\bibitem{Kampf:2011ty}
Karol Kampf and Jiri Novotny.
\newblock {Resonance saturation in the odd-intrinsic parity sector of
  low-energy QCD}.
\newblock {\em Phys. Rev. D}, 84:014036, 2011.

\bibitem{Charpak:1962zz}
G.~Charpak, F.~J.~M. Farley, and R.~L. Garwin.
\newblock {A New Measurement of the Anomalous Magnetic Moment of the Muon}.
\newblock {\em Phys. Lett.}, 1:16, 1962.

\bibitem{Bailey:1968rxd}
J.~Bailey, W.~Bartl, G.~Von~Bochmann, R.~C.~A. Brown, F.~J.~M. Farley,
  H.~Joestlein, E.~Picasso, and R.~W. Williams.
\newblock {Precision measurement of the anomalous magnetic moment of the muon}.
\newblock {\em Phys. Lett. B}, 28:287--290, 1968.

\bibitem{BAILEY19791}
J.~Bailey, K.~Borer, F.~Combley, H.~Drumm, C.~Eck, F.J.M. Farley, J.H. Field,
  W.~Flegel, P.M. Hattersley, F.~Krienen, F.~Lange, G.~Lebée, E.~McMillan,
  G.~Petrucci, E.~Picasso, O.~Rúnolfsson, W.~{von Rüden}, R.W. Williams, and
  S.~Wojcicki.
\newblock Final report on the cern muon storage ring including the anomalous
  magnetic moment and the electric dipole moment of the muon, and a direct test
  of relativistic time dilation.
\newblock {\em Nuclear Physics B}, 150:1--75, 1979.

\bibitem{Muong-2:2006rrc}
G.~W. Bennett et~al.
\newblock {Final Report of the Muon E821 Anomalous Magnetic Moment Measurement
  at BNL}.
\newblock {\em Phys. Rev. D}, 73:072003, 2006.

\bibitem{Muong-2:2021ojo}
B.~Abi et~al.
\newblock {Measurement of the Positive Muon Anomalous Magnetic Moment to 0.46
  ppm}.
\newblock {\em Phys. Rev. Lett.}, 126(14):141801, 2021.

\bibitem{Aoyama:2020ynm}
T.~Aoyama et~al.
\newblock {The anomalous magnetic moment of the muon in the Standard Model}.
\newblock {\em Phys. Rept.}, 887:1--166, 2020.

\bibitem{Colangelo:2018mtw}
Gilberto Colangelo, Martin Hoferichter, and Peter Stoffer.
\newblock {Two-pion contribution to hadronic vacuum polarization}.
\newblock {\em JHEP}, 02:006, 2019.

\bibitem{Davier:2019can}
M.~Davier, A.~Hoecker, B.~Malaescu, and Z.~Zhang.
\newblock {A new evaluation of the hadronic vacuum polarisation contributions
  to the muon anomalous magnetic moment and to
  $\mathbf{\boldsymbol\alpha(m_Z^2)}$}.
\newblock {\em Eur. Phys. J. C}, 80(3):241, 2020.
\newblock [Erratum: Eur.Phys.J.C 80, 410 (2020)].

\bibitem{Keshavarzi:2019abf}
Alexander Keshavarzi, Daisuke Nomura, and Thomas Teubner.
\newblock {$g-2$ of charged leptons, $\alpha (M^2_Z)$ , and the hyperfine
  splitting of muonium}.
\newblock {\em Phys. Rev. D}, 101(1):014029, 2020.

\bibitem{Ce:2022kxy}
Marco C\`e et~al.
\newblock {Window observable for the hadronic vacuum polarization contribution
  to the muon g-2 from lattice QCD}.
\newblock {\em Phys. Rev. D}, 106(11):114502, 2022.

\bibitem{Borsanyi:2020mff}
Sz. Borsanyi et~al.
\newblock {Leading hadronic contribution to the muon magnetic moment from
  lattice QCD}.
\newblock {\em Nature}, 593(7857):51--55, 2021.

\bibitem{Alexandrou:2022amy}
C.~Alexandrou et~al.
\newblock {Lattice calculation of the short and intermediate time-distance
  hadronic vacuum polarization contributions to the muon magnetic moment using
  twisted-mass fermions, arxiv: 2206.15084 [hep-lat]}.

\bibitem{Miranda:2020wdg}
J.~A. Miranda and P.~Roig.
\newblock {New $\tau$-based evaluation of the hadronic contribution to the
  vacuum polarization piece of the muon anomalous magnetic moment}.
\newblock {\em Phys. Rev. D}, 102:114017, 2020.

\bibitem{Hoid:2020xjs}
Bai-Long Hoid, Martin Hoferichter, and Bastian Kubis.
\newblock {Hadronic vacuum polarization and vector-meson resonance parameters
  from $e^+e^-\rightarrow \pi ^0\gamma$}.
\newblock {\em Eur. Phys. J. C}, 80(10):988, 2020.

\bibitem{Benayoun:2021ody}
Maurice Benayoun, Luigi DelBuono, and Friedrich Jegerlehner.
\newblock {BHLS$_2$ upgrade: $\tau $ spectra, muon HVP and the [$\pi ^0,~\eta
  ,~{\eta ^\prime }$] system}.
\newblock {\em Eur. Phys. J. C}, 82(2):184, 2022.

\bibitem{Yi:2021ccc}
Jing-Yu Yi, Zhong-Yu Wang, and C.~W. Xiao.
\newblock {Study of the pion vector form factor and its contribution to the
  muon g-2}.
\newblock {\em Phys. Rev. D}, 104(11):116017, 2021.

\bibitem{Hoferichter:2021wyj}
Martin Hoferichter and Thomas Teubner.
\newblock {Mixed Leptonic and Hadronic Corrections to the Anomalous Magnetic
  Moment of the Muon}.
\newblock {\em Phys. Rev. Lett.}, 128(11):112002, 2022.

\bibitem{Colangelo:2022jxc}
G.~Colangelo et~al.
\newblock {Prospects for precise predictions of $a_\mu$ in the Standard Model,
  arxiv:2203.15810 [hep-ph] }.

\bibitem{Qin:2020udp}
Wen Qin, Ling-Yun Dai, and Jorge Portoles.
\newblock {Two and three pseudoscalar production in $e^+e^-$ annihilation and
  their contributions to $(g-2)_\mu$}.
\newblock {\em JHEP}, 03:092, 2021.

\bibitem{Dai:2013joa}
L.~Y. Dai, J.~Portoles, and O.~Shekhovtsova.
\newblock {Three pseudoscalar meson production in $e^+ e^-$ annihilation}.
\newblock {\em Phys. Rev. D}, 88:056001, 2013.

\bibitem{Benayoun:2012wc}
M.~Benayoun, P.~David, L.~DelBuono, and F.~Jegerlehner.
\newblock {An Update of the HLS Estimate of the Muon g-2}.
\newblock {\em Eur. Phys. J. C}, 73:2453, 2013.

\bibitem{Benayoun:2015gxa}
M.~Benayoun, P.~David, L.~DelBuono, and F.~Jegerlehner.
\newblock {Muon $g-2$ estimates: can one trust effective Lagrangians and global
  fits?}
\newblock {\em Eur. Phys. J. C}, 75(12):613, 2015.

\bibitem{Rosell:2004mn}
I.~Rosell, J.~J. Sanz-Cillero, and A.~Pich.
\newblock {Quantum loops in the resonance chiral theory: The Vector
  form-factor}.
\newblock {\em JHEP}, 08:042, 2004.

\bibitem{Masjuan:2008fv}
P.~Masjuan, S.~Peris, and J.~J. Sanz-Cillero.
\newblock {Vector Meson Dominance as a first step in a systematic
  approximation: The Pion vector form-factor}.
\newblock {\em Phys. Rev. D}, 78:074028, 2008.

\bibitem{Gasser:1982ap}
J.~Gasser and H.~Leutwyler.
\newblock {Quark Masses}.
\newblock {\em Phys. Rept.}, 87:77--169, 1982.

\bibitem{Akhmetshin:2002vj}
R.~R. Akhmetshin et~al.
\newblock {Study of the process $e^+ e^- \to K^0_L K^0_S$ in the CM energy
  range 1.05-GeV to 1.38-GeV with CMD-2}.
\newblock {\em Phys. Lett. B}, 551:27--34, 2003.

\bibitem{Achasov:2006bv}
M.~N. Achasov et~al.
\newblock {Experimental study of the reaction $e^+ e^- \to K_S K_L$ in the
  energy range s**(1/2) = 1.04-GeV divided by 1.38-GeV}.
\newblock {\em J. Exp. Theor. Phys.}, 103(5):720--727, 2006.

\bibitem{Ivanov:1982cr}
P.~m. Ivanov, L.~m. Kurdadze, M.~yu. Lelchuk, E.~v. Pakhtusova, V.~a. Sidorov,
  A.~n. Skrinsky, A.~g. Chilingarov, Yu.~m. Shatunov, B.~a. Shvarts, and S.~i.
  Eidelman.
\newblock {MEASUREMENTS OF THE FORM-FACTOR OF THE NEUTRAL KAON FROM 1.06-GEV TO
  1.40-GEV}.
\newblock {\em JETP Lett.}, 36:112--115, 1982.

\bibitem{Mane:1980ep}
F.~Mane, D.~Bisello, J.~C. Bizot, J.~Buon, A.~Cordier, and B.~Delcourt.
\newblock {Study of the Reaction $e^+ e^- \to K^0_S K^0_L$ in the Total Energy
  Range 1.4-{GeV} to 2.18-{GeV} and Interpretation of the $K^+$ and $K^0$
  Form-factors}.
\newblock {\em Phys. Lett. B}, 99:261--264, 1981.

\bibitem{CMD-2:2003gqi}
R.~R. Akhmetshin et~al.
\newblock {Reanalysis of hadronic cross-section measurements at CMD-2}.
\newblock {\em Phys. Lett. B}, 578:285--289, 2004.

\bibitem{Achasov:2000am}
M.~N. Achasov et~al.
\newblock {Measurements of the parameters of the $\phi(1020)$ resonance through
  studies of the processes $e^+ e^- \to K^+ K^-$, $K_SK_L$, and $\pi^+ \pi^-
  \pi^0$}.
\newblock {\em Phys. Rev. D}, 63:072002, 2001.

\bibitem{CMD-3:2016nhy}
E.~A. Kozyrev et~al.
\newblock {Study of the process $e^+ e^- \to K^0_{S}K^0_{L}$ in the
  center-of-mass energy range 1004--1060 MeV with the CMD-3 detector at the
  VEPP-2000 $e^+ e^-$ collider}.
\newblock {\em Phys. Lett. B}, 760:314--319, 2016.

\bibitem{BaBar:2014uwz}
J.~P. Lees et~al.
\newblock {Cross sections for the reactions $e^+ e^-\to K_S^0 K_L^0$, $K_S^0
  K_L^0 \pi^+\pi^-$, $K_S^0 K_S^0 \pi^+\pi^-$, and $K_S^0 K_S^0 K^+K^-$ from
  events with initial-state radiation}.
\newblock {\em Phys. Rev. D}, 89(9):092002, 2014.

\bibitem{BESIII:2021yam}
Medina Ablikim et~al.
\newblock {Cross section measurement of $e^{+}e^{-} \to K_{S}^{0}K_{L}^{0}$ at
  $\sqrt{s}=2.00-3.08~{GeV}$}.
\newblock {\em Phys. Rev. D}, 104(9):092014, 2021.

\bibitem{Achasov:2000zd}
M.~N. Achasov et~al.
\newblock {Experimental study of the processes $e^+ e^- \to \phi \to \eta
  \gamma, \pi^0 \gamma$ at VEPP-2M}.
\newblock {\em Eur. Phys. J. C}, 12:25--33, 2000.

\bibitem{Achasov:2003ed}
M.~N. Achasov et~al.
\newblock {Experimental study of the $e^+ e^- \to \pi^0 \gamma$ process in the
  energy region s**1/2 = 0.60-GeV - 0.97-GeV}.
\newblock {\em Phys. Lett. B}, 559:171--178, 2003.

\bibitem{SND:2016drm}
M.~N. Achasov et~al.
\newblock {Study of the reaction $e^+e^- \to \pi^0\gamma$ with the SND detector
  at the VEPP-2M collider}.
\newblock {\em Phys. Rev. D}, 93(9):092001, 2016.

\bibitem{Achasov:2018ujw}
M.~N. Achasov et~al.
\newblock {Measurement of the $e^+e^- \to \pi^0\gamma$ cross section in the
  energy range 1.075-2 GeV at SND}.
\newblock {\em Phys. Rev. D}, 98(11):112001, 2018.

\bibitem{CMD-2:2004ahv}
R.~R. Akhmetshin et~al.
\newblock {Study of the processes $e^+ e^- \to \eta \gamma, \pi^0 \gamma \to 3
  \gamma$ in the c.m. energy range 600-MeV to 1380-MeV at CMD-2}.
\newblock {\em Phys. Lett. B}, 605:26--36, 2005.

\bibitem{CMD-2:2001dnv}
R.~R. Akhmetshin et~al.
\newblock {Study of the process $e^+ e^- \to \eta \gamma$ in center-of-mass
  energy range 600-MeV to 1380-MeV at CMD-2}.
\newblock {\em Phys. Lett. B}, 509:217--226, 2001.

\bibitem{Achasov:2006dv}
M.~N. Achasov et~al.
\newblock {Study of the $e^+ e^- \to \eta \gamma$ process with SND detector at
  the VEPP-2M e+ e- collider}.
\newblock {\em Phys. Rev. D}, 74:014016, 2006.

\bibitem{Achasov:2013eli}
M.~N. Achasov et~al.
\newblock {Study of the process $e^+e^-\to\eta\gamma$ in the center-of-mass
  energy range 1.07--2.00 GeV}.
\newblock {\em Phys. Rev. D}, 90(3):032002, 2014.

\bibitem{Zyla:2020zbs}
P.~A. Zyla et~al.
\newblock {Review of Particle Physics}.
\newblock {\em PTEP}, 2020(8):083C01, 2020.

\bibitem{Leutwyler:1997yr}
H.~Leutwyler.
\newblock {On the 1/N expansion in chiral perturbation theory}.
\newblock {\em Nucl. Phys. B Proc. Suppl.}, 64:223--231, 1998.

\bibitem{Kaiser:1998ds}
Roland Kaiser and H.~Leutwyler.
\newblock {Pseudoscalar decay constants at large N(c)}.
\newblock In {\em {Workshop on Methods of Nonperturbative Quantum Field
  Theory}}, pages 15--29, 6 1998.

\bibitem{Guo:2015xva}
Xu-Kun Guo, Zhi-Hui Guo, Jose~Antonio Oller, and Juan~Jose Sanz-Cillero.
\newblock {Scrutinizing the $\eta$-$\eta'$ mixing, masses and pseudoscalar
  decay constants in the framework of U(3) chiral effective field theory}.
\newblock {\em JHEP}, 06:175, 2015.

\bibitem{Gao:2022xqz}
Rui Gao, Zhi-Hui Guo, J.~A. Oller, and Hai-Qing Zhou.
\newblock {Axion-meson mixing in light of recent lattice $\eta$-$\eta'$
  simulations and their two-photon couplings within $U(3)$ chiral theory,
  arxiv: 2211.02867[hep-ph]}.

\bibitem{Arteaga:2022xxy}
Saray Arteaga, Ling-Yun Dai, Adolfo Guevara, and Pablo Roig.
\newblock {Tension between $e^+e^-\to\eta\pi^-\pi^+$ and
  $\tau\to\eta\pi^-\pi^0\nu_\tau$ data and nonstandard interactions}.
\newblock {\em Phys. Rev. D}, 106(9):096016, 2022.

\bibitem{Dai:2017tew}
Ling-Yun Dai, Xian-Wei Kang, Ulf-G. Mei\ss{}ner, Xin-Ying Song, and De-Liang
  Yao.
\newblock {Amplitude analysis of the anomalous decay
  $\eta'\to\pi^+\pi^-\gamma$}.
\newblock {\em Phys. Rev. D}, 97(3):036012, 2018.

\bibitem{Scherer:2002tk}
Stefan Scherer.
\newblock {Introduction to chiral perturbation theory}.
\newblock {\em Adv. Nucl. Phys.}, 27:277, 2003.

\bibitem{Dai:2019lmj}
Ling-Yun Dai, Javier Fuentes-Mart\'\i{}n, and Jorge Portol\'es.
\newblock {Scalar-involved three-point Green functions and their
  phenomenology}.
\newblock {\em Phys. Rev. D}, 99(11):114015, 2019.

\bibitem{Wess:1971yu}
J.~Wess and B.~Zumino.
\newblock {Consequences of anomalous Ward identities}.
\newblock {\em Phys. Lett. B}, 37:95--97, 1971.

\bibitem{Witten:1983tw}
Edward Witten.
\newblock {Global Aspects of Current Algebra}.
\newblock {\em Nucl. Phys. B}, 223:422--432, 1983.

\bibitem{Hoferichter:2014vra}
Martin Hoferichter, Bastian Kubis, Stefan Leupold, Franz Niecknig, and
  Sebastian~P. Schneider.
\newblock {Dispersive analysis of the pion transition form factor}.
\newblock {\em Eur. Phys. J. C}, 74:3180, 2014.

\bibitem{Ruiz-Femenia:2003jdx}
P.~D. Ruiz-Femenia, A.~Pich, and J.~Portoles.
\newblock {Odd intrinsic parity processes within the resonance effective theory
  of QCD}.
\newblock {\em JHEP}, 07:003, 2003.

\bibitem{Chen:2012vw}
Yun-Hua Chen, Zhi-Hui Guo, and Han-Qing Zheng.
\newblock {Study of $\eta$-$\eta'$ mixing from radiative decay processes}.
\newblock {\em Phys. Rev. D}, 85:054018, 2012.

\bibitem{ParticleDataGroup:2020ssz}
P.~A. Zyla et~al.
\newblock {Review of Particle Physics}.
\newblock {\em PTEP}, 2020(8):083C01, 2020.

\bibitem{Cirigliano:2003yq}
V.~Cirigliano, G.~Ecker, H.~Neufeld, and A.~Pich.
\newblock {Meson resonances, large N(c) and chiral symmetry}.
\newblock {\em JHEP}, 06:012, 2003.

\bibitem{Guo:2009hi}
Zhi-Hui Guo and Juan~Jose Sanz-Cillero.
\newblock {pi pi-scattering lengths at O(p**6) revisited}.
\newblock {\em Phys. Rev. D}, 79:096006, 2009.

\bibitem{Niecknig:2012sj}
Franz Niecknig, Bastian Kubis, and Sebastian~P. Schneider.
\newblock {Dispersive analysis of $\omega \to 3\pi$ and $\phi \to 3\pi$
  decays}.
\newblock {\em Eur. Phys. J. C}, 72:2014, 2012.

\bibitem{Schneider:2012ez}
Sebastian~P. Schneider, Bastian Kubis, and Franz Niecknig.
\newblock {The $\omega \to \pi^0 \gamma^*$ and $\phi \to \pi^0 \gamma^*$
  transition form factors in dispersion theory}.
\newblock {\em Phys. Rev. D}, 86:054013, 2012.

\bibitem{Danilkin:2014cra}
I.~V. Danilkin, C.~Fern\'andez-Ram\'\i{}rez, P.~Guo, V.~Mathieu, D.~Schott,
  M.~Shi, and A.~P. Szczepaniak.
\newblock {Dispersive analysis of $\omega/\phi\rightarrow 3\pi,\pi\gamma*$}.
\newblock {\em Phys. Rev. D}, 91(9):094029, 2015.

\bibitem{Albaladejo:2017hhj}
M.~Albaladejo and B.~Moussallam.
\newblock {Extended chiral Khuri-Treiman formalism for $\eta\to 3\pi$ and the
  role of the $a_0(980)$, $f_0(980)$ resonances}.
\newblock {\em Eur. Phys. J. C}, 77(8):508, 2017.

\bibitem{Isken:2017dkw}
Tobias Isken, Bastian Kubis, Sebastian~P. Schneider, and Peter Stoffer.
\newblock {Dispersion relations for $\eta '\rightarrow \eta \pi \pi $}.
\newblock {\em Eur. Phys. J. C}, 77(7):489, 2017.

\bibitem{Colangelo:2018jxw}
Gilberto Colangelo, Stefan Lanz, Heinrich Leutwyler, and Emilie Passemar.
\newblock {Dispersive analysis of $\eta \rightarrow 3 \pi $}.
\newblock {\em Eur. Phys. J. C}, 78(11):947, 2018.

\bibitem{Yao:2020bxx}
De-Liang Yao, Ling-Yun Dai, Han-Qing Zheng, and Zhi-Yong Zhou.
\newblock {A review on partial-wave dynamics with chiral effective field theory
  and dispersion relation}.
\newblock {\em Rept. Prog. Phys.}, 84(7):076201, 2021.

\bibitem{Omnes:1958hv}
R.~Omnes.
\newblock {On the Solution of certain singular integral equations of quantum
  field theory}.
\newblock {\em Nuovo Cim.}, 8:316--326, 1958.

\bibitem{Guerrero:1997ku}
Francisco Guerrero and Antonio Pich.
\newblock {Effective field theory description of the pion form-factor}.
\newblock {\em Phys. Lett. B}, 412:382--388, 1997.

\bibitem{Efron:1979bxm}
B.~Efron.
\newblock {Bootstrap Methods: Another Look at the Jackknife}.
\newblock {\em Annals Statist.}, 7(1):1--26, 1979.

\bibitem{James:1975dr}
F.~James and M.~Roos.
\newblock {Minuit: A System for Function Minimization and Analysis of the
  Parameter Errors and Correlations}.
\newblock {\em Comput. Phys. Commun.}, 10:343--367, 1975.

\bibitem{GomezDumm:2012dpx}
Daniel Gomez~Dumm and Pablo Roig.
\newblock {Resonance Chiral Lagrangian analysis of $\tau^- \to \eta^{(\prime)}
  \pi^- \pi^0 \nu_\tau$ decays}.
\newblock {\em Phys. Rev. D}, 86:076009, 2012.

\bibitem{BaBar:2013jqz}
J.~P. Lees et~al.
\newblock {Precision measurement of the $e^+e^- \to K^+K^-(\gamma)$ cross
  section with the initial-state radiation method at BABAR}.
\newblock {\em Phys. Rev. D}, 88(3):032013, 2013.

\bibitem{Knecht:2001xc}
Marc Knecht and Andreas Nyffeler.
\newblock {Resonance estimates of O(p**6) low-energy constants and QCD short
  distance constraints}.
\newblock {\em Eur. Phys. J. C}, 21:659--678, 2001.

\bibitem{Nugent:2013hxa}
I.M. Nugent, T.~Przedzinski, P.~Roig, O.~Shekhovtsova, and Z.~Was.
\newblock {Resonance chiral Lagrangian currents and experimental data for
  $\tau^-\to\pi^{-}\pi^{-}\pi^{+}\nu_{\tau}$}.
\newblock {\em Phys. Rev. D}, 88:093012, 2013.

\bibitem{BaBar:2012bdw}
J.~P. Lees et~al.
\newblock {Precise Measurement of the $e^+ e^- \to \pi^+\pi^- (\gamma)$ Cross
  Section with the Initial-State Radiation Method at BABAR}.
\newblock {\em Phys. Rev. D}, 86:032013, 2012.

\bibitem{KLOE:2008fmq}
F.~Ambrosino et~al.
\newblock {Measurement of $\sigma(e^+ e^- \to \pi^+ \pi^- \gamma(\gamma))$ and
  the dipion contribution to the muon anomaly with the KLOE detector}.
\newblock {\em Phys. Lett. B}, 670:285--291, 2009.

\bibitem{KLOE:2010qei}
F.~Ambrosino et~al.
\newblock {Measurement of $\sigma(e^+ e^- \to \pi^+ \pi^-)$ from threshold to
  0.85 GeV$^2$ using Initial State Radiation with the KLOE detector}.
\newblock {\em Phys. Lett. B}, 700:102--110, 2011.

\bibitem{KLOE:2012anl}
D.~Babusci et~al.
\newblock {Precision measurement of $\sigma(e^+e^-\rightarrow
  \pi^+\pi^-\gamma)/ \sigma(e^+e^-\rightarrow \mu^+\mu^-\gamma)$ and
  determination of the $\pi^+\pi^-$ contribution to the muon anomaly with the
  KLOE detector}.
\newblock {\em Phys. Lett. B}, 720:336--343, 2013.

\bibitem{KLOE-2:2017fda}
A.~Anastasi et~al.
\newblock {Combination of KLOE
  $\sigma\big(e^+e^-\rightarrow\pi^+\pi^-\gamma(\gamma)\big)$ measurements and
  determination of $a_{\mu}^{\pi^+\pi^-}$ in the energy range $0.10 < s < 0.95$
  GeV$^2$}.
\newblock {\em JHEP}, 03:173, 2018.

\bibitem{SND:2020nwa}
M.~N. Achasov et~al.
\newblock {Measurement of the $e^+e^- \to\pi^+\pi^- $ process cross section
  with the SND detector at the VEPP-2000 collider in the energy region
  $0.525<\sqrt{s}<0.883$ GeV}.
\newblock {\em JHEP}, 01:113, 2021.

\bibitem{BESIII:2015equ}
M.~Ablikim et~al.
\newblock {Measurement of the $e^+ e^- \to \pi^+ \pi^-$ cross section between
  600 and 900 MeV using initial state radiation}.
\newblock {\em Phys. Lett. B}, 753:629--638, 2016.
\newblock [Erratum: Phys.Lett.B 812, 135982 (2021)].

\bibitem{Xiao:2017dqv}
T.~Xiao, S.~Dobbs, A.~Tomaradze, Kamal~K. Seth, and G.~Bonvicini.
\newblock {Precision Measurement of the Hadronic Contribution to the Muon
  Anomalous Magnetic Moment}.
\newblock {\em Phys. Rev. D}, 97(3):032012, 2018.

\bibitem{CMD-2:2005mvb}
V.~M. Aul'chenko et~al.
\newblock {Measurement of the pion form-factor in the range 1.04-GeV to
  1.38-GeV with the CMD-2 detector}.
\newblock {\em JETP Lett.}, 82:743--747, 2005.

\bibitem{Aulchenko:2006dxz}
V.~M. Aul'chenko et~al.
\newblock {Measurement of the e+ e- ---\ensuremath{>} pi+ pi- cross section
  with the CMD-2 detector in the 370 - 520-MeV c.m. energy range}.
\newblock {\em JETP Lett.}, 84:413--417, 2006.

\bibitem{CMD-2:2006gxt}
R.~R. Akhmetshin et~al.
\newblock {High-statistics measurement of the pion form factor in the rho-meson
  energy range with the CMD-2 detector}.
\newblock {\em Phys. Lett. B}, 648:28--38, 2007.

\bibitem{DM2:1988xqd}
D.~Bisello et~al.
\newblock {The Pion Electromagnetic Form-factor in the Timelike Energy Range
  1.35-{GeV} $\le \sqrt{s} \le$ 2.4-{GeV}}.
\newblock {\em Phys. Lett. B}, 220:321--327, 1989.

\bibitem{Barkov:1985ac}
L.~M. Barkov et~al.
\newblock {Electromagnetic Pion Form-Factor in the Timelike Region}.
\newblock {\em Nucl. Phys. B}, 256:365--384, 1985.

\bibitem{CMD-3:2023alj}
F.~V. Ignatov et~al.
\newblock {Measurement of the $e^+e^-\to\pi^+\pi^-$ cross section from
  threshold to 1.2 GeV with the CMD-3 detector, arxiv: 2302.08834 [hep-ex]}.

\bibitem{Achasov:2007kg}
M.~N. Achasov et~al.
\newblock {Measurement of the $e^+ e^- \to K^+K^-$ process cross-section in the
  energy range s**(1/2) = 1.04 - 1.38 GeV with the SND detector in the
  experiment at VEPP-2M e+e- collider}.
\newblock {\em Phys. Rev. D}, 76:072012, 2007.

\bibitem{Achasov:2016lbc}
M.~N. Achasov et~al.
\newblock {Measurement of the $\mathbf{e^+e^-\to K^+K^-}$ cross section in the
  energy range $\mathbf{\sqrt{s}=1.05-2.0}$ GeV}.
\newblock {\em Phys. Rev. D}, 94(11):112006, 2016.

\bibitem{CMD-2:2008fsu}
R.~R. Akhmetshin et~al.
\newblock {Measurement of $e^+e^- \to \phi \to K^+K^-$ cross section with the
  CMD-2 detector at VEPP-2M Collider}.
\newblock {\em Phys. Lett. B}, 669:217--222, 2008.

\bibitem{Kozyrev:2017agm}
E.~A. Kozyrev et~al.
\newblock {Study of the process $e^+e^- \to K^+K^-$ in the center-of-mass
  energy range 1010--1060\textasciitilde{}MeV with the CMD-3 detector}.
\newblock {\em Phys. Lett. B}, 779:64--71, 2018.

\bibitem{BESIII:2018ldc}
M.~Ablikim et~al.
\newblock {Measurement of $e^{+} e^{-} \rightarrow K^{+} K^{-}$ cross section
  at $\sqrt{s} = 2.00 - 3.08$ GeV}.
\newblock {\em Phys. Rev. D}, 99(3):032001, 2019.

\bibitem{CMD-2:1999chh}
R.~R. Akhmetshin et~al.
\newblock {Measurement of $\phi$ meson parameters in K$^0_L$ K$^0_S$ decay mode
  with CMD-2}.
\newblock {\em Phys. Lett. B}, 466:385, 1999.
\newblock [Erratum: Phys.Lett.B 508, 217--218 (2001)].

\bibitem{Gourdin:1969dm}
M.~Gourdin and E.~De~Rafael.
\newblock {Hadronic contributions to the muon g-factor}.
\newblock {\em Nucl. Phys. B}, 10:667--674, 1969.

\bibitem{Jegerlehner:2017gek}
Friedrich Jegerlehner.
\newblock {\em {The Anomalous Magnetic Moment of the Muon}}, volume 274.
\newblock Springer, Cham, 2017.

\bibitem{Jegerlehner:2009ry}
Fred Jegerlehner and Andreas Nyffeler.
\newblock {The Muon g-2}.
\newblock {\em Phys. Rept.}, 477:1--110, 2009.

\bibitem{Sturm:2013uka}
Christian Sturm.
\newblock {Leptonic contributions to the effective electromagnetic coupling at
  four-loop order in QED}.
\newblock {\em Nucl. Phys. B}, 874:698--719, 2013.

\bibitem{BESIII:2021wib}
M.~Ablikim et~al.
\newblock {Measurement of the Cross Section for $e^{+}e^{-}\to$Hadrons at
  Energies from 2.2324 to 3.6710~GeV}.
\newblock {\em Phys. Rev. Lett.}, 128(6):062004, 2022.

\bibitem{Blum:2023vlm}
Thomas Blum, Norman Christ, Masashi Hayakawa, Taku Izubuchi, Luchang Jin,
  Chulwoo Jung, Christoph Lehner, and Cheng Tu.
\newblock {Hadronic light-by-light contribution to the muon anomaly from
  lattice QCD with infinite volume QED at physical pion mass}.
\newblock 4 2023.

\bibitem{Dai:2014zta}
Ling-Yun Dai and Michael~R. Pennington.
\newblock {Comprehensive amplitude analysis of $\gamma\gamma \rightarrow
  \pi^+\pi^-, \pi^0\pi^0$ and $\overline{K} K$ below 1.5 GeV}.
\newblock {\em Phys. Rev. D}, 90(3):036004, 2014.

\bibitem{Dai:2014lza}
Ling-Yun Dai and M.~R. Pennington.
\newblock {Two photon couplings of the lightest isoscalars from BELLE data}.
\newblock {\em Phys. Lett. B}, 736:11--15, 2014.

\bibitem{Passarino:1978jh}
G.~Passarino and M.~J.~G. Veltman.
\newblock {One Loop Corrections for e+ e- Annihilation Into mu+ mu- in the
  Weinberg Model}.
\newblock {\em Nucl. Phys. B}, 160:151--207, 1979.

\bibitem{Dai:2017ont}
Ling-Yun Dai, Johann Haidenbauer, and Ulf-G Mei\ss{}ner.
\newblock {Antinucleon-nucleon interaction at next-to-next-to-next-to-leading
  order in chiral effective field theory}.
\newblock {\em JHEP}, 07:078, 2017.

\bibitem{Yang:2022qoy}
Qin-He Yang, Ling-Yun Dai, Di~Guo, Johann Haidenbauer, Xian-Wei Kang, and
  Ulf-G. Mei\ss{}ner.
\newblock {New insights into the oscillation of the nucleon electromagnetic
  form factors, arxiv: 2206.01494 [nucl-th]}.

\end{thebibliography}

\end{document}